%% file: ms.tex
\documentclass[12pt,preprint]{emulateapj}
\slugcomment{{\sc Accepted to ApJS:} April 23, 2008} 
\usepackage{lscape}

\newcommand{\mapiii}{MAPPINGS\,\textsc{iii}}
\newcommand{\mapii}{MAPPINGS\,\textsc{ii}}
\newcommand{\kms}{\ifmmode {\rm\,km~s^{-1}} \else ${\rm\,km\,s^{-1}}$\fi}
\slugcomment{}
\shorttitle{The MAPPINGS {\sc III} Shock Model Library}
\shortauthors{Allen et al.}

\begin{document}

\title{The MAPPINGS {\sc III} Library of Fast Radiative Shock Models}

\author{Mark G. Allen }
\affil{Observatoire de Strasbourg UMR 7550 
       Strasbourg 67000, France}
\email{allen@astro.u-strasbg.fr}

\author{Brent A. Groves}
\affil{Sterrewacht Leiden, Leiden University, Neils Bohrweg 2, Leiden 2333-CA
       The Netherlands}
\email{brent@strw.leidenuniv.nl} 

\author{Michael A. Dopita }
\affil{Research School of Astronomy and Astrophysics, 
       Australian National University, Cotter Road , Weston Creek ACT 2611, 
       Australia }

\author{Ralph S. Sutherland }
\affil{Research School of Astronomy and Astrophysics, 
       Australian National University, Cotter Road , Weston Creek ACT 2611, 
       Australia }

\author{Lisa J. Kewley}
\affil{Institute for Astronomy,  University of Hawaii, 2680 Woodlawn Drive
Manoa, HI 96822. USA}

\begin{abstract}
We present a new library of fully-radiative shock models calculated with the
\mapiii\ shock and photoionization code.  The library consists of grids of
models with shock velocities in the range $v_{s}$=100-1000\ km\ s$^{-1}$ and
magnetic parameters B/$\sqrt{\rm n}$ of 10$^{-4}$-10 $\mu$G\ cm$^{3/2}$ for
five different atomic abundance sets, and for a pre-shock density of 1.0\
cm$^{-3}$. Additionally, Solar abundance model grids have been calculated for
densities of 0.01, 0.1, 10, 100, and 1000\ cm$^{-3}$ with the same range in
$v_{s}$ and B/$\sqrt{\rm n}$.  Each model includes
components of both the radiative shock and its photoionized precursor, ionized
by the EUV and soft X-ray radiation generated in the radiative gas.  We
present the details of the ionization structure, the column densities,
and the luminosities of the shock and its precursor. Emission line ratio predictions
are separately given for the shock and its precursor as well as for the
composite shock+precursor structure to facilitate comparison with observations
in cases where the shock and its precursor are not resolved. Emission line
ratio grids for shock and shock+precursor are presented on standard line ratio
diagnostic diagrams, and we compare these grids to observations of radio
galaxies and a sample of AGN and star forming galaxies from the Sloan Digital
Sky Survey.
This library is available online, along with a suite of tools to
enable the analysis of the shocks and the easy creation of emission
line ratio diagnostic diagrams. These models represent a significant increase in
parameter space coverage over previously available models, and
therefore provide a unique tool in the diagnosis of emission by shocks.
\end{abstract}


\section{Introduction}

Supersonic motions are a common phenomenon in our complex and fascinating
Universe. The kinetic energy of such motions will almost inevitably be
eventually dissipated through radiative shocks. Cloud-cloud collisions, the
expansion of \ion{H}{2} regions into the surrounding interstellar medium,
outflows from young stellar objects, supernova blast waves, outflows from
active and starburst galaxies, and collisions between galaxies are all
examples of astrophysical situations in which radiative shock waves provide an
important component of the total energy budget and may determine the line
emission spectrum.

In this paper we do not consider very slow (molecular) shocks or indeed, the
faster atomic shocks for which the theory has been well-developed by such
authors as \citet{Dopita77} or \citet{Raymond79}. Rather, we consider only
fast shocks, where the ionizing radiation generated by the cooling of hot
gas behind the shock front generates a strong radiation field of extreme
ultraviolet and soft X-ray photons, which leads to significant photoionizing
effects. The detailed theory of steady-flow photoionizing (or auto-ionizing)
shocks was developed by \cite{sutherland1993} and \cite{dopita1995,dopita1996}
(hereafter DS95 and DS96). A detailed text-book development of the theory 
of shocks is given in \cite{dopita2003}.

\begin{deluxetable}{lrrrrr}
\tablecolumns{6}
\tablewidth{0pc}
\tablecaption{Abundance Sets (by number wrt Hydrogen) \label{abund_table}}
\tablehead{
\colhead{Element}    &
\colhead{Solar}    &
\colhead{Solar $\times$2}    &
\colhead{dopita2005}    &
\colhead{LMC}    &
\colhead{SMC}    }
\startdata
 H    &    0.00 &   0.00 &   0.00 &    0.00  &  0.00 \\
 He   &   -1.01 &  -1.01 &  -1.01 &   -1.05  & -1.09 \\
 C    &   -3.44 &  -3.14 &  -4.11 &   -3.96  & -4.24 \\
 N    &   -3.95 &  -3.65 &  -4.42 &   -4.86  & -5.37 \\
 O    &   -3.07 &  -2.77 &  -3.56 &   -3.65  & -3.97 \\
 Ne   &   -3.91 &  -3.61 &  -3.91 &   -4.39  & -4.73 \\
 Na   &         &        &  -6.35 &   -4.85  & -5.92 \\
 Mg   &   -4.42 &  -4.12 &  -5.12 &   -4.53  & -5.01 \\
 Al   &   -5.53 &  -5.23 &  -7.31 &   -4.28  & -5.60 \\
 Si   &   -4.45 &  -4.15 &  -5.49 &   -5.29  & -4.69 \\
 S    &   -4.79 &  -4.49 &  -5.01 &   -7.23  & -5.41 \\
 Cl   &         &        &  -6.70 &          & -7.30 \\
 Ar   &   -5.44 &  -5.14 &  -5.44 &   -5.71  & -6.29 \\
 Ca   &   -5.88 &  -5.58 &  -8.16 &   -6.03  & -6.16 \\
 Fe   &   -4.63 &  -4.33 &  -6.55 &   -4.77  & -5.11 \\
 Ni   &         &        &  -7.08 &   -6.04  & -6.14 \\
\enddata 
\end{deluxetable}

In photoionizing shocks, the flux of the ionizing radiation emitted
by the shock increases in approximate proportion to the energy flux through the
shock ($\propto v_{\rm s}^{3}$). The ratio of this flux in advance of
the shock to the
pre-shock density, classified as the ionization parameter, determines the
velocity of the photoionization front that is driven into the pre-shock gas. At
low values of the ionization parameter, the velocity of the photoionization
front is lower than the velocity of the shock and the ionizing photons are
absorbed in the immediate vicinity of the shock front. The effect of this is
to change the ionization state of the gas feeding across the shock front.  As
the velocity of the shock increases, the emitted ionizing flux, and therefore
the velocity of the photoionization front, increases rapidly. At shock
velocities above a certain limit ($v_{\rm s} \approx$ 170 km\ s$^{-1}$), the
ionization front velocity exceeds that of the shock (and is supersonic
with respect to the pre-shock gas) and the photoionization
front detaches from the shock front as an R-Type ionization 
front\footnote{ R-Type ionization fronts are defined as being supersonic with
respect to the gas ahead of the front. (subsonic ionization fronts are
classified as D-Type) see \citet{mckee1980}}.  This front expands to form a
precursor \ion{H}{2} region ahead of the shock. At the highest shock
velocities, the photoionized precursor emission may come to dominate the
optical emission of the shock, and the global radiative shock spectrum
provides a rich mixture of emission lines from both high- and low-ionization
species.

\begin{figure}[t]
\includegraphics[scale=0.5]{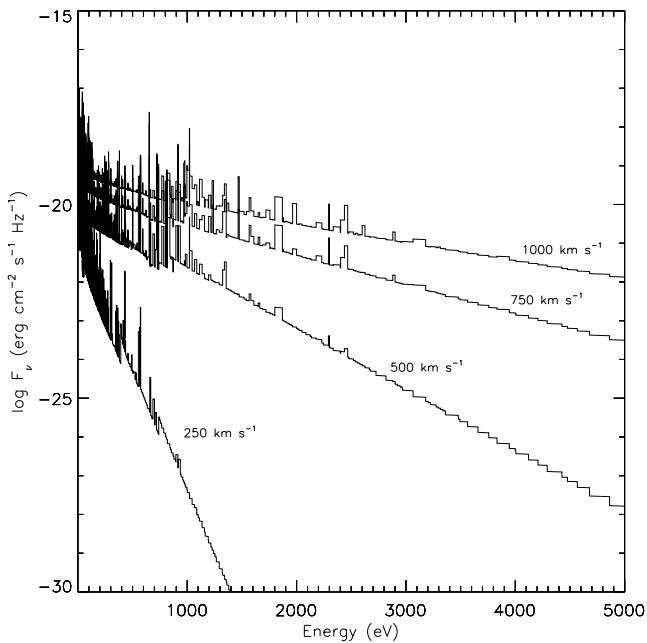}
\caption{Ionizing spectra generated by $v_s$=250, 500, 750 and 1000
km\ s$^{-1}$ shock models 
with n=1.0 cm$^{-3}$ and solar abundance. \label{specfig1}}
\end{figure}

\begin{figure}[t]
\includegraphics[scale=0.5]{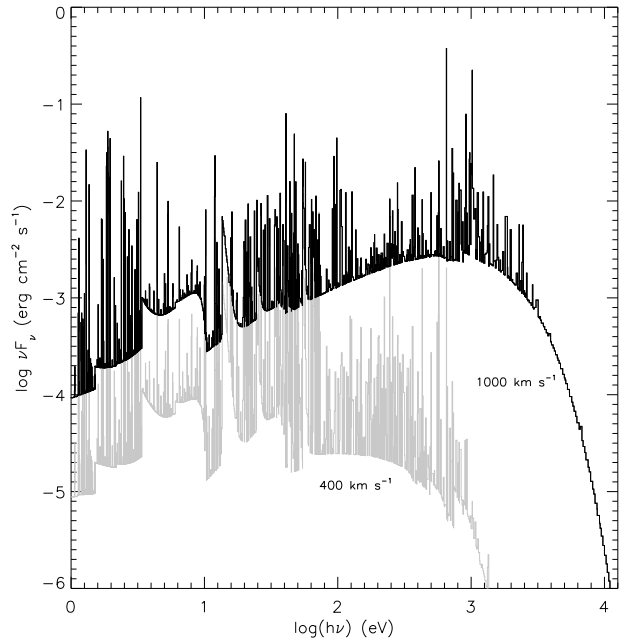}
\caption{Ionizing spectra of $v_s$=400 and 1000  km\ s$^{-1}$ shock
models shown on  $\nu F_{\nu}$ scale.  \label{specfig2}}
\end{figure}

Appreciable photoionization effects are also produced in the gas behind the
shock front, near to the recombination region of the shock. However, in this
shock region the velocity of the ionization front is much slower due to the
gas compression through the shock front and the ionization front stays trapped
in the recombination region to much higher shock velocities.  When magnetic
effects are negligible, the compression of the gas in the shocked region is
proportional to the square of the Mach number in the pre-shock gas, ${\cal M}
= v_{\rm s}/c_{\rm II}$, where $c_{\rm II}$ is the sound speed in the
pre-shock gas. When the magnetic field pressure in the post-shock gas
dominates over the gas pressure, the compression factor is determined by the
Alfv\'en Mach Number, ${\cal M}_{\rm A}$, which is the ratio of the shock
velocity to the Alfv\'en velocity; $v_{\rm A} = ({\rm B}^2 /4\pi \rho)^{1/2}$,
where B is the transverse component of the pre-shock magnetic field, and
$\rho$ is the pre-shock mass density. Thus the presence of magnetic
fields act to limit the compression through the shock. In order to account for
this effect in computations of the shock structure, DS96 developed the concept
of a magnetic parameter; B/$\sqrt{\rm n}$, where n is the pre-shock particle
number density.

\begin{figure*}
\centering
\includegraphics[scale=0.80]{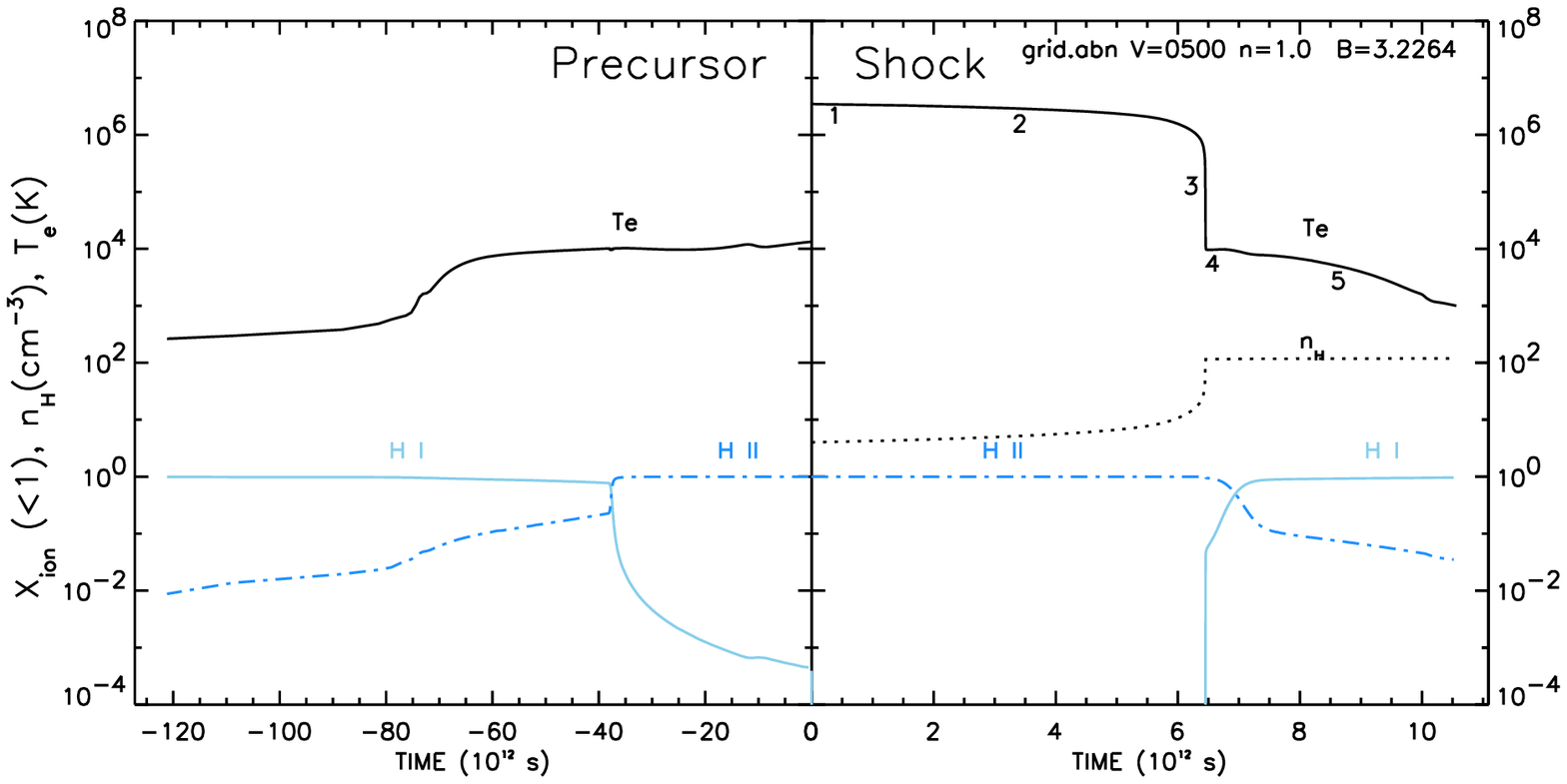}[t]
\caption{The hydrogen ionization structure, temperature profile and density
profile of a shock and precursor.  The vertical axis represents i) the
ionization fraction X$_{ion}$ (values $\le 1$) for the ionization structure
curves, ii) the electron temperature profile (T$_e$) in degrees Kelvin, shown
as the upper line plotted in the figure, and iii) the hydrogen density
(n$_{\rm H}$) in cm$^{-3}$, shown by the dotted line in the shock region
(right) panel of the figure. As n$_{\rm H}$ is constant in the precursor
(left) it is not plotted.  The horizontal axis represents the time since
passage of the shock front, with positive values for the shock structure shown
in the right of the figure, and negative values for the precursor region where
the shock is yet to arrive. The numerical labels indicated on the temperature
profile of the shock correspond to different regions of the shock structure as
described in the text.  The model plotted here is our fiducial model with
solar abundances, equipartition magnetic field, a precursor density of n$_{\rm
H}=1$ cm$^{-3}$, and shock velocity of $v_{\rm s}$=500 km\ s$^{-1}$, as
labelled in the upper right corner of the figure.  See the electronic edition
of the Journal for a color version of this figure.
\label{Hstruct_eg}}
\end{figure*}

The grid of low density photoionizing shock models described in DS96 has
proved to be a valuable resource to the astronomical community for assessing
the role of photoionizing shocks in a range of astrophysical objects. This
utility is increased by their availability in electronic form from the AAS
CD-ROM Series, Vol 7. \citep[see][]{leitherer1996}.  These models have most
often been applied in studies of the narrow line regions (NLR) of active
galaxies, developing the debate on the relative roles of shock and/or
photoionization excitation of the NLR, and the question of the radiative
versus mechanical energy output of active galactic nuclei (AGN). Other uses of
the models have included studies of ionized gas around high velocity clouds
\citep{fox2004}, and the intergalactic medium \citep{shull2003}.

In DS95 these models were applied to optical line ratios observed in
NLRs. They found that Seyferts have values close to the shock + precursor
predictions and that Low Ionization Emission Line Region (LINER) galaxies fall within 
the range of the shock-only models. In
\cite{allen1998} we highlighted the importance of UV line ratios which help
separate shock and photoionization models, and we defined a set of UV line
ratio diagnostic diagrams. These UV diagnostics were applied to individual
objects M~87 \citep{dopita1997}, NGC~5728 and NGC~5643 \citep{evans1999} and
NGC~1068 \citep{groves2004}.  NGC~2992 was also investigated in detail with
optical diagnostics and other multi-wavelength data \citep{allen1999}.

The DS95 model grid has also been extensively used by other authors in studies
of AGN. \cite{best2000} and \cite{Inskip02} used optical and UV line ratios to
show that shocks associated with radio sources in $z\sim1$ 3CR radio galaxies
can dominate the kinematics and ionization as the radio source expands through
the interstellar medium of the host galaxy. These models have also application
to the analysis of emission lines in high redshift radio galaxies
\citep{villar1997, maxfield2002, debreuck2000,  villar2003, iwamuro2003, Reuland07}, and in
studies of individual low redshift active galaxies including NGC~4151
\citep{nelson2000}, M~51 \citep{bradley2004}, M~87 \citep{sabra2003}, Mkn~78
\citep{whittle2005}, 3C~299 \citep{feinstein1999}, NGC~2110 and NGC~5929
\citep{ferruit1999}.

Studies of individual objects make it obvious that higher dimensional modeling
of the physical structures and radiation fields associated with shocks is
necessary in order to be able to more closely model these complex physical
systems, and to be able to draw deeper conclusions. Indeed the 1-D,
steady-flow nature of these models is acknowledged as their greatest
limitation, since all fast shocks are subject to thermal instability, and the
generation of secondary shocks within the cooling zone. To address these
issues, much effort is being put into higher dimensional numerical simulations
of radiative shocks, in particular in the area of understanding the role of
local thermal and dynamical instabilities on both the shock structure and on
the emergent spectrum \citep{sutherland2003a}.  However, given the
complexities inherent in their computation, and the large amount of
supercomputer time required, complete grids of model predictions from full 3-D
models are some time off, and until then, modeling will most likely favour
simulations of individual and specific cases.

Thus, the simpler 1-D models remain a useful tool for comparing with
observations of both individual objects and in the investigation of the mean
parameters of groups of objects observed in surveys. Furthermore, these models
serve as a stepping stone towards higher dimensional models.  For example,
under some simplifying assumptions the radiative properties of 1-D models may
be mapped onto those of 3-D hydrodynamical models. In particular, thermally
unstable shocks of a given velocity, $v_{s}$, behave similarly to steady-flow
shocks with shock velocities $\sim \frac{2}{3}v_{s}$ \citep{sutherland2003a}.
Bearing in mind these effects thermal instability, we
advise that care should be taken in the use of 1-D models to derive physical
parameters in individual cases, or at least allow for this factor of 2/3.

In this paper we present a new library of 1-D steady flow photoionizing shock
models, calculated with an updated version of the shock modeling code,
\mapiii. The library is an improvement upon and extension of the DS95 \&
DS96 models, and includes a range of chemical abundance sets, pre-shock
densities from 0.01 to 1000 cm$^{-3}$, velocities up to 1000 km\ s$^{-1}$,
magnetic fields from 10$^{-10}$ to 10$^{-4}$ G, and magnetic parameters
(B/$\sqrt{\rm n}$) from 10$^{-4}$ to 100 $\mu$G\ cm$^{3/2}$. These new
model grids therefore
supersede the previous models, and are designed to be of maximum utility
to observers for comparing observations with photoionizing shock
models. 

As well as reiterating the technique used in MAPPINGS to create the
shock models, we discuss in detail the radiation fields generated by
shocks, and the resulting ionization, density and temperature
structure of shocks for the range of parameters considered. 
We also present the resulting model grids on a range of ultraviolet,
optical and infrared emission-line diagnostic diagrams, discussing the
effects of the various parameters upon the grids, and demonstrate
their usefulness and applicability by comparing these grids with
emission-line galaxy sample from the Sloan Digital Sky Survey (SDSS).

\begin{figure*}[htb]
\centering
\includegraphics[scale=0.7]{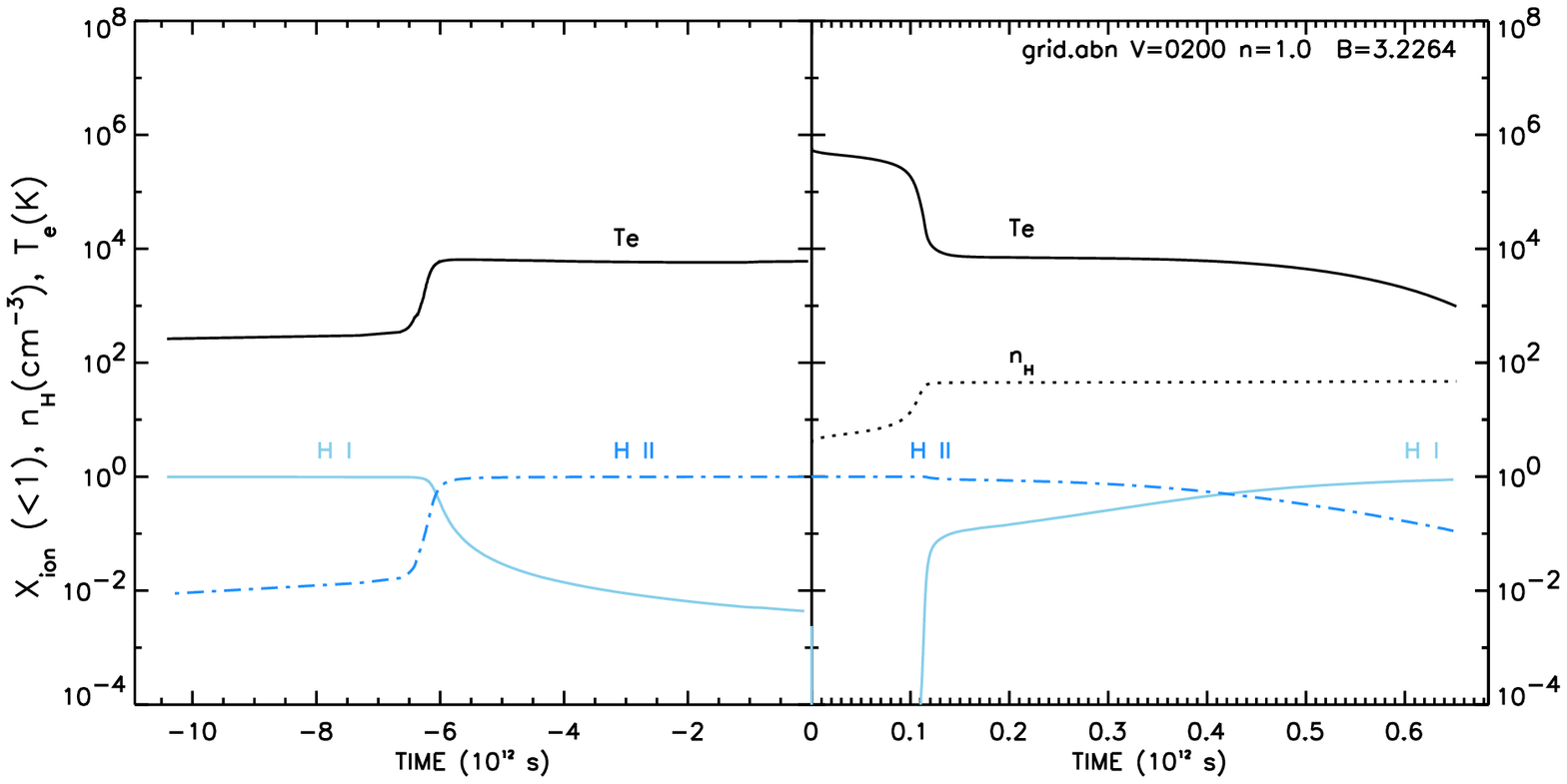} \\
\includegraphics[scale=0.7]{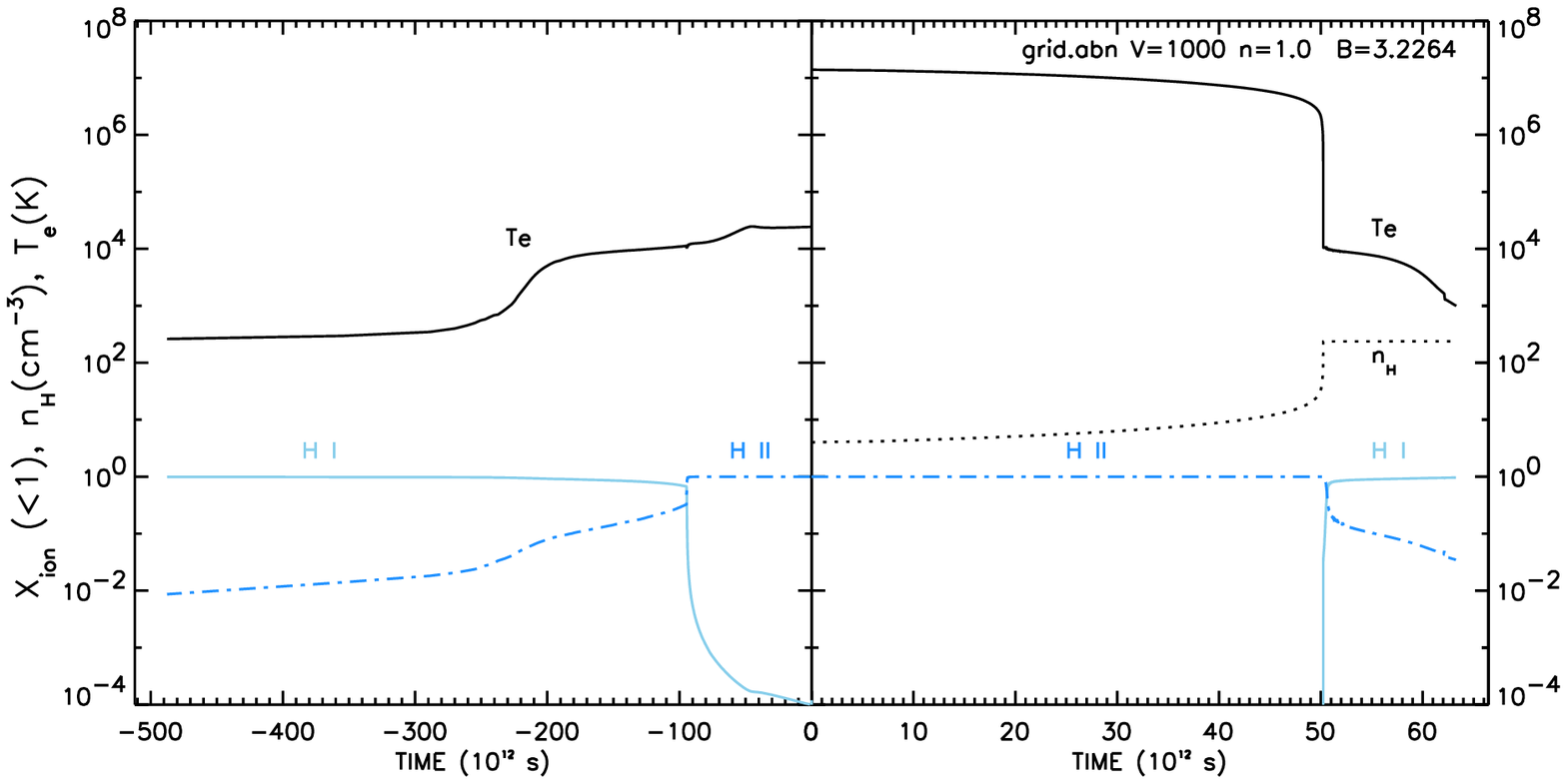} \\

\caption{The hydrogen ionization structure, temperature profile and
density profile of a shock and precursor for
models with shock veocities of  $v_{\rm s}$ = 200 and 1000
km\ s$^{-1}$ and equipartition magnetic field. The axes are as
described in Figure~\ref{Hstruct_eg}.
See the electronic edition of the Journal for a color version of this
figure.\label{Hstruct_V} } 
\end{figure*}

\begin{figure*}[htb]
\centering
\includegraphics[scale=0.7]{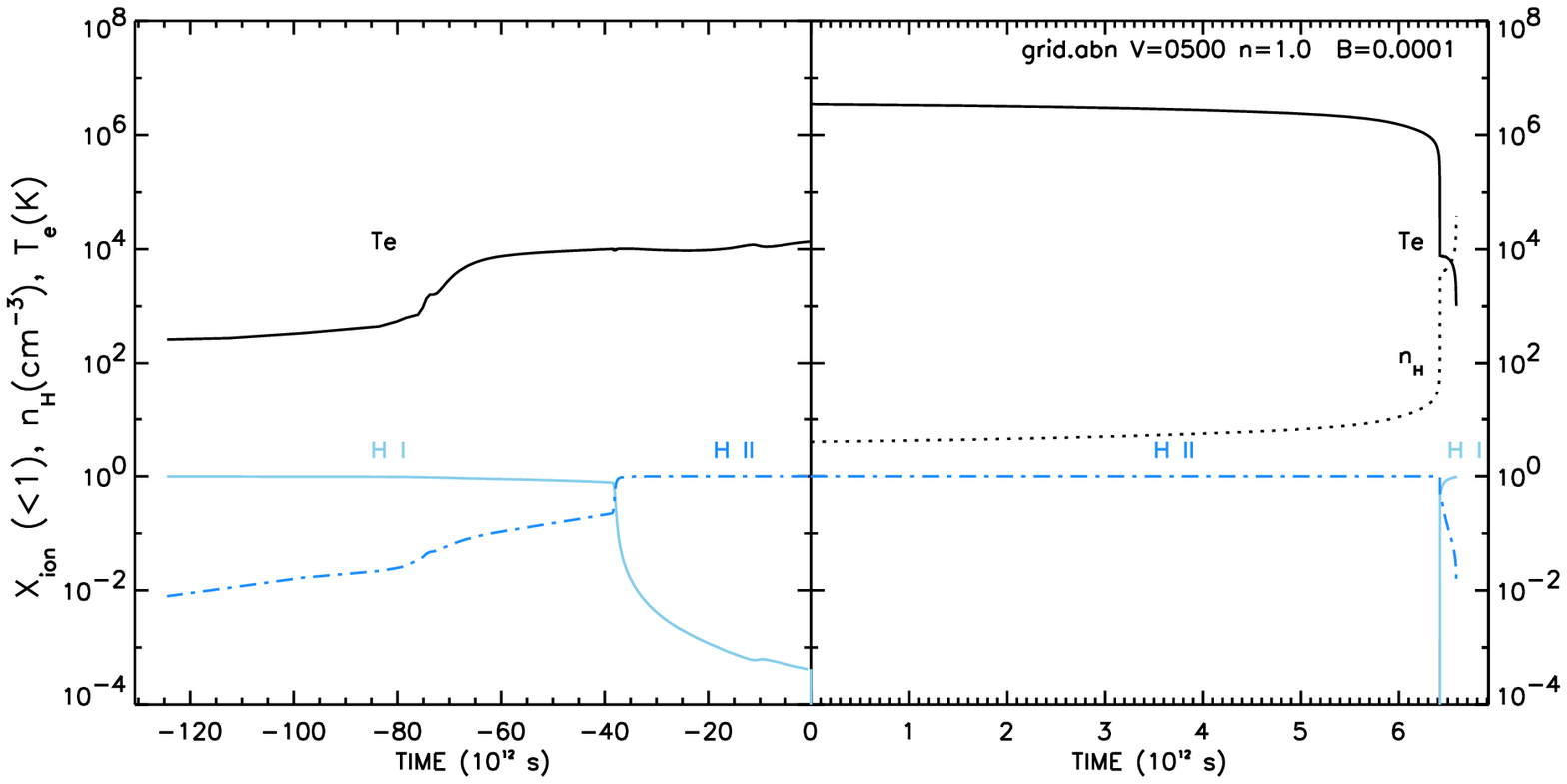} \\
\includegraphics[scale=0.7]{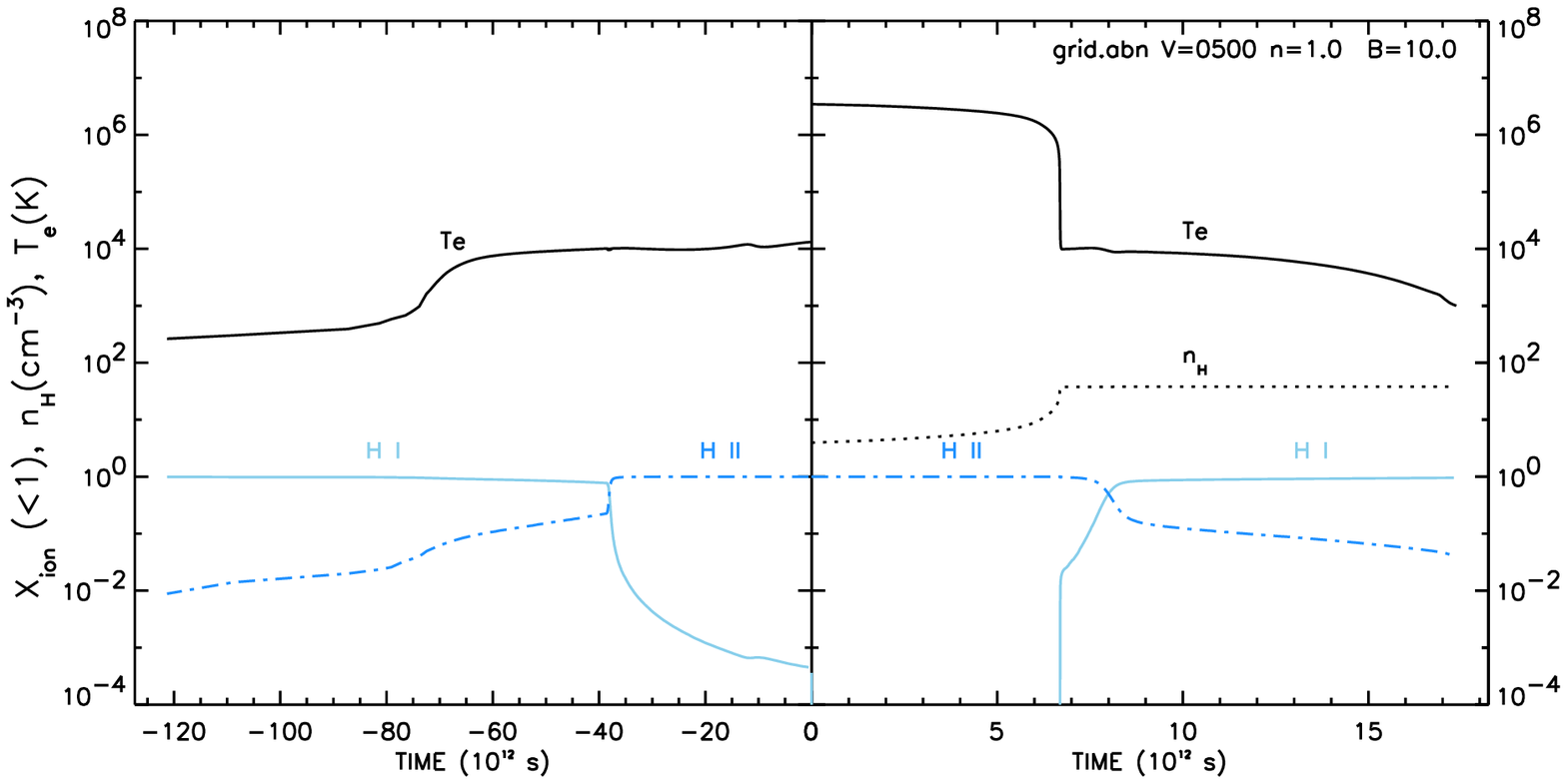} \\
\caption{The hydrogen ionization structure, temperature profile and
density profile of  
models with different magnetic parameter.  Models of $v_{\rm s} = 500$
km\ s$^{-1}$ are shown for the magnetic parameters of 0.0001 and 10
$\mu$G\ cm$^{3/2}$.  
The axes are the same as for Figure~\ref{Hstruct_eg}. See the
electronic edition of the Journal for  
a color version of this figure.
\label{Hstruct_B}}
\end{figure*}

We introduce the \mapiii\ online shock library, the
main access point for the results of the shock models, which includes
both the resulting emission lines and column depths from the models,
as well as the full results of the \mapiii\ models for detailed
analysis. Also included as part of the library are the analysis tools
used within this work, providing a simple and easy mechanism to
determine the shock diagnostic power of any emission-line ratios.

\section{Modeling Technique}

The models presented here have been calculated with the \mapiii\ shock and
photoionization modeling code, version {\sc III}q. This is an updated version
of the \mapii\ code which was described in
\cite{sutherland1993b}. The main enhancements of version III include the use
of a higher resolution radiation vector, and a more stable scheme for choosing
time-steps which allows computation of higher velocity shocks. Other
improvements include the explicit inclusion of all the He~{\sc II} emission
lines, and a corrected computation of the neutral oxygen emission. Further details
on the use of and the chronology of the development of MAPPINGS models may be
found at the MAPPINGS Online web
pages\footnote{\url{http://www.ifa.hawaii.edu/\~{}kewley/Mappings}}. Various
improvements to the calculation of photoionization processes have been
implemented in version III \citep{groves2004a,groves2007} although the dust
effects discussed there have not been considered in these shock models, which
are fundamentally dust-free. This approximation is probably a physical one,
since in the faster shocks dust will be effectively destroyed by grain-grain
collisions, through both shattering and spallation, and by thermal
sputtering. The physics of these processes have been discussed in detail
\citet{Draine79, Dwek92, Jones94, Jones96, Pineau97} and \citet{perna2002}.
We have in
any case run models with depleted abundances to facilitate comparison with the
earlier DS96 high-velocity shock modeling, so our different model grids can be
used to provide some estimate of the effect of dust depletion on the output
spectra.

\begin{figure*}[t]
\centering
\includegraphics[scale=0.6]{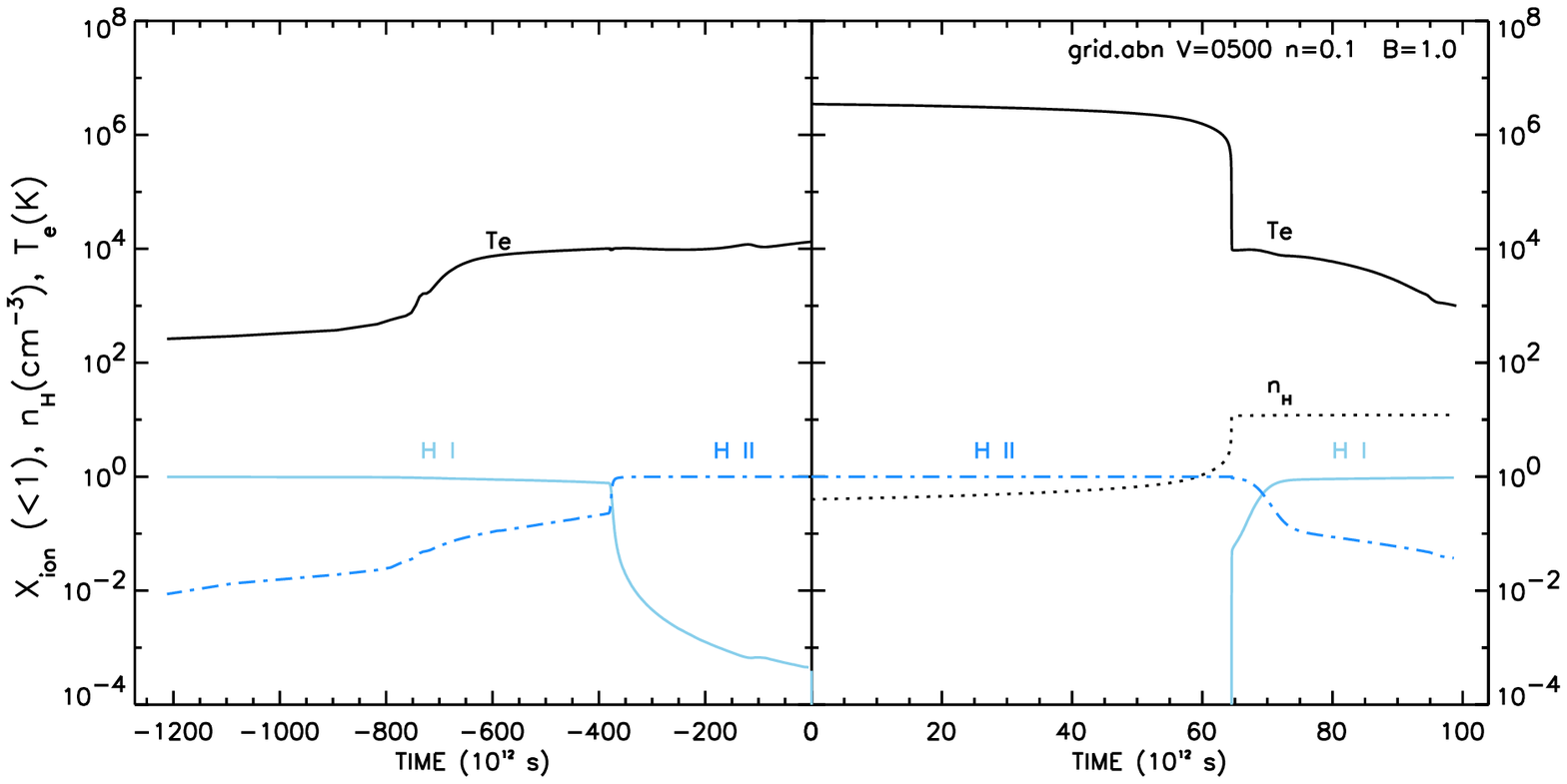} \\
\includegraphics[scale=0.6]{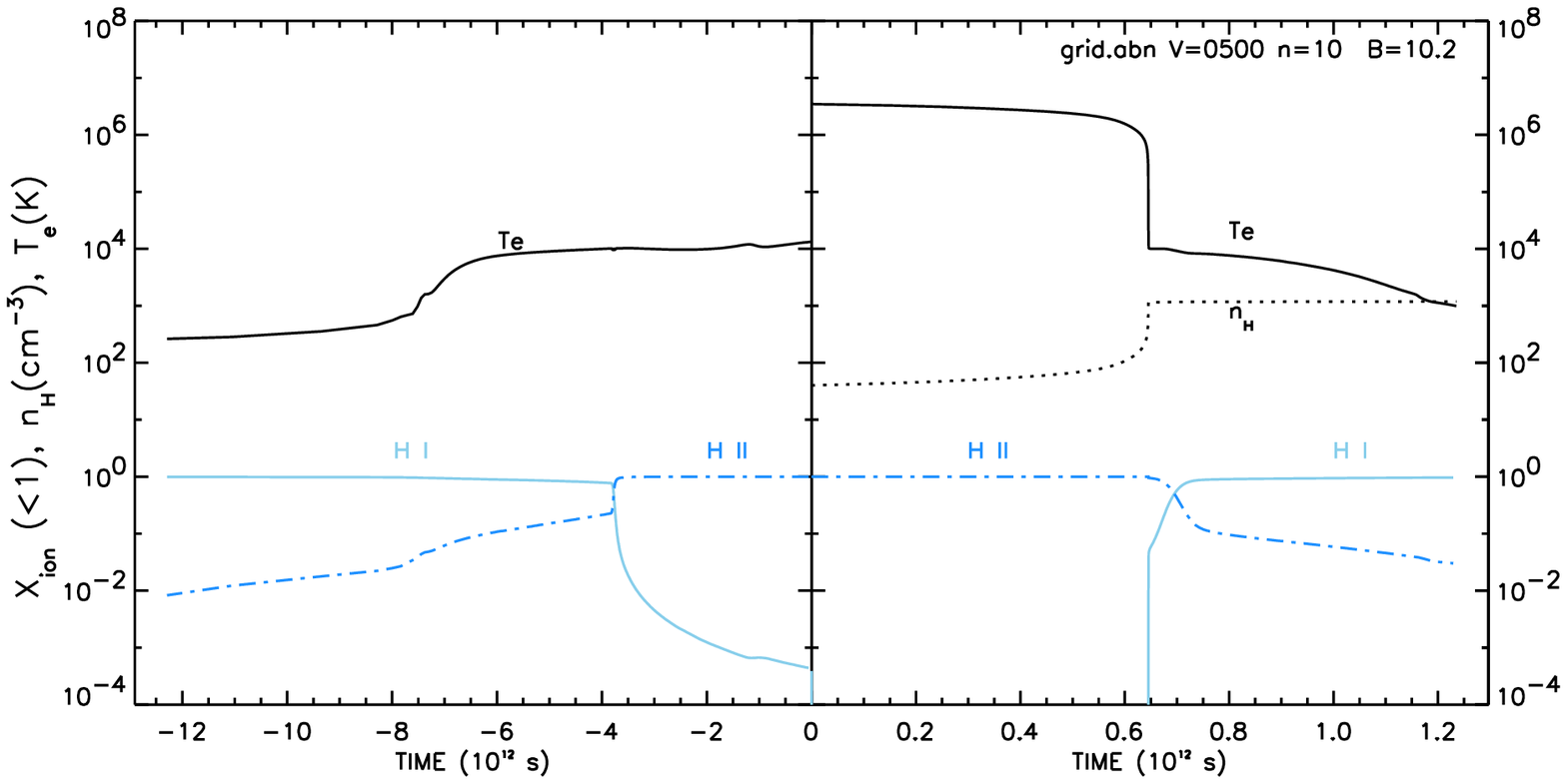} \\
\includegraphics[scale=0.6]{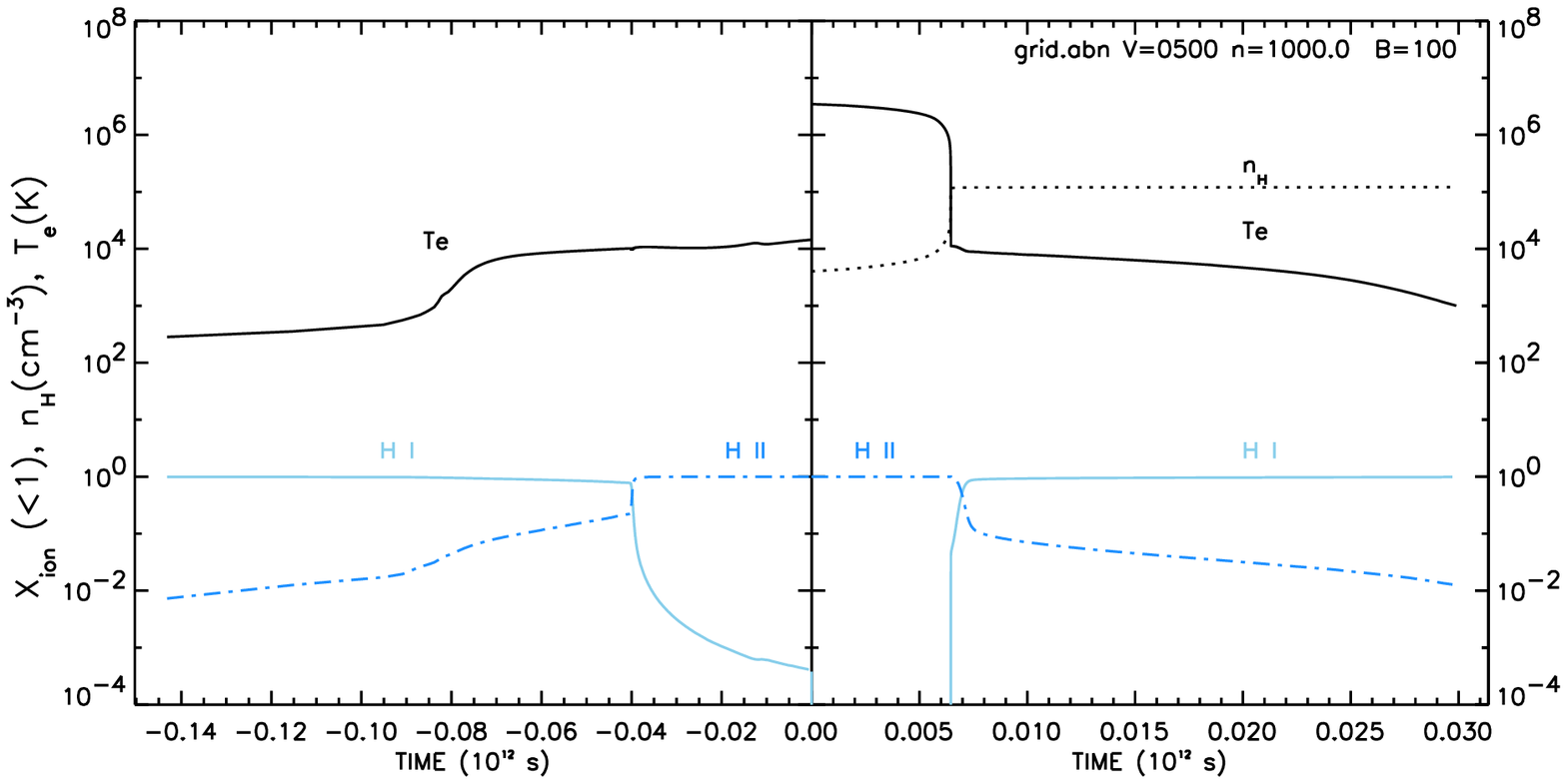} \\
\caption{The hydrogen ionization structure, temperature profile and density
profile of models with different pre-shock density. Models of $v_{\rm s} =
500$ km\ s$^{-1}$ and equipartition magnetic field are shown for pre-shock
densities of 0.1, 10 and 1000 cm$^{-3}$. The axes are the same as for
Figure~\ref{Hstruct_eg}.  See the electronic edition of the Journal for a
color version of this figure.
\label{Hstruct_dens} }
\end{figure*}

\begin{figure*}[t]
\centering
\includegraphics[scale=0.6]{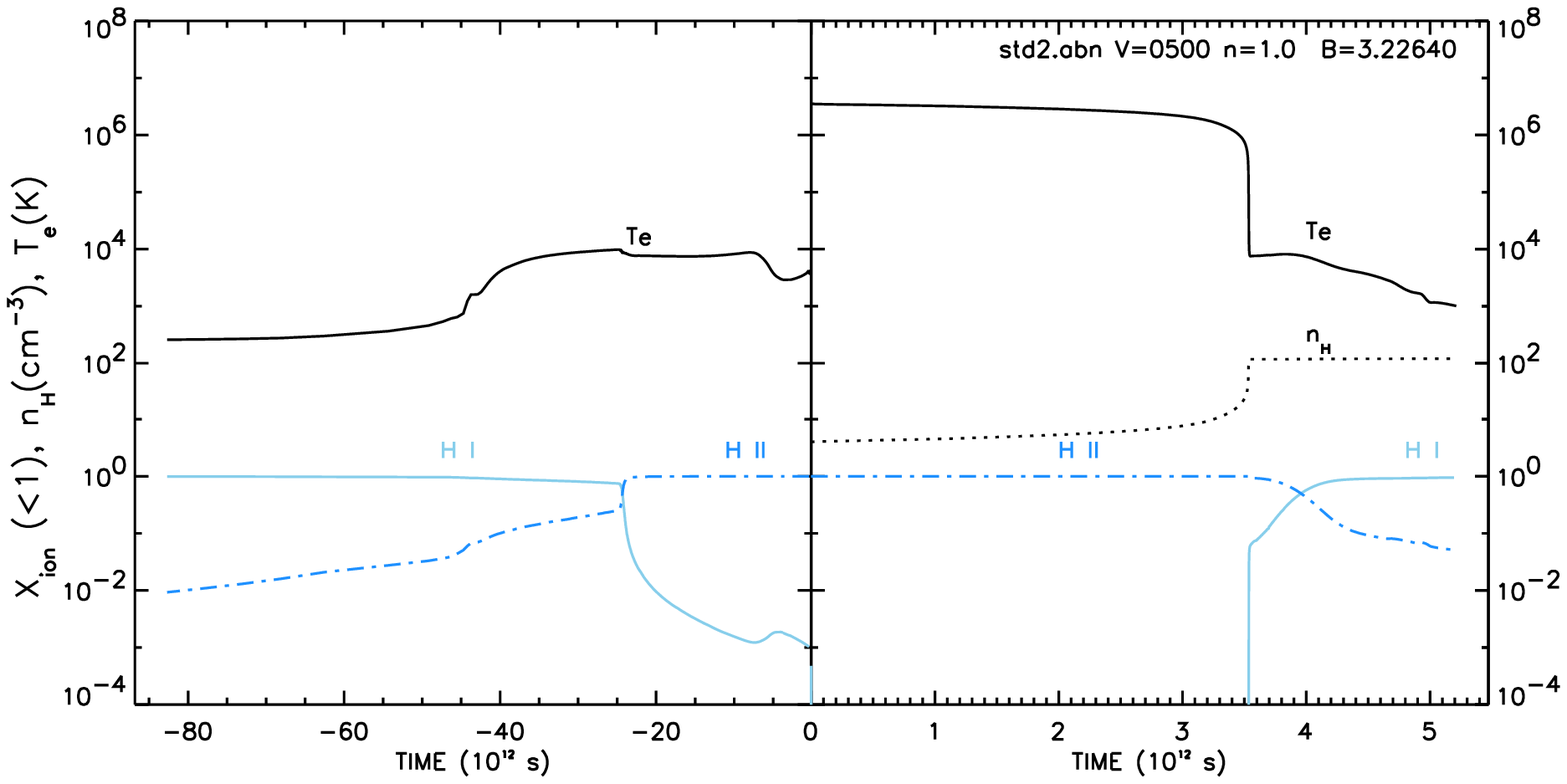} \\
\includegraphics[scale=0.6]{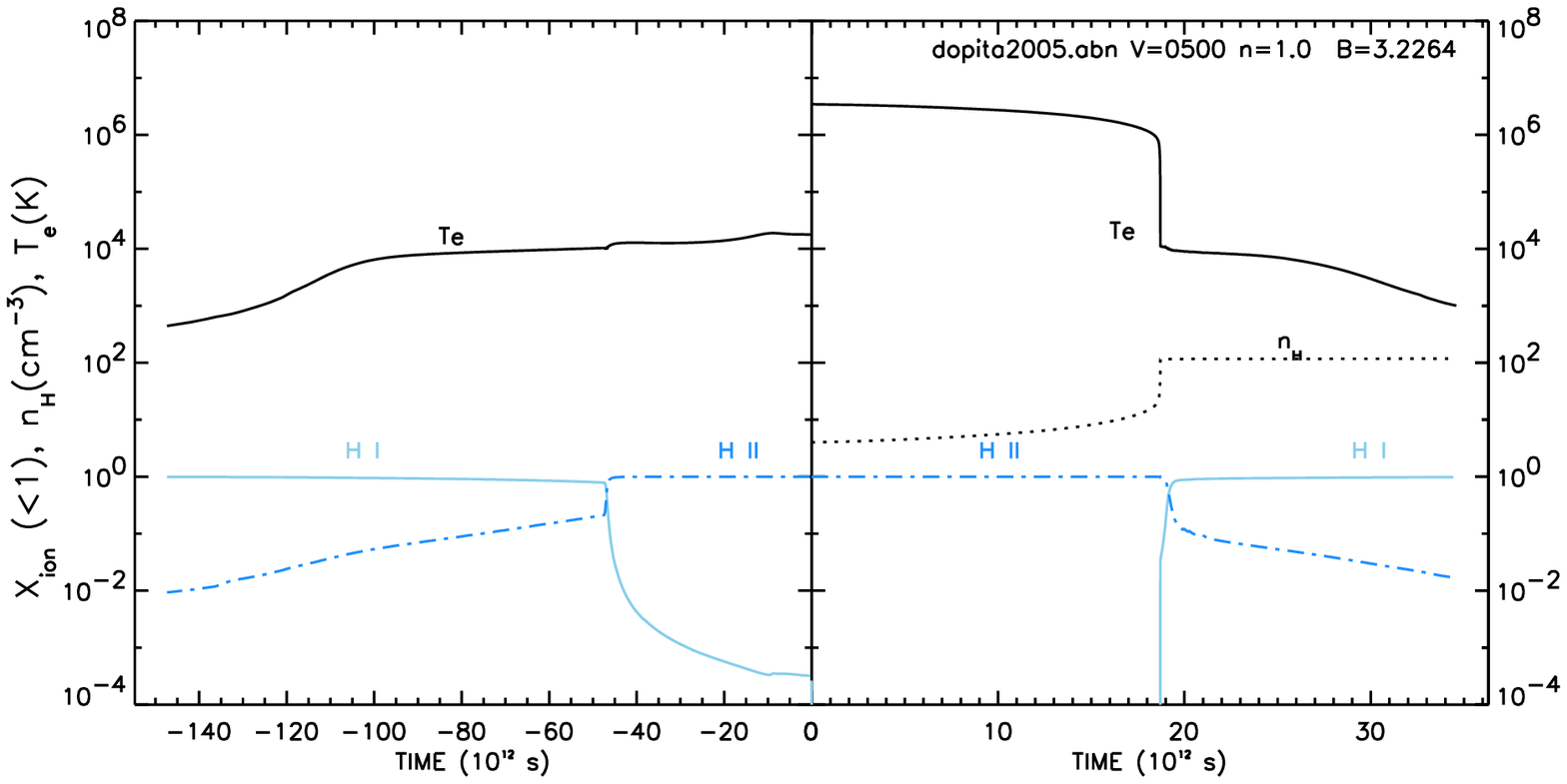} \\
\includegraphics[scale=0.6]{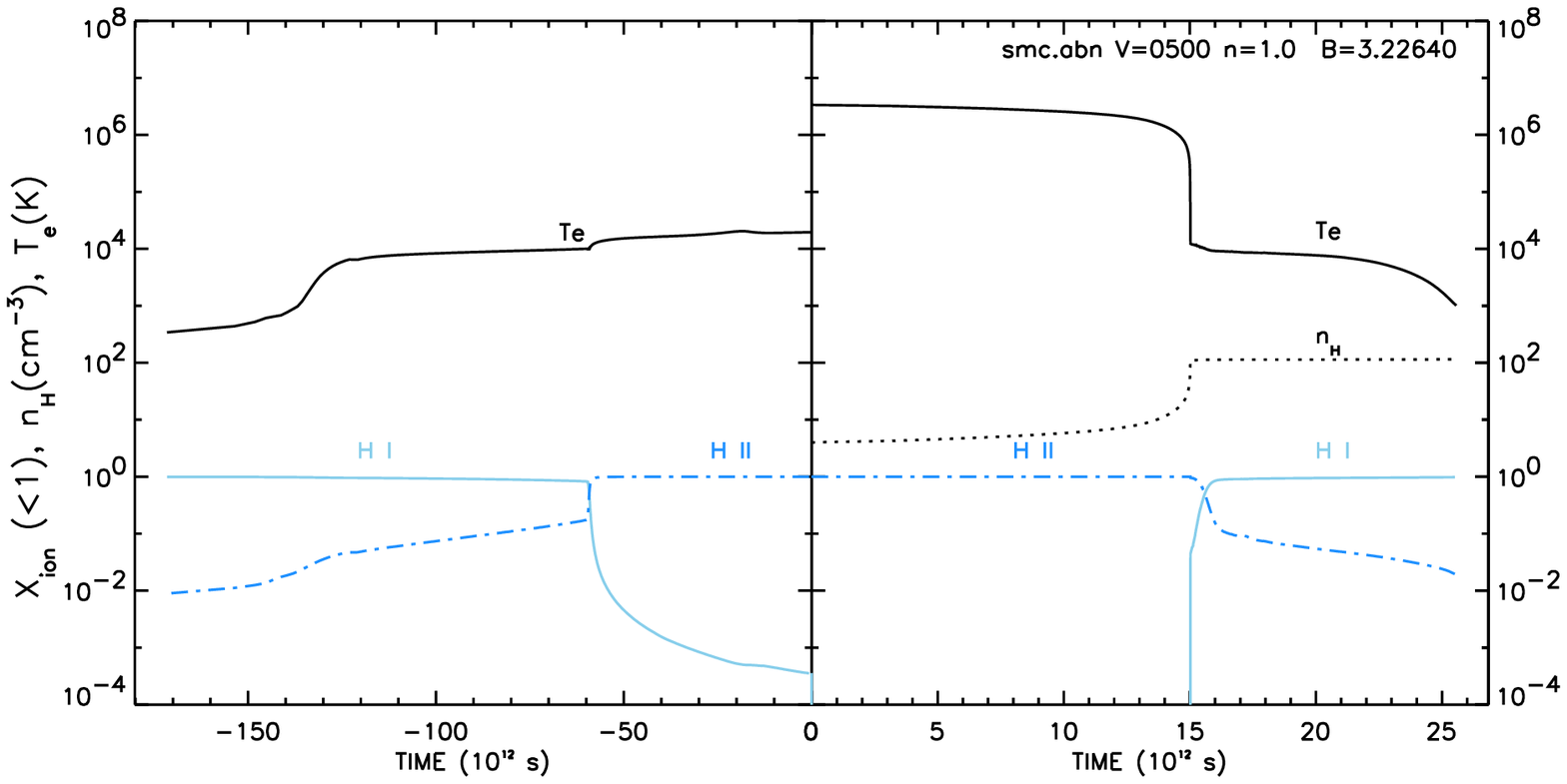} \\
\caption{The hydrogen ionization structure, temperature profile and density
profile of models with different atomic abundances.  Models of $v_{\rm s} =
500$ km\ s$^{-1}$ and equipartition magnetic field are shown for the 
2$\times$solar, dopita2005 and SMC abundance sets.  The axes are the
same as for Figure~\ref{Hstruct_eg}.  See the electronic edition of
the Journal for a color version of this figure.
\label{Hstruct_abund}}
\end{figure*}

\mapiii\ uses exactly the same computational recipe for calculating the flow
solution of these one-dimensional radiative shock models as described in
detail in DS96 for \mapii . Briefly, the hydrodynamics of the flow
are derived from the Rankine-Hugoniot jump conditions including the magnetic
field terms. The four Rankine-Hugoniot conditions, equations \ref{rh1} -
\ref{rh4}, represent the conservation of mass, conservation of momentum, the
condition that the magnetic field is locked into the ionized plasma, and
conservation of energy (see \cite{heng2007} for a generalization of these conditions). 
These equations connect two points in the flow at
times $t$ and $t_0$ with density, velocity, pressure, internal energy, and
transverse magnetic field $\rho$, $V$,$P$, $U$, B, and $\rho_0$,
$V_0$,$P_0$, $U_0$, B$_0$, respectively. $\bar{\Lambda}$ is the mean cooling
rate of the plasma over the time step. As set out in DS96, by choosing the
subscripted flow variables as the point immediately in front of the shock, the
equations can be reduced to a quadratic expression for the flow velocity. By
choosing appropriate fractions of the geometric mean of the plasma timescales,
the physical conditions in the plasma can be followed smoothly.

\begin{equation}
\label{rh1}
\rho v=\rho_0 v_0
\end{equation}

\begin{equation}
\label{rh2}
\rho v^2 + P + \frac{{\rm B}^2}{8\pi} =\rho_0 v_0^2 + P_0 + \frac{{\rm
B}_0^2}{8\pi} 
\end{equation}

\begin{equation}
\label{rh3}
\frac{{\rm B}}{\rho} = \frac{{\rm B}_0}{\rho_0}
\end{equation}

\begin{equation}
\label{rh4}
\frac{v^2}{2} + U + \frac{P}{\rho} + \frac{{\rm B}^2}{4\pi} +
\bar{\Lambda}(t - t_0) 
= \frac{v_0^2}{2} + U_0 + \frac{P_0}{\rho_0} + \frac{{\rm B}_0^2}{4\pi}
\end{equation}

The main calculation to be performed at each time step in these shock models
is that of the ionization balance and the corresponding mean cooling rate,
$\bar{\Lambda}$. The cooling and radiative emission is calculated using a very
large atomic data set which allows treatment of all ionization stages of
cosmically abundant elements up to fully ionized nickel. Using this atomic
data set, the rate equations for non-equilibrium ionization, recombination,
excitation, and radiative transfer and cooling are solved at each time step of
the flow. Details of the various collisional and radiative processes included 
in MAPPINGS are described in \citet{sutherland1993}.

Modeling of all fast radiative shocks is necessarily an iterative process
because the detailed structure of the shock depends on the ionization state of
the precursor gas entering the shock front. To calculate a fully
self-consistent model of a radiative shock a number of iterations are
required. Firstly an initial shock model is calculated using an estimate of
the ionization state of the precursor. Then the photoionized precursor is
calculated using the ionizing radiation field generated by the shock. This
process is then repeated, updating the ionization state of the precursor gas
at each iteration.  The models presented here employ four such iterations,
which was found to be sufficient to allow the temperature and ionization state
of the precursor gas to stabilize at a constant value.

In the final iteration of each model, the computation of the shock structure
is allowed to proceed until the gas has cooled to 1000K, below which no further significant 
emission in the considered species is produced.  The computation of
the precursor structure is terminated when the ionized fraction of hydrogen
falls to below 1 percent.

Precursor components were computed for each individual model, in contrast to
DS96 who computed precursors only for each value of the shock velocity. The
precursor ionization does not generally depend on the magnetic parameter but
here we chose to compute each individual precursor to ensure full self
consistency between the shock and its precursor.

\begin{figure*}
\centering
\includegraphics[scale=0.6]{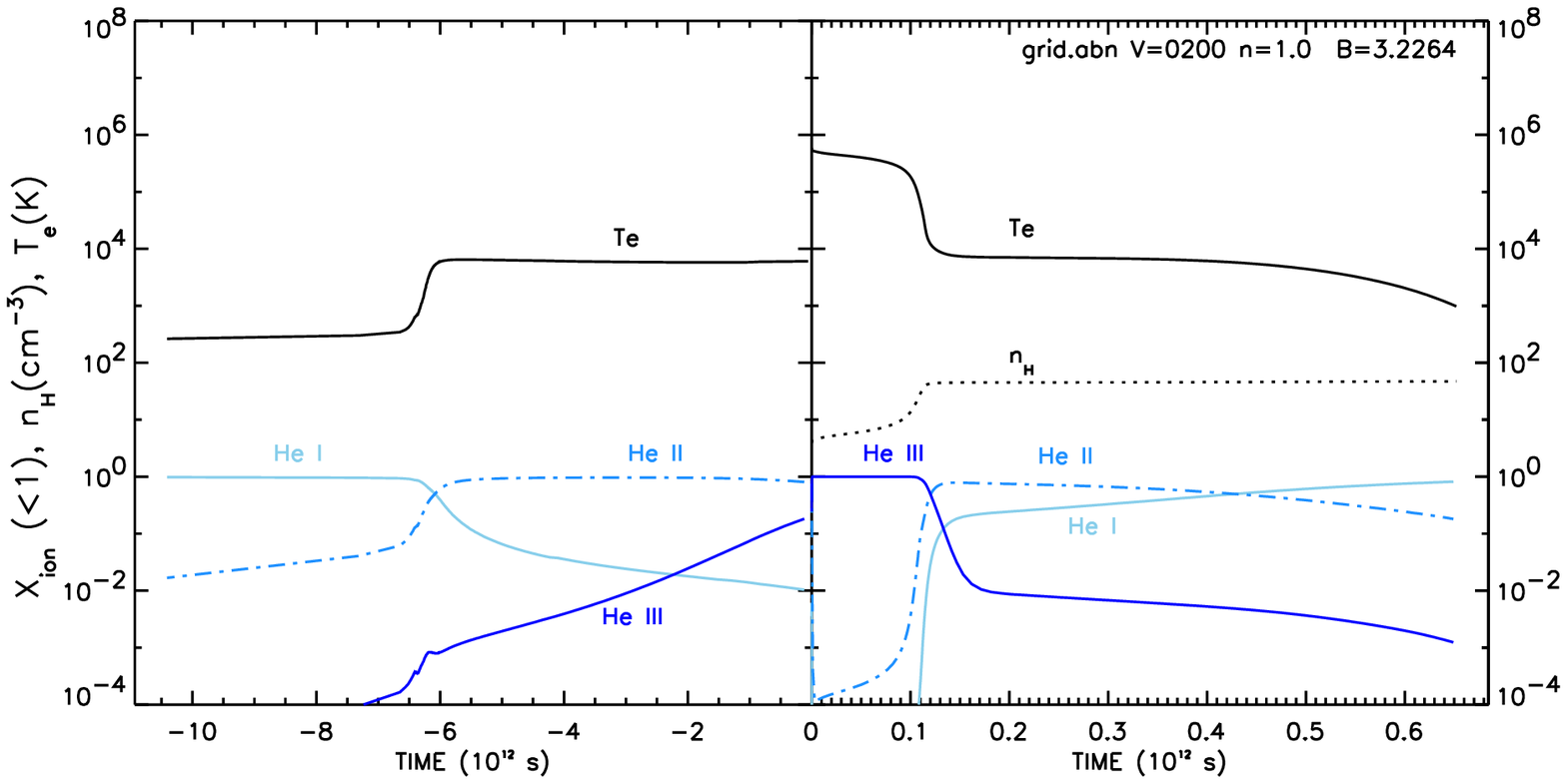} \\
\includegraphics[scale=0.6]{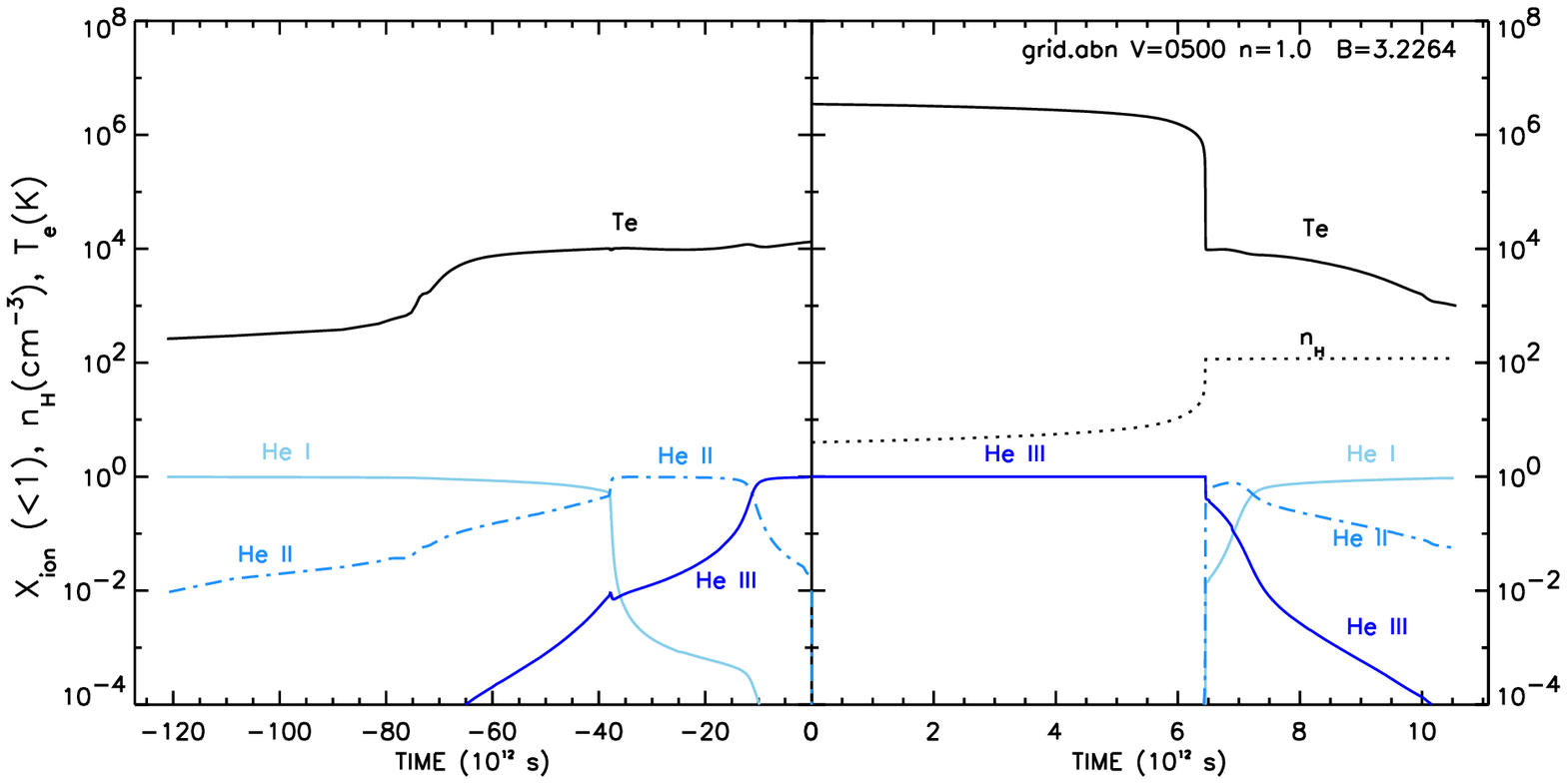} \\
\includegraphics[scale=0.6]{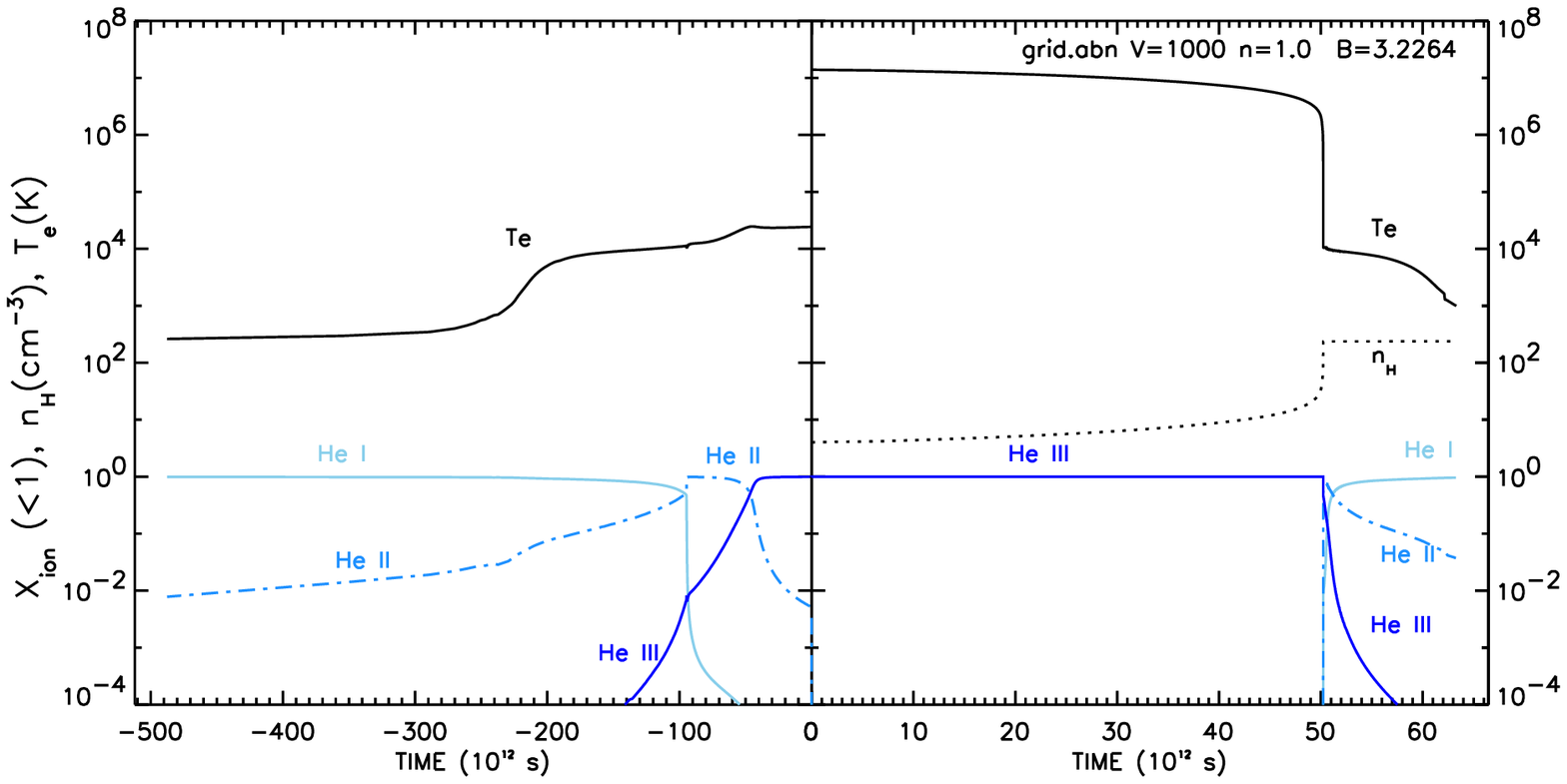} \\
\caption{The helium ionization structure, temperature profile and density
profile of models with solar abundance, equipartition magnetic field for shock
velocities of $v_{\rm s}$ = 200, 500, and 1000 km\ s$^{-1}$.  The axes are the
same as for Figure~\ref{Hstruct_eg}.  See the electronic edition of the
Journal for a color version of this figure.
 \label{Hestruct_fig}} 
\end{figure*}

\begin{figure*}
\centering
\includegraphics[scale=0.6]{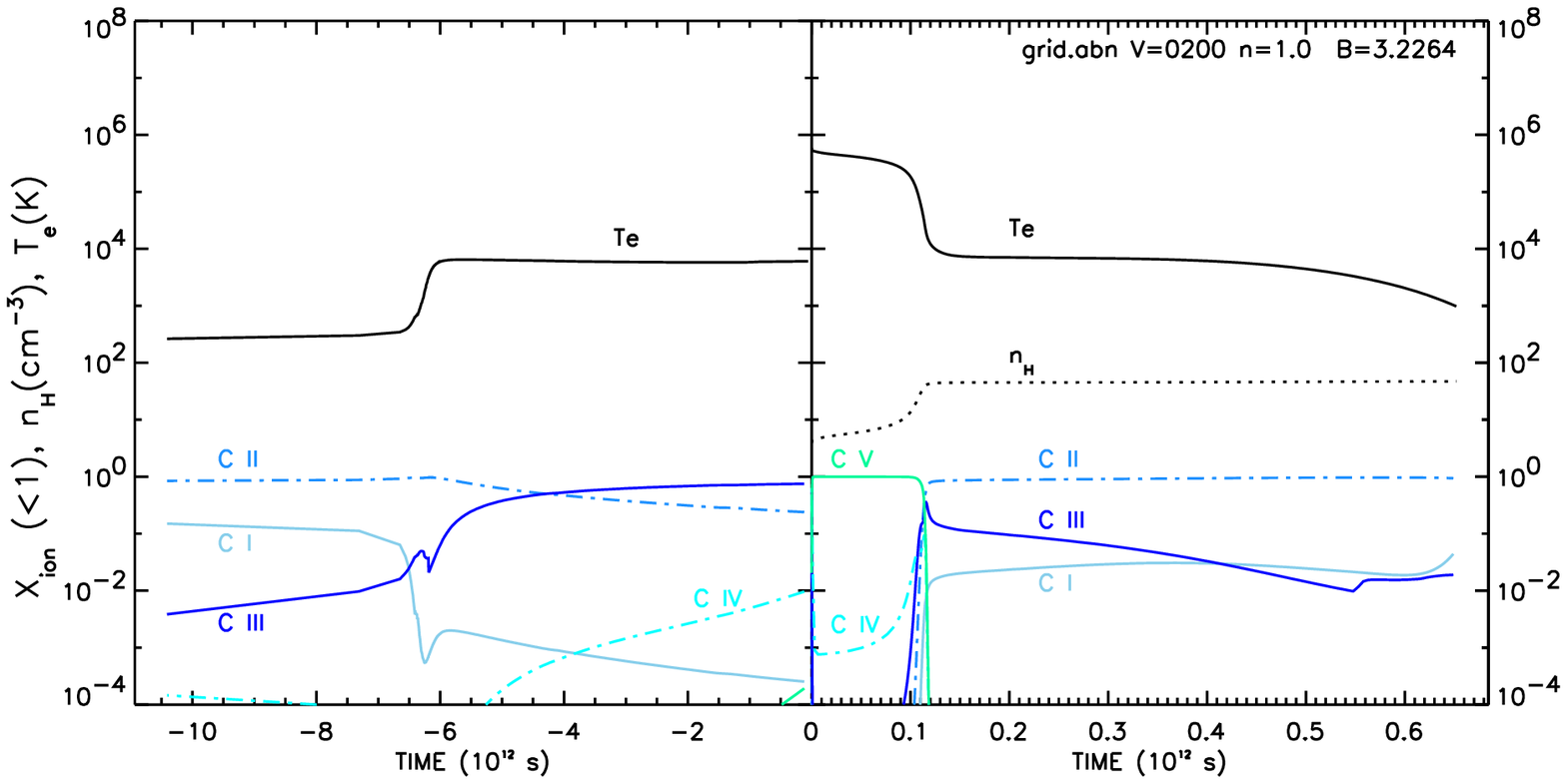} \\
\includegraphics[scale=0.6]{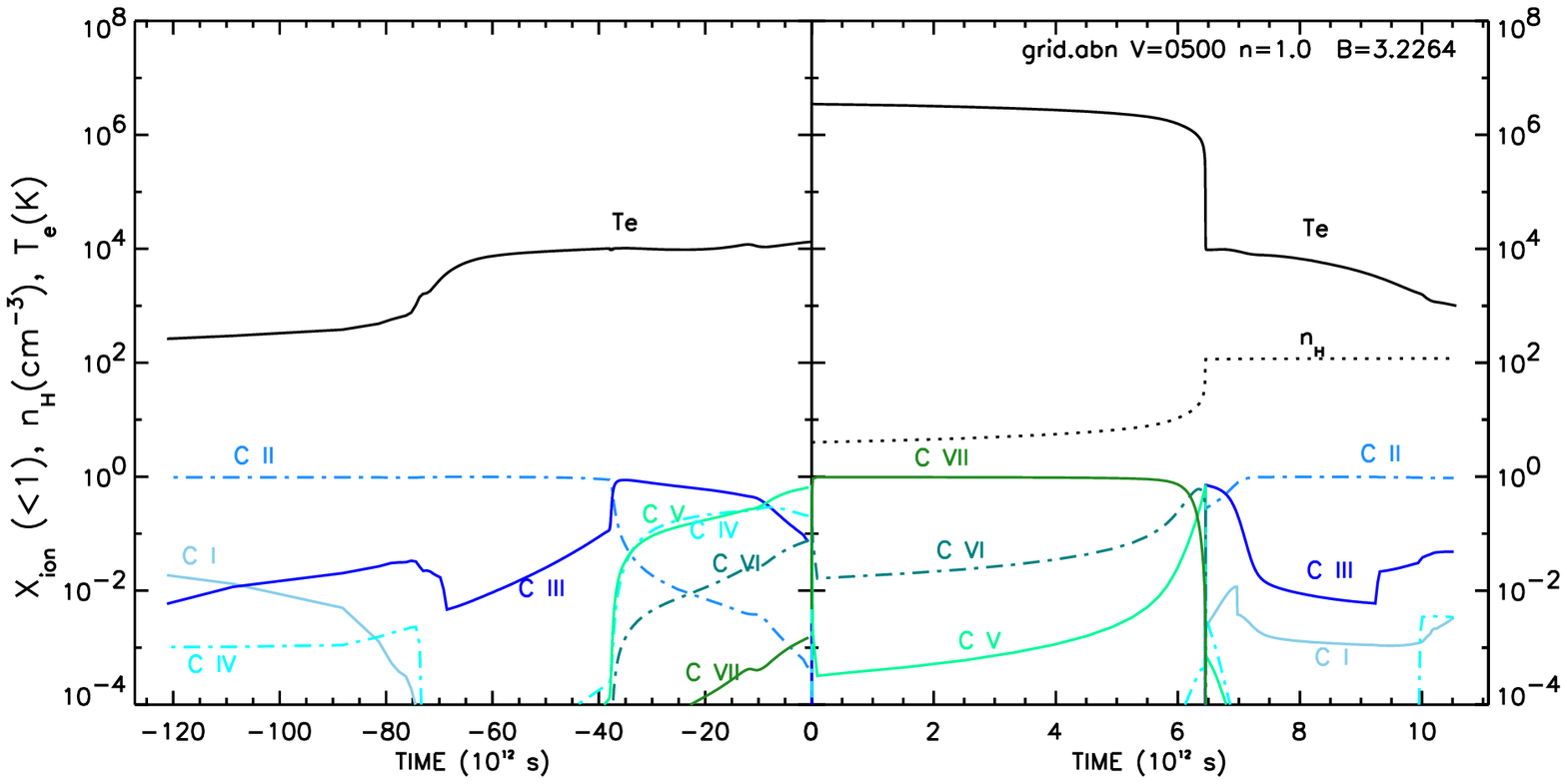} \\
\includegraphics[scale=0.6]{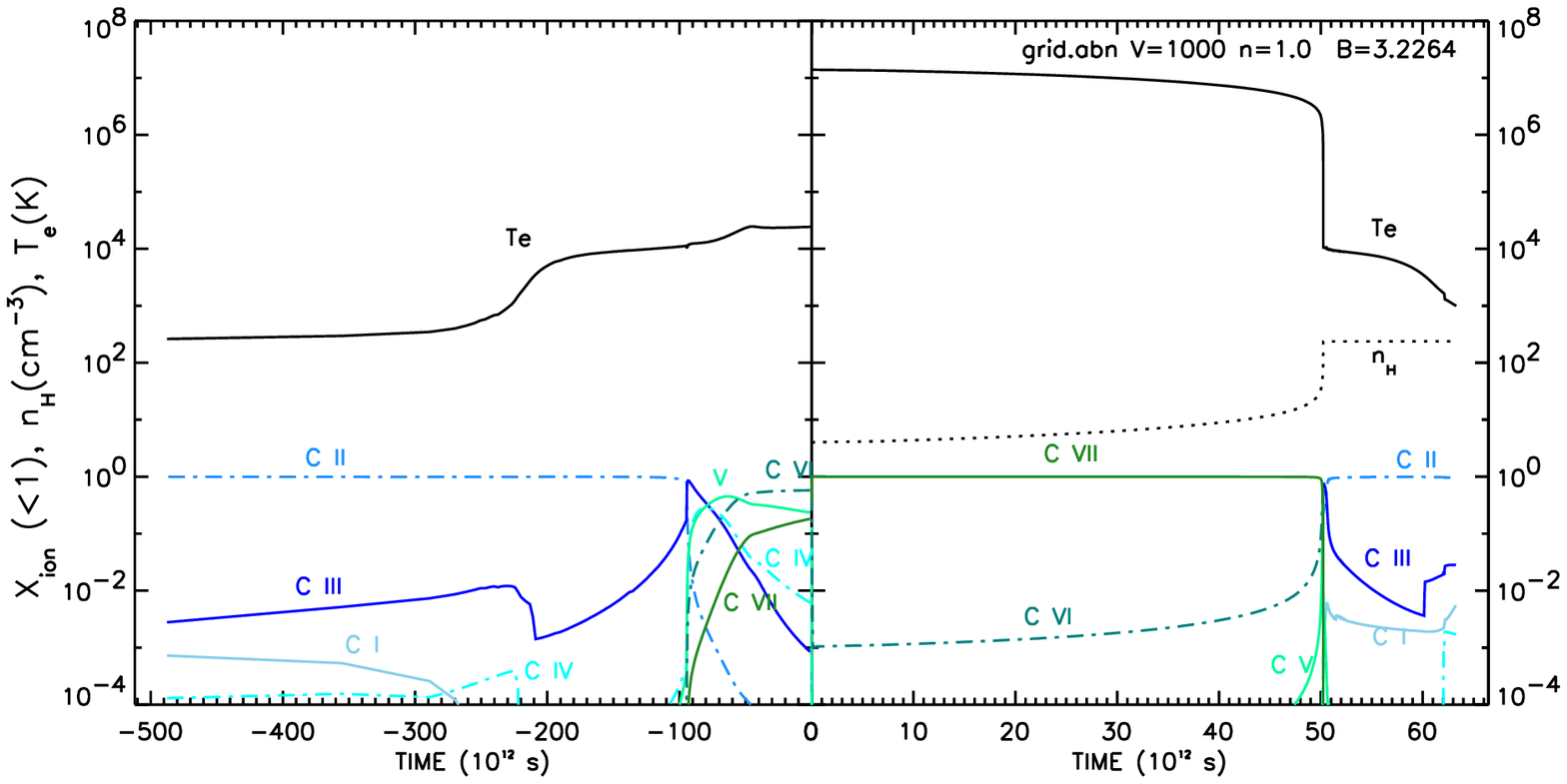} \\
\caption{The carbon ionization structure, temperature profile and density
profile of models with solar abundance, equipartition magnetic field for shock
velocities of $v_{\rm s}$ = 200, 500, and 1000 km\ s$^{-1}$.  The axes are the
same as for Figure~\ref{Hstruct_eg}.  See the electronic edition of the
Journal for a color version of this figure.
\label{Cstruct_fig}}
\end{figure*}                

\begin{figure*}
\centering
\includegraphics[scale=0.6]{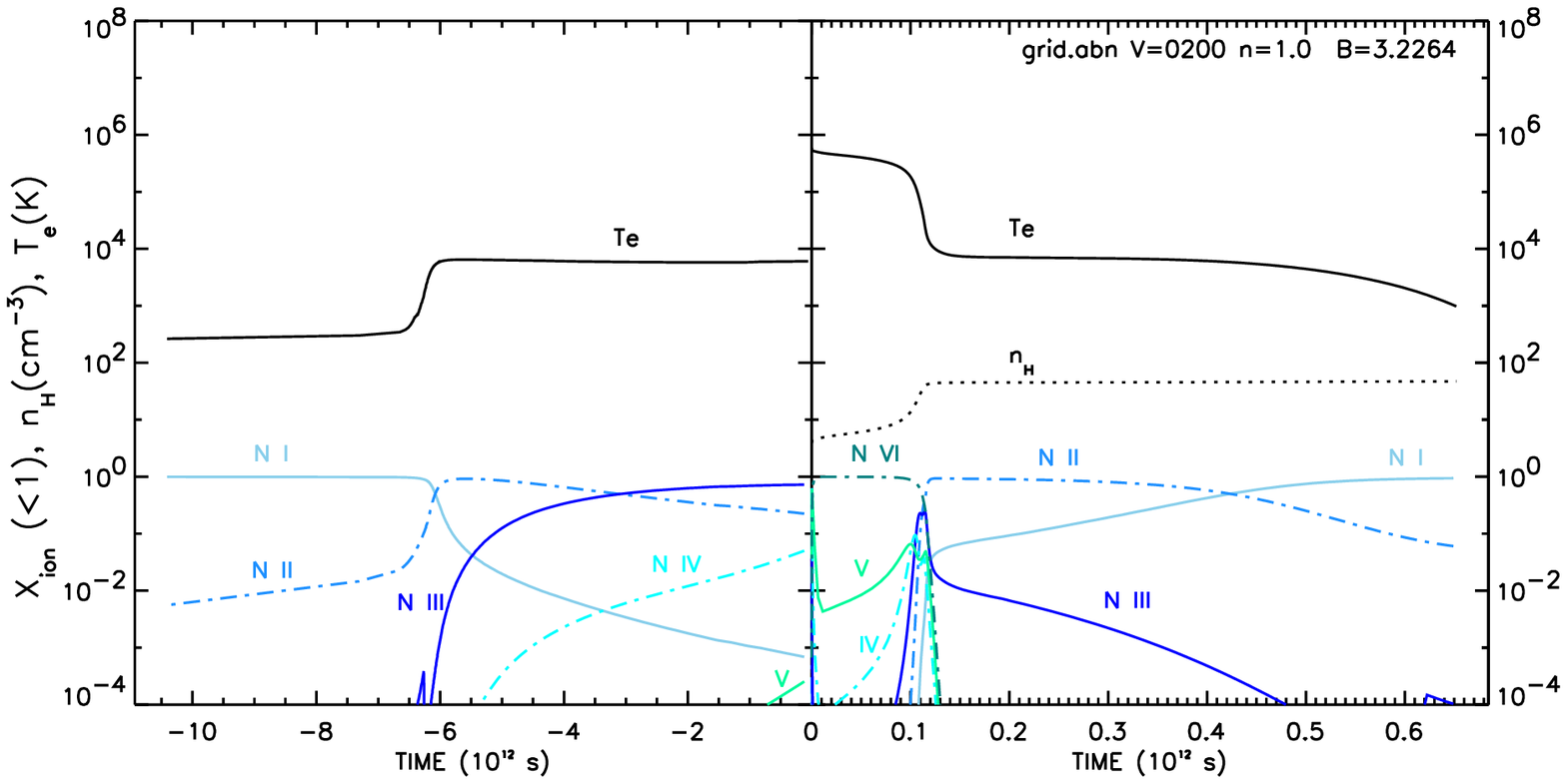} \\
\includegraphics[scale=0.6]{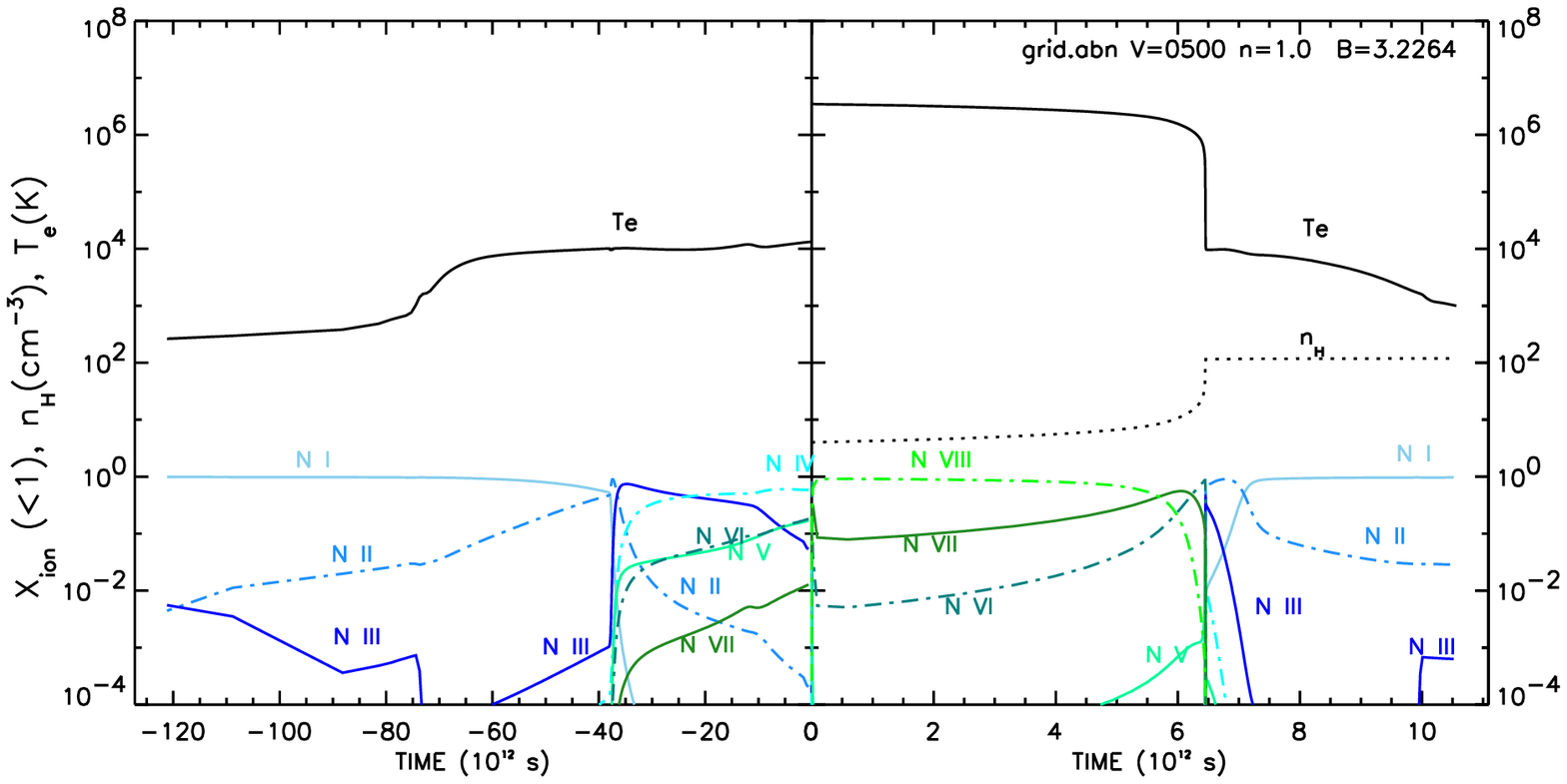} \\
\includegraphics[scale=0.6]{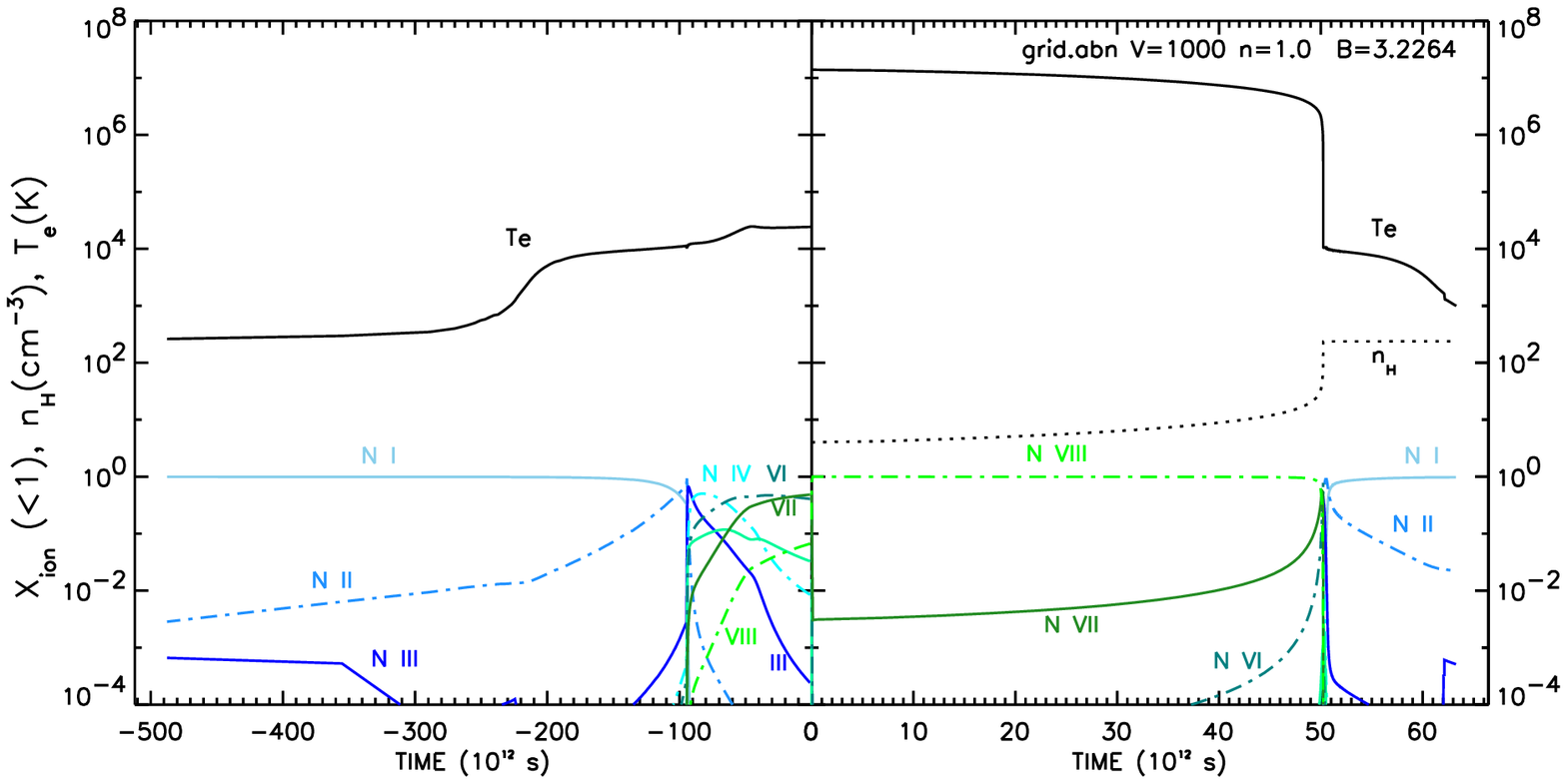} \\
\caption{The nitrogen ionization structure, temperature profile and density
profile of models with solar abundance, equipartition magnetic field for shock
velocities of $v_{\rm s}$ = 200, 500, and 1000 km\ s$^{-1}$.  The axes are the
same as for Figure~\ref{Hstruct_eg}.  See the electronic edition of the
Journal for a color version of this figure.
\label{Nstruct_fig} }
\end{figure*}                

\begin{figure*}
\centering
\includegraphics[scale=0.6]{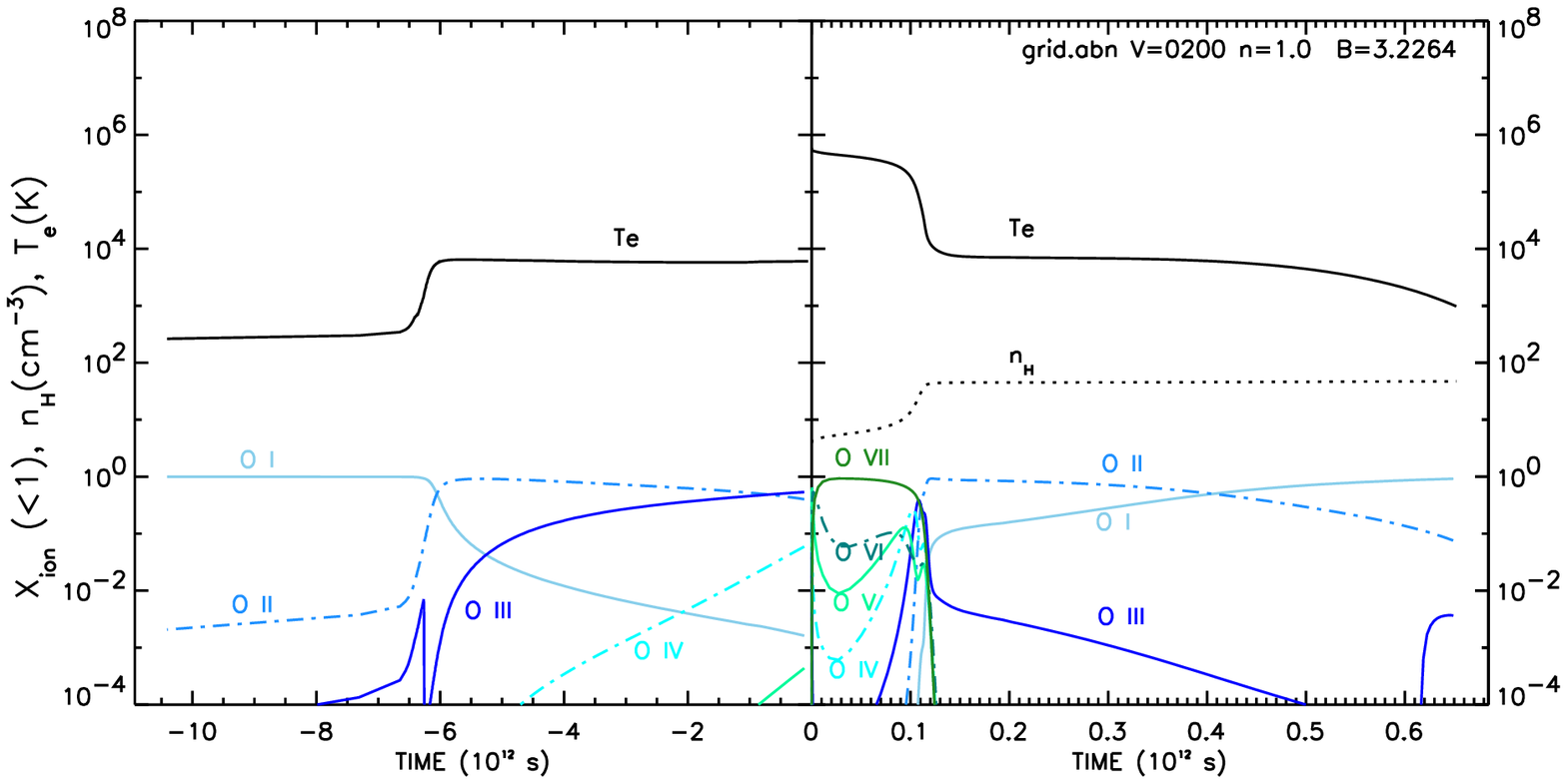} \\
\includegraphics[scale=0.6]{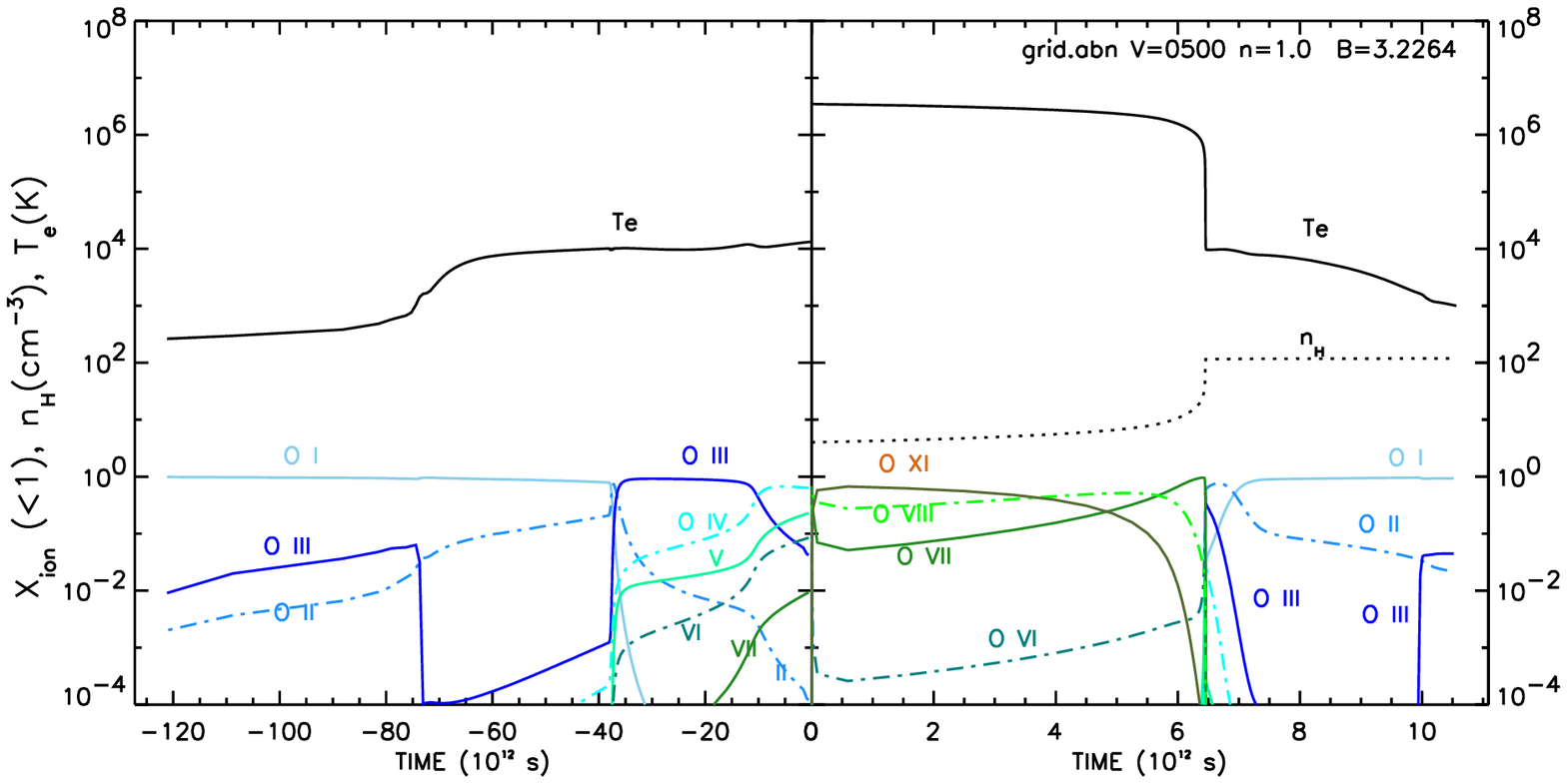} \\
\includegraphics[scale=0.6]{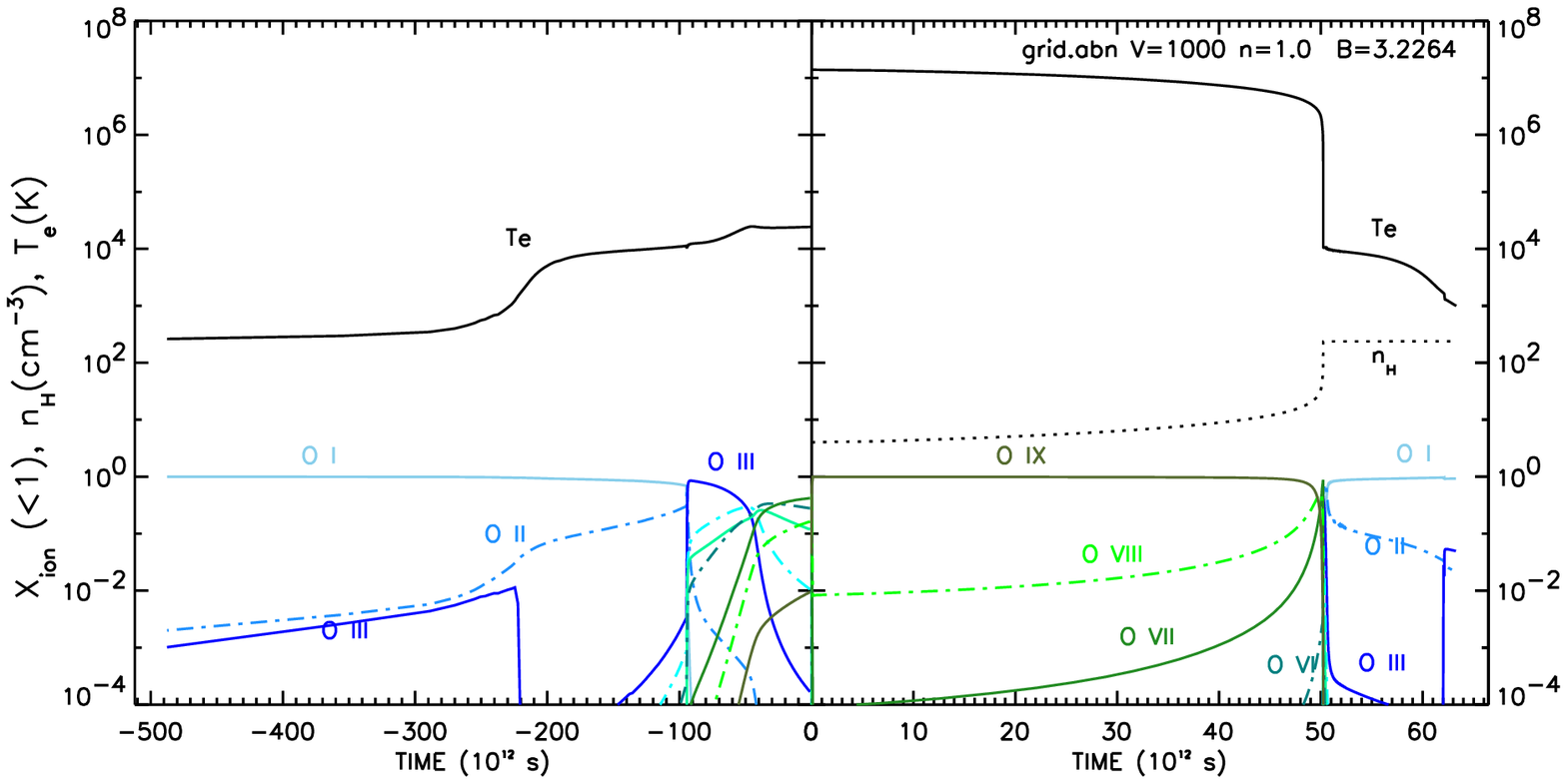} \\
\caption{The oxygen ionization structure, temperature profile and density
profile of models with solar abundance, equipartition magnetic field for shock
velocities of $v_{\rm s}$ = 200, 500, and 1000 km\ s$^{-1}$.  The axes are the
same as for Figure~\ref{Hstruct_eg}.  See the electronic edition of the
Journal for a color version of this figure.
\label{Ostruct_fig}}
\end{figure*}                

\begin{figure*}
\centering
\includegraphics[scale=0.6]{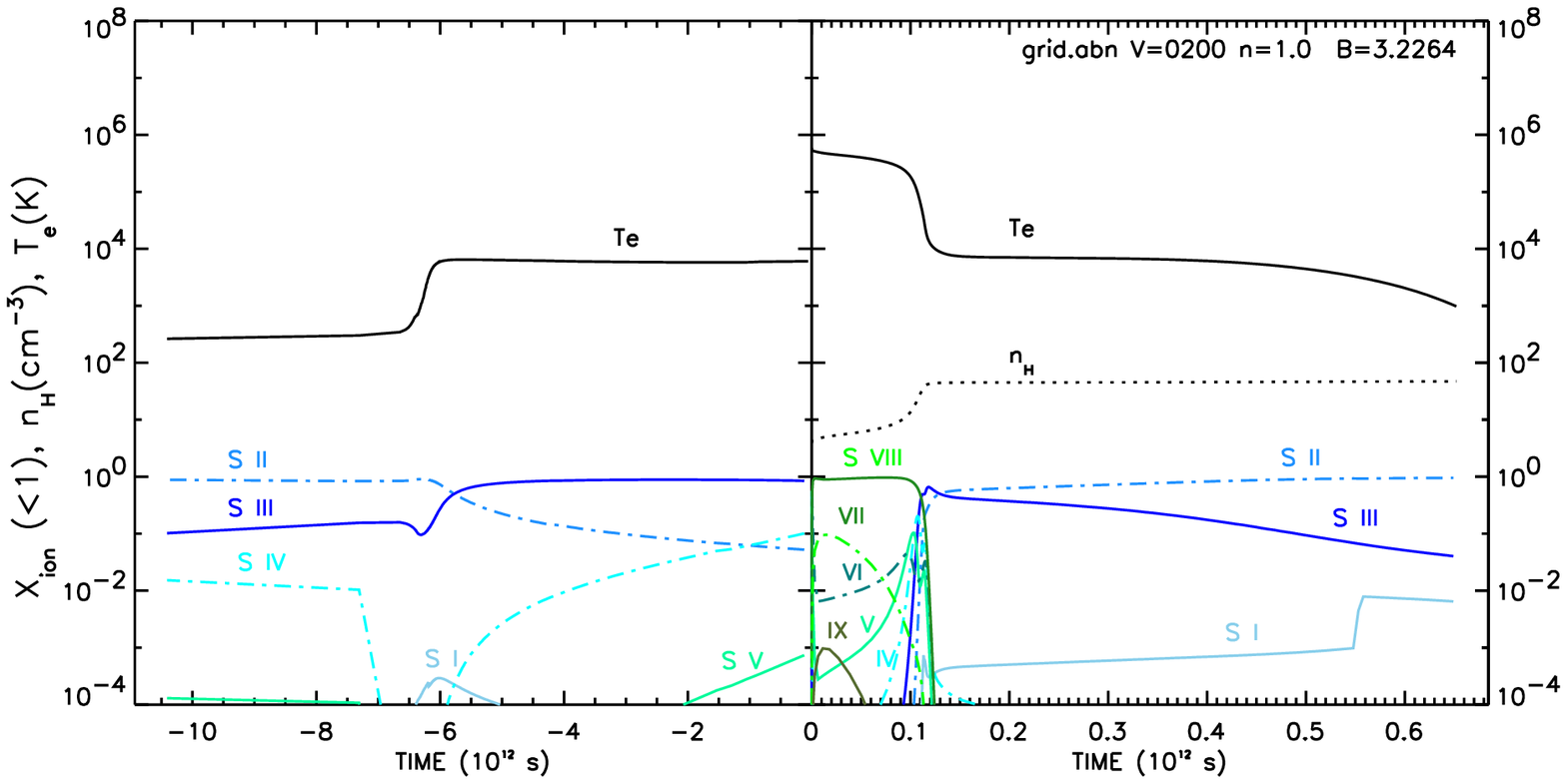} \\
\includegraphics[scale=0.6]{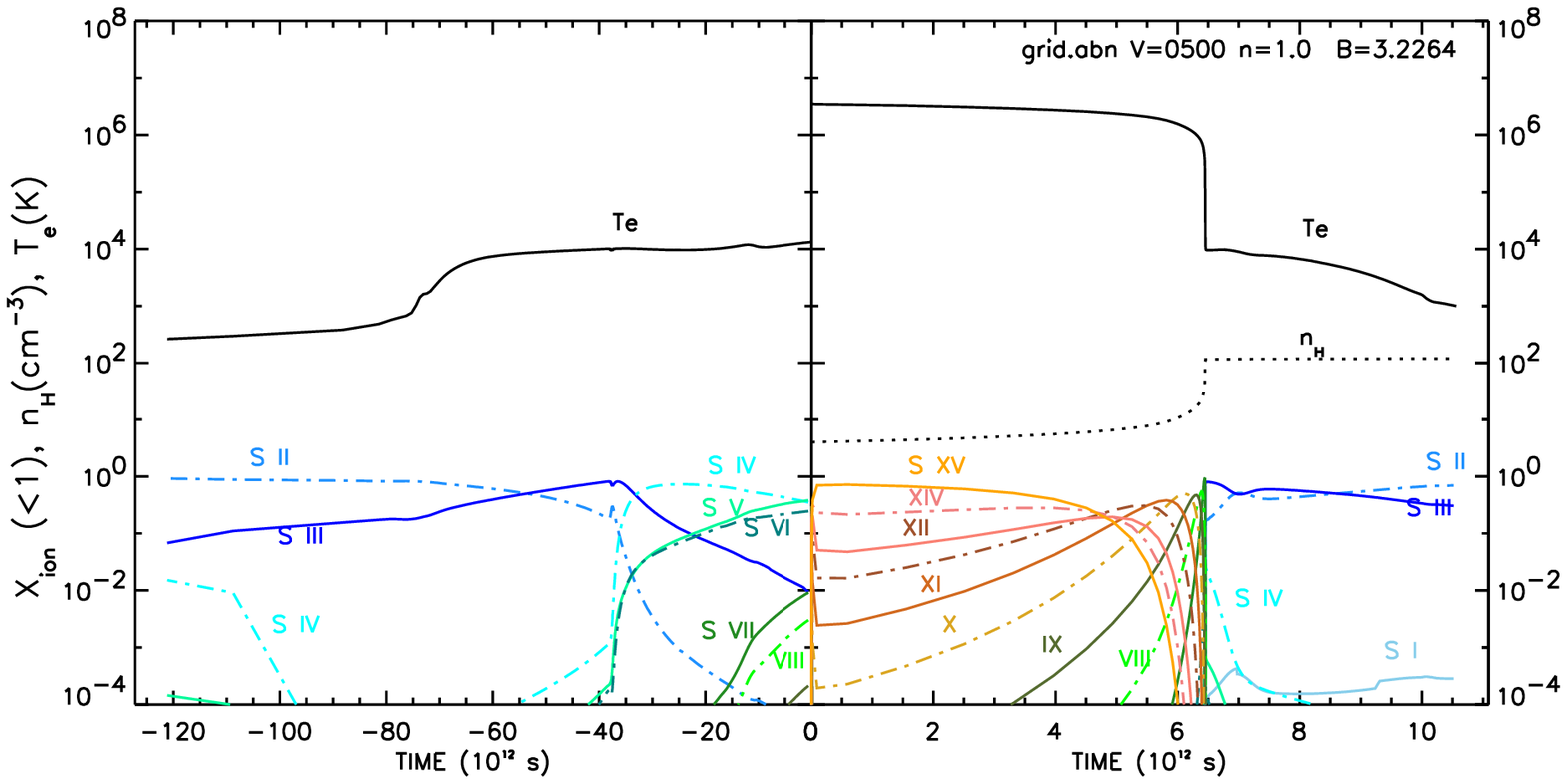} \\
\includegraphics[scale=0.6]{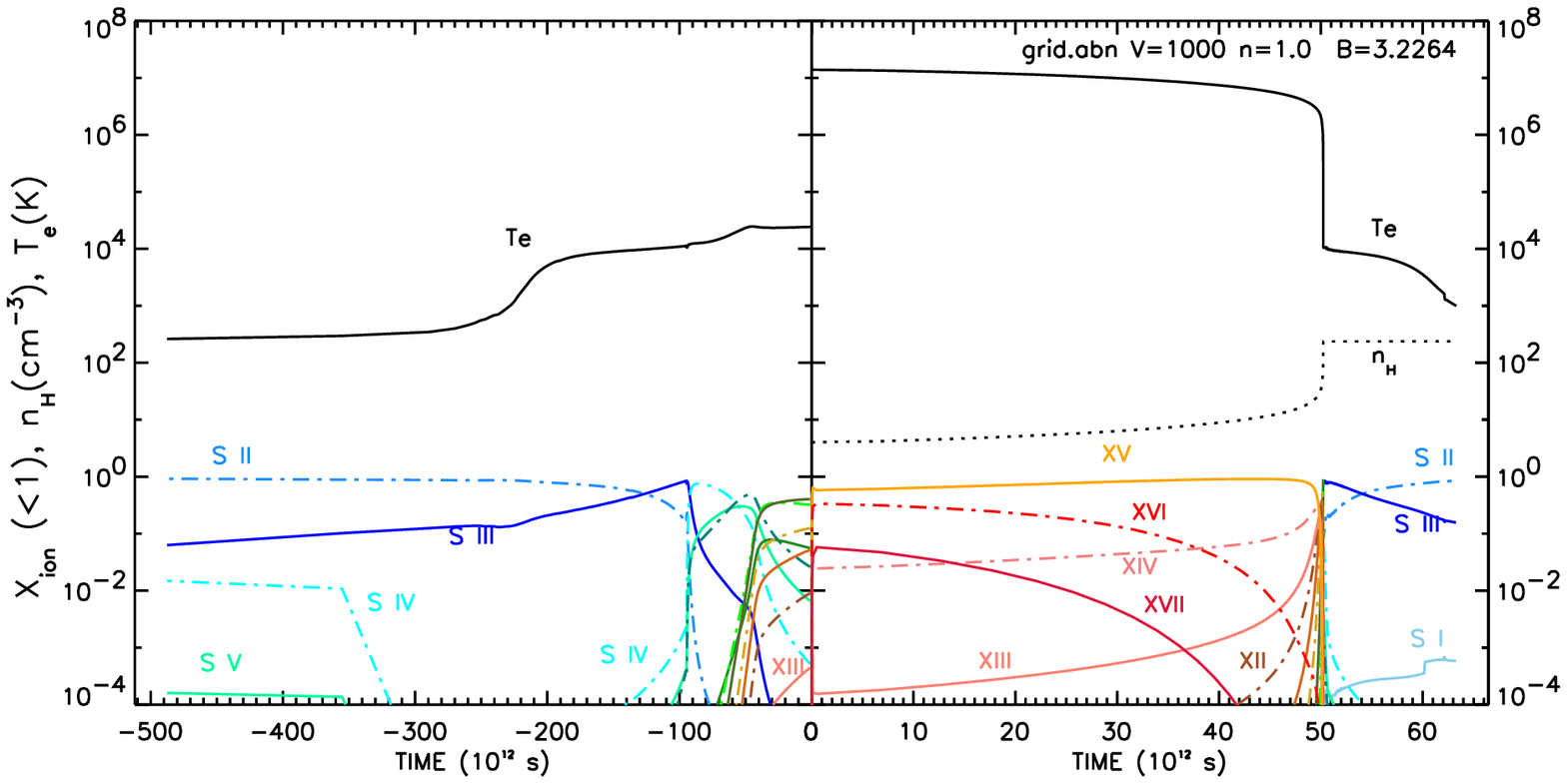} \\
\caption{The sulfur ionization structure, temperature profile and density
profile of models with solar abundance, equipartition magnetic field for shock
velocities of $v_{\rm s}$ = 200, 500, and 1000 km\ s$^{-1}$.  The axes are the
same as for Figure~\ref{Hstruct_eg}.  See the electronic edition of the
Journal for a color version of this figure.
\label{Sstruct_fig} }
\end{figure*}                

\subsection{Model Grid Input Parameters}\label{sec:input}

Each individual model in our shock library is defined by five physical
parameters; the pre-shock density, n, the shock velocity, $v_{s}$, the
pre-shock transverse magnetic field, B and the set of atomic abundances. The
library itself is organized into two main groups of model grids:

\begin{itemize}
\item First, complete grids of models calculated for five different atomic
  abundance sets; (depleted) Solar and 2$\times$Solar as used by
  DS96 and which are based upon the older \citet{Anders89}
  abundance set;  a Solar abundance set based upon the  \citet{Asplund05}
  abundances as listed in \citet{dopita2005}, which is referred to
  here as the `dopita2005' abundances; and an LMC and an
  SMC abundance set as given by \citet{Russell92}. The abundances of the
  individual elements (by number with respect to hydrogen) are listed for each
  of these abundance sets in Table~\ref{abund_table}. Each of these model
  grids assume a fixed pre-shock density of n=1
  cm$^{-3}$, and consists of a set of models with shock velocities
  covering the range 100 up to 1000 km 
  s$^{-1}$ in steps of 25 km s$^{-1}$, and magnetic fields of 10$^{-4}$, 0.5, 1.0, 2.0,
  3.23, 4.0, 5.0 and 10.0 $\mu$G.  As the pre-shock density is unity, the
  corresponding magnetic parameters are B/$\sqrt{\rm n}=10^{-4}$, 0.5, 1.0,
  2.0, 3.23, 4.0, 5.0 and 10.0 $\mu$G\ cm$^{3/2}$. 

\item Second, grids of models with solar abundance (as used by DS96),
calculated for densities of 0.01, 0.1, 1.0, 10, 100 and 1000 cm$^{-3}$. As
with the first set, each of these model grids covers shock velocities of 100
up to 1000 km\ s$^{-1}$ in steps of 25 km\ s$^{-1}$, and have the same
magnetic parameters of 10$^{-4}$, 0.5, 1.0, 2.0, 3.23, 4.0, 5.0 and 10.0
$\mu$G\ cm$^{3/2}$.  However, as the magnetic field values required to obtain
these magnetic parameters are of course different for each density, additional
models were computed in order to be able to also
compare models of different 
densities with the same transverse magnetic field. These are B$\sim10^{-3}$,
$\sim10^{-2}$, $\sim10^{-1}$, 1.0, 10 and 100 $\mu$G, calculated for
each density.

\end{itemize}

The magnetic field, B, and magnetic parameter, B/$\sqrt{\rm n}$, values are chosen
so as to cover the extremes expected in the ISM, while also sampling more
finely the magnetic field strengths which are near equipartition. Under
equipartition conditions the magnetic pressure is equal to the thermal
pressure, and the Alfv\'en speed is approximately equal to the gas sound
speed. Pressure equipartition occurs for ${\rm B}_{0}^{2}/4\pi \sim
n_0kT_0$ where 
B$_{0}$ is the transverse magnetic field, and n$_0$, $T_0$ are the pre-shock
densities and temperatures \citep{dopita2003}. This condition is satisfied for
magnetic parameters B$/\sqrt{\rm n} \sim 3-5$. The value B$/\sqrt{\rm n} \sim 3.23$
was chosen as the nominal equipartition value.
 
\begin{figure*}
\centering
\includegraphics[scale=0.6]{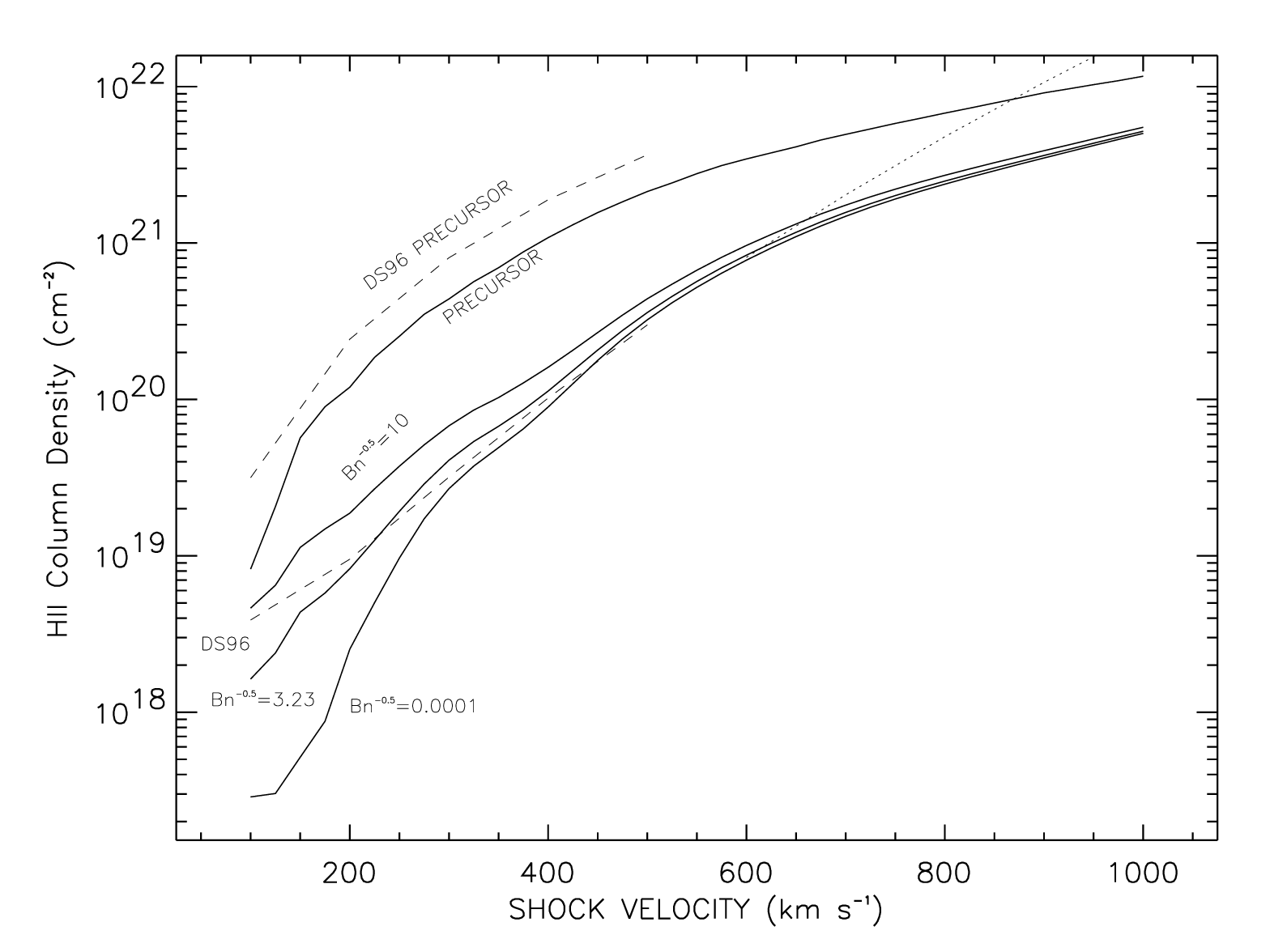} \\
\caption{Integrated hydrogen column densities for the shock and the precursor
structures. The column densities of the shock structure are shown for magnetic
parameters B/$\sqrt{\rm n}$=0.0001, 3.23 and 10. The curve labelled `PRECURSOR'
shows the column density of the precursor gas which does not depend on
the magnetic parameter. The dashed curves show the shock and precursor column
densities for the DS96 models, and the dotted curve shows the extrapolation of
the column density scaling relation for the shock column density of
DS96. \label{coldens_fig}}
\end{figure*}                

\begin{figure}
\centering
\includegraphics[scale=0.60]{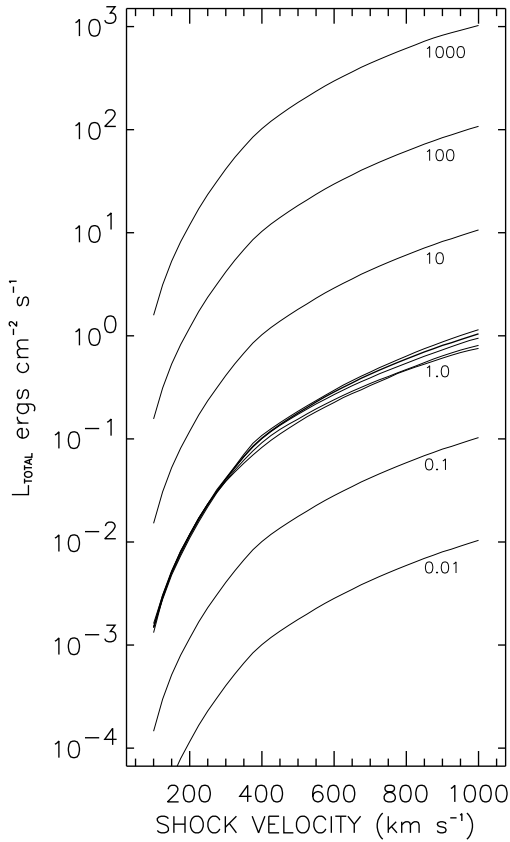} 
\caption{Total radiative fluxes of shocks as a function of shock velocity and 
various values of the pre-shock densities. 
The bold line is for the standard model having a solar abundance set, a
density n$=1$ cm$^{-3}$, and an equipartition pre-shock magnetic field. The
thinner lines near the bold line show the  luminosities of the n$=1$ cm$^{-3}$,
equipartition models with abundances of twice solar, LMC and
SMC abundance sets. \label{lum_compare1_fig}}
\end{figure}

\begin{figure}
\centering
\includegraphics[scale=0.60]{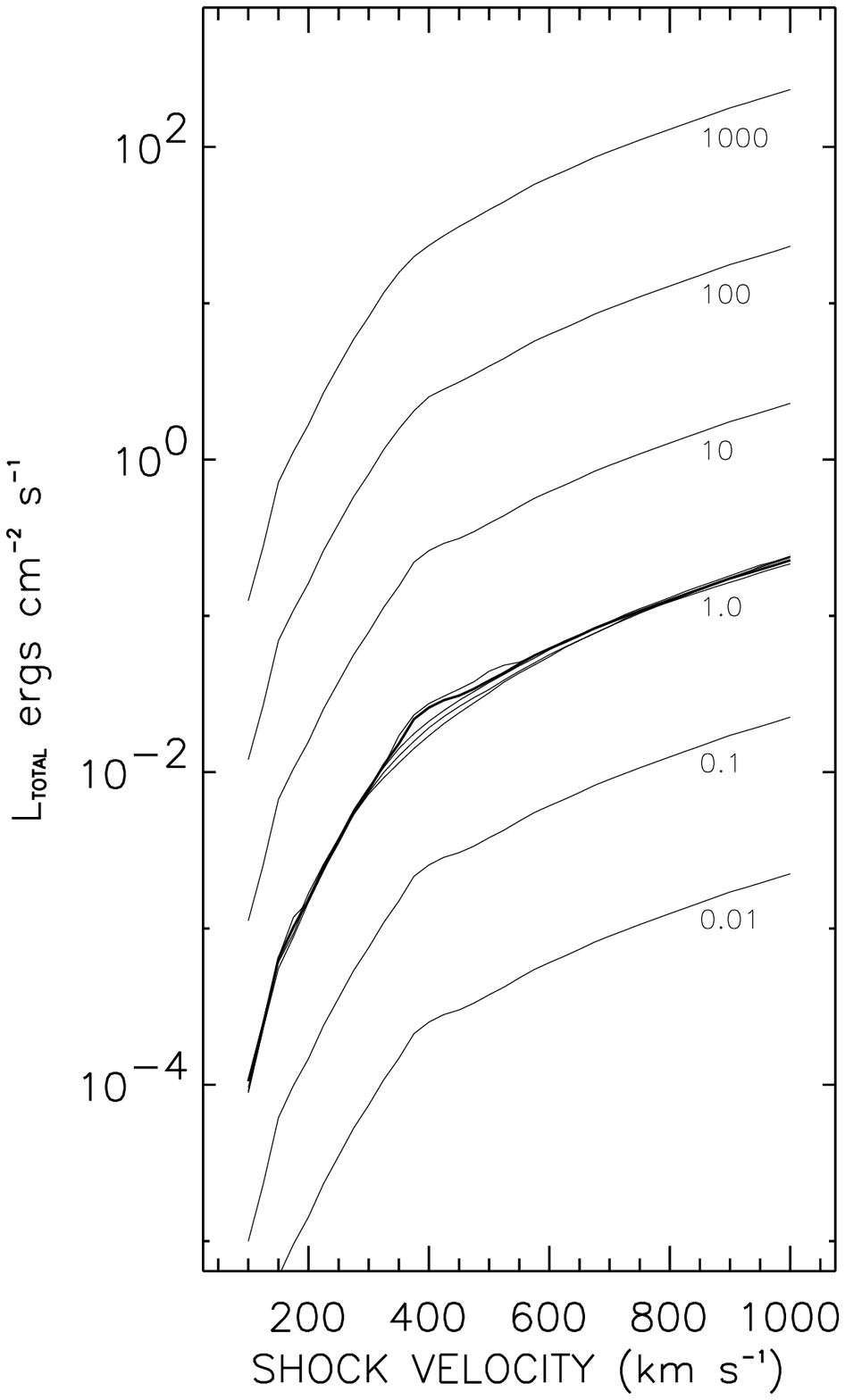} 
\caption{As for Figure \ref{lum_compare1_fig}, but showing the
precursor total radiative fluxes. 
\label{lum_compare2_fig}}
\end{figure}

The input parameters for the all of the models are provided in
Table~\ref{param_table}. Each row of the table represents a velocity sequence
of models at a given abundance, density and magnetic field. The table is
organized into ten sets of models. The first five sets of models represent the
model grids for the five different abundances, and the following 5 sets of
models are the solar abundance models for different input densities and
magnetic fields. The number and range of model input parameters allows the
construction of various 2-D, or higher dimensional, grids of models.

\section{Ionizing Radiation Generated by the Shock}\label{sec:ionrad}

The ionizing radiation produced in the cooling zone behind the shock front
shock is mostly composed of thermal bremsstrahlung (free-free) continuum and
resonance lines arising from many different elements and ionic stages. The
underlying exponential shape of the continuum is emphasized in
Figure~\ref{specfig1} where we show the ionizing spectra generated in the
n=1.0 cm$^{-3}$ solar abundance shock models. As can be seen, 
higher velocity shocks result in
harder and more luminous ionizing spectra, with the spectral slope in
the log-linear plot clearly flattening with increasing velocity. Note
that the histogram nature of the figure also reveals the high energy
(frequency) resolution of the MAPPINGS spectral vector.
Figure \ref{specfig2} shows the ionizing spectra of
400 and 1000 km\ s$^{-1}$ shocks on a $\nu F_{\nu}$ scale, illustrating both
the stronger UV fluxes and harder X-ray spectra generated in high velocity
shocks. This figure also shows a prominent low-temperature bound-free
continuum of hydrogen, produced in the cool, partially-ionized zone of the
recombination region of the shock, and the strong hydrogen two-photon
continuum produced mostly by the down-conversion of Ly$\alpha$ photons trapped
in this same region of the shock structure. Also present, though to a much
weaker scale, is the bound--free continuum arising from the heavier elements,
with the helium continuum the most obvious.

\begin{figure*}[htb]            
\centering  
\includegraphics[scale=0.95]{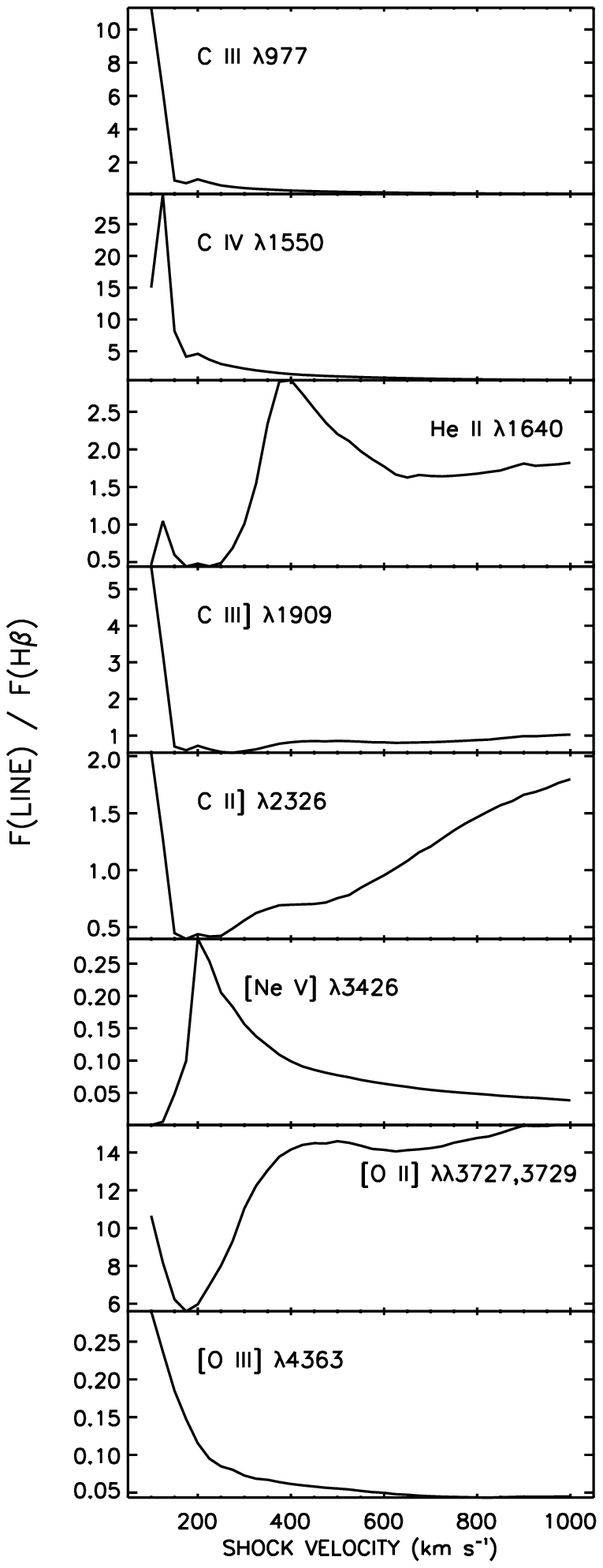} 
\includegraphics[scale=0.95]{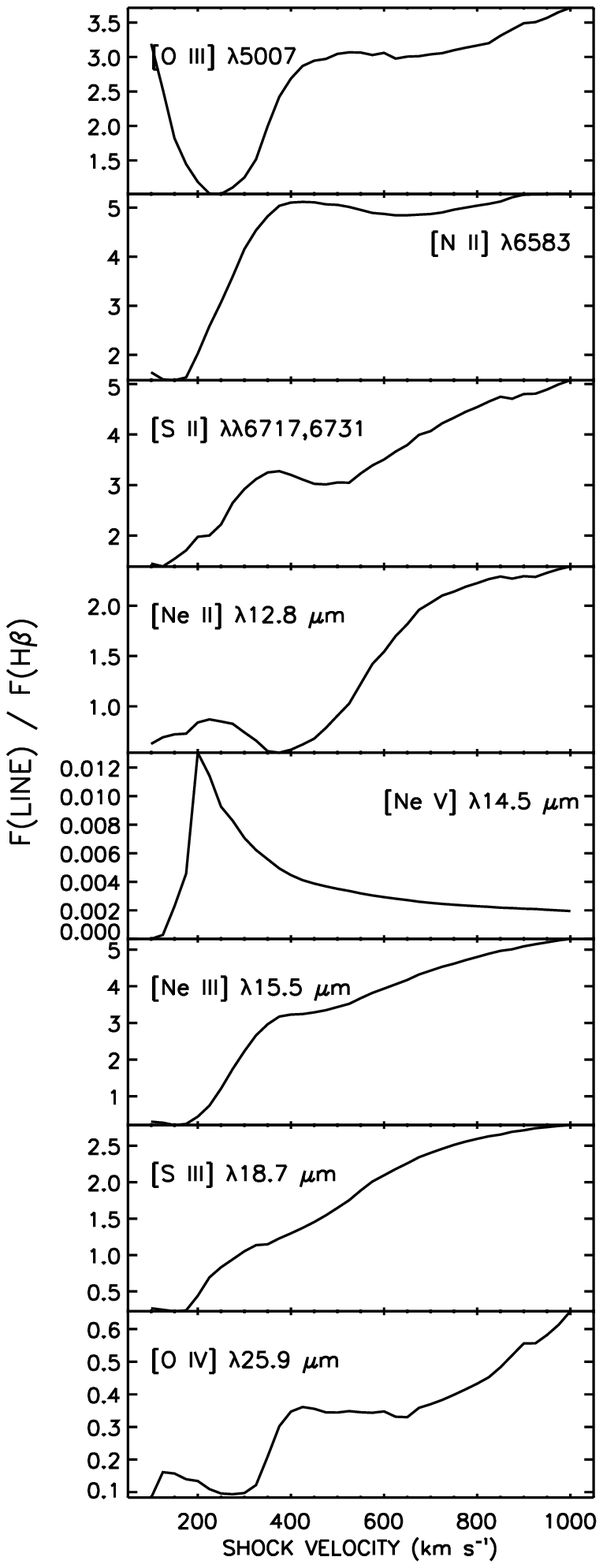} 
\caption{Line strength relative to H$\beta$ for 16 strong emission
lines between the Lyman Limit and 26$\mu$m as a function of shock
velocity. These are
computed for the shock component only, and are for the fiducial model with
solar abundance, n$=1.0$ cm$^{-3}$ and B$=3.23~ \mu$G. \label{M_n1_be_s_llf}}
\end{figure*}

\begin{figure*}[htb]              
\centering
\includegraphics[scale=0.95]{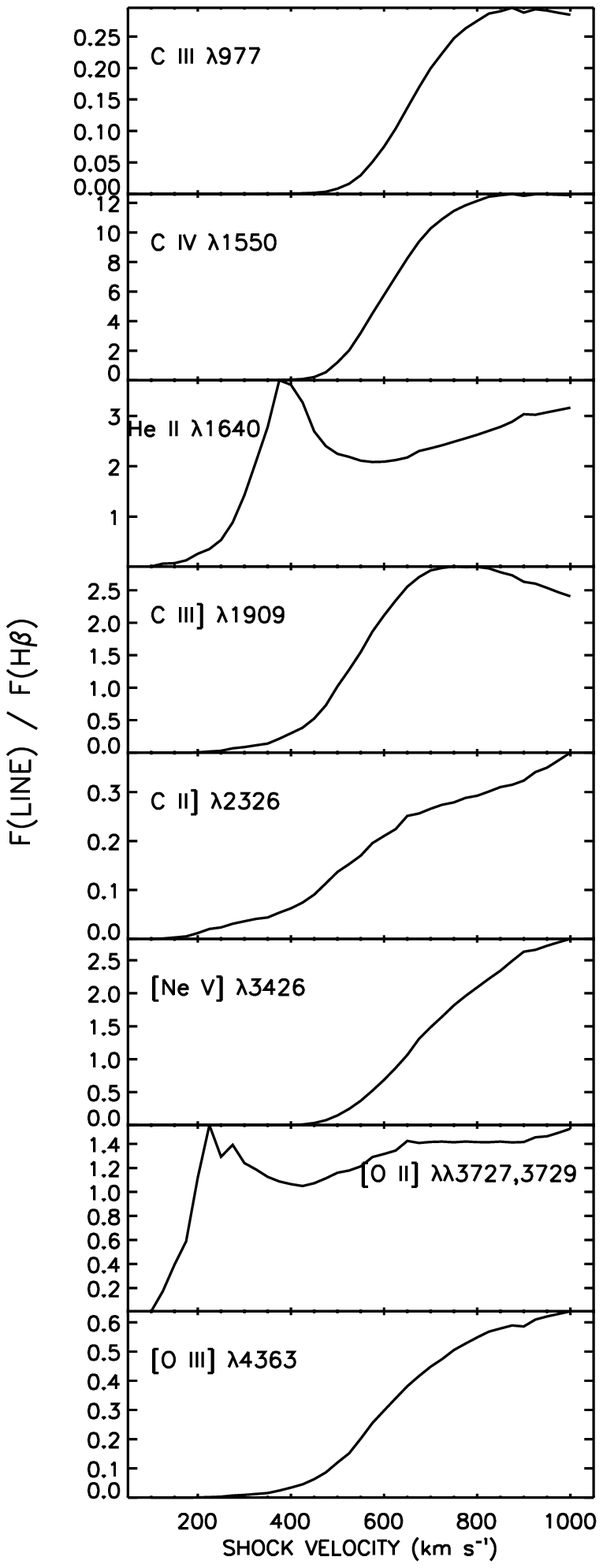} 
\includegraphics[scale=0.95]{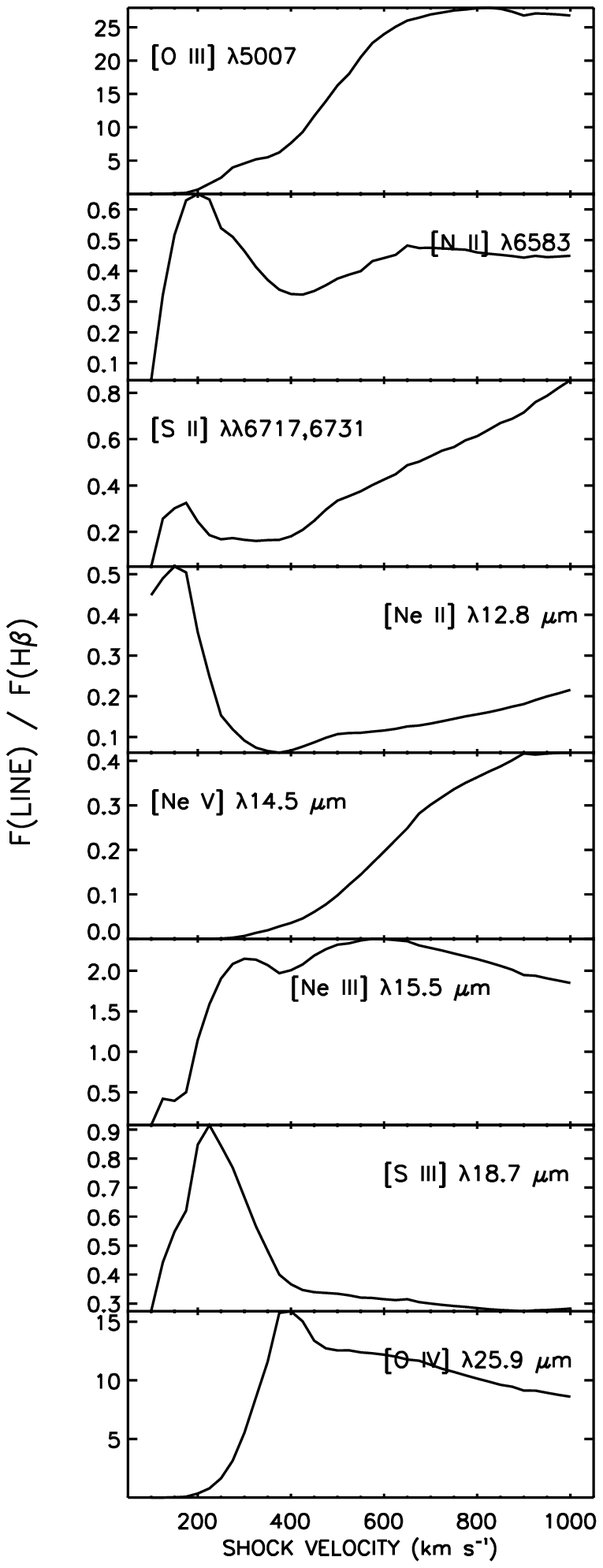} 
\caption{As Figure \ref{M_n1_be_s_llf} but for the precursor component
of the models only. \label{M_n1_be_p_llf} } 
\end{figure*}

\begin{figure*}[htb]              
\centering
\includegraphics[scale=0.95]{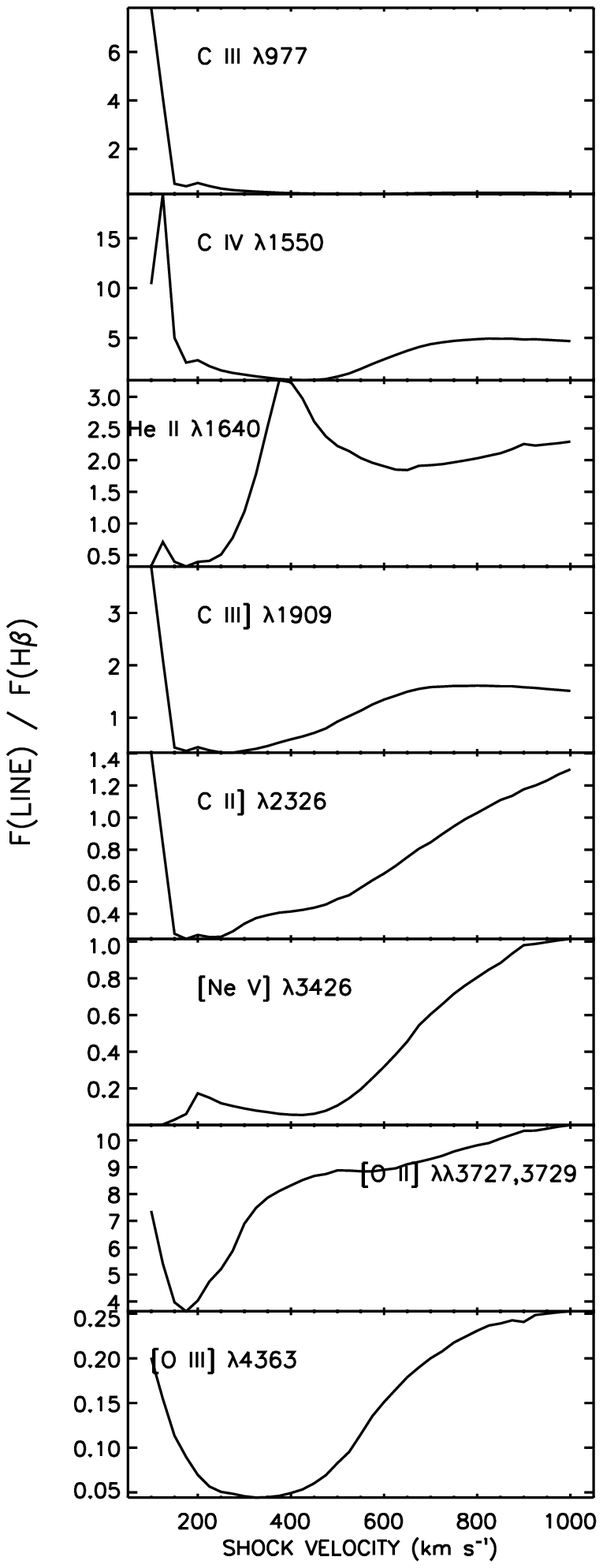} 
\includegraphics[scale=0.95]{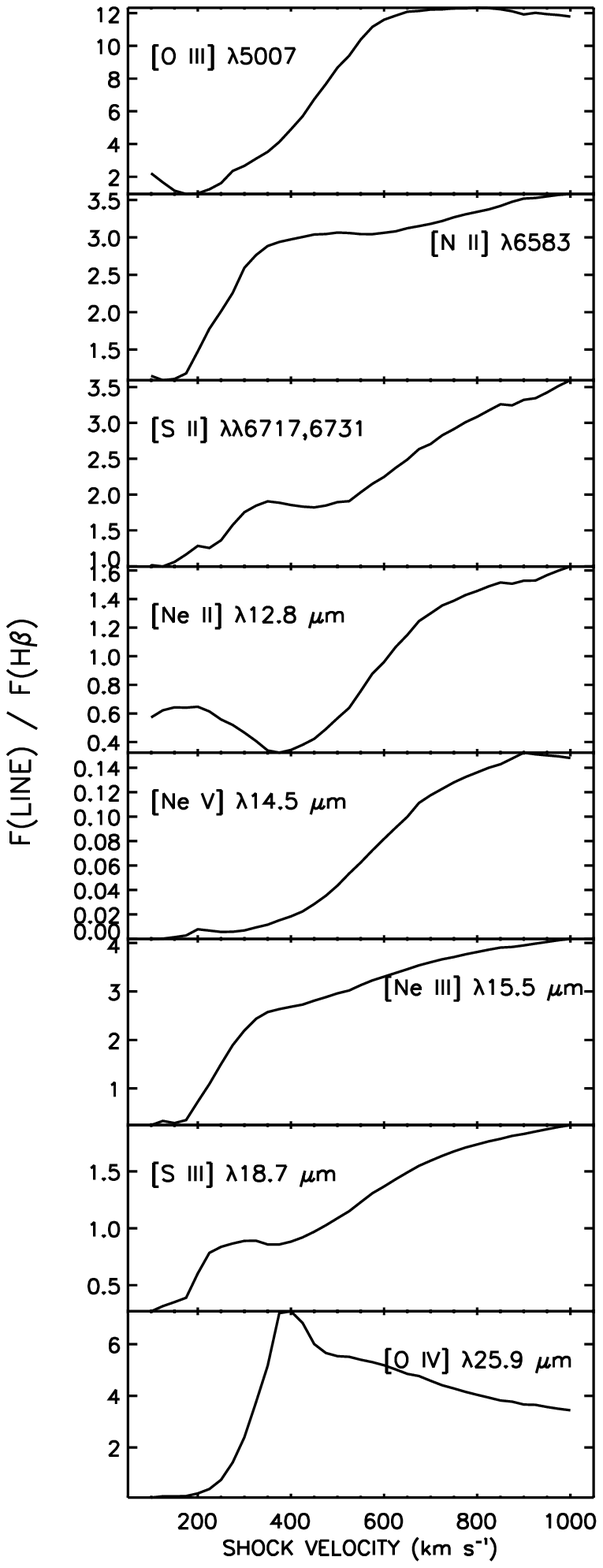} 
\caption{As Figure \ref{M_n1_be_s_llf} but for the the full
shock+precursor model. \label{M_n1_be_sp_llf}} 
\end{figure*}                

The strength of the ionizing field is a strong function of the shock velocity,
but does not significantly depend on either the atomic abundance or the
magnetic field. This is because the ionizing field is dominated by the
bremsstrahlung radiation, whose strength is controlled by the temperature and
density of the radiative zone, which is determined by the shock velocity and
pre-shock density. As magnetic field support in this cooling zone is
negligible, it has little affect on the emission, and, as hydrogen dominates
both electron and ion numbers for both the bremsstrahlung and free-bound
emission, changes in metallicity have only a small impact.

\begin{figure*}              
\centering
\includegraphics[scale=0.6]{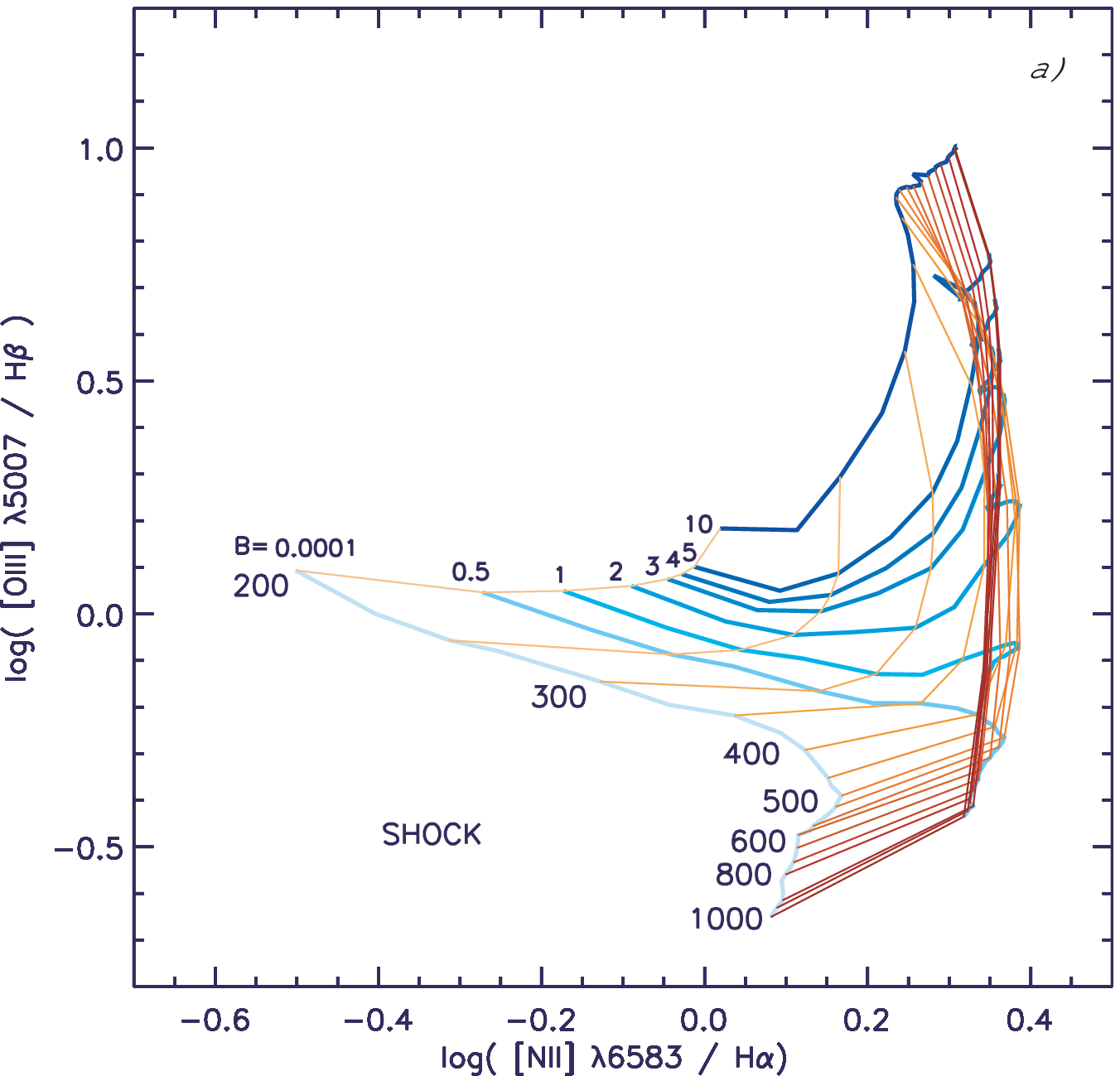}
\includegraphics[scale=0.6]{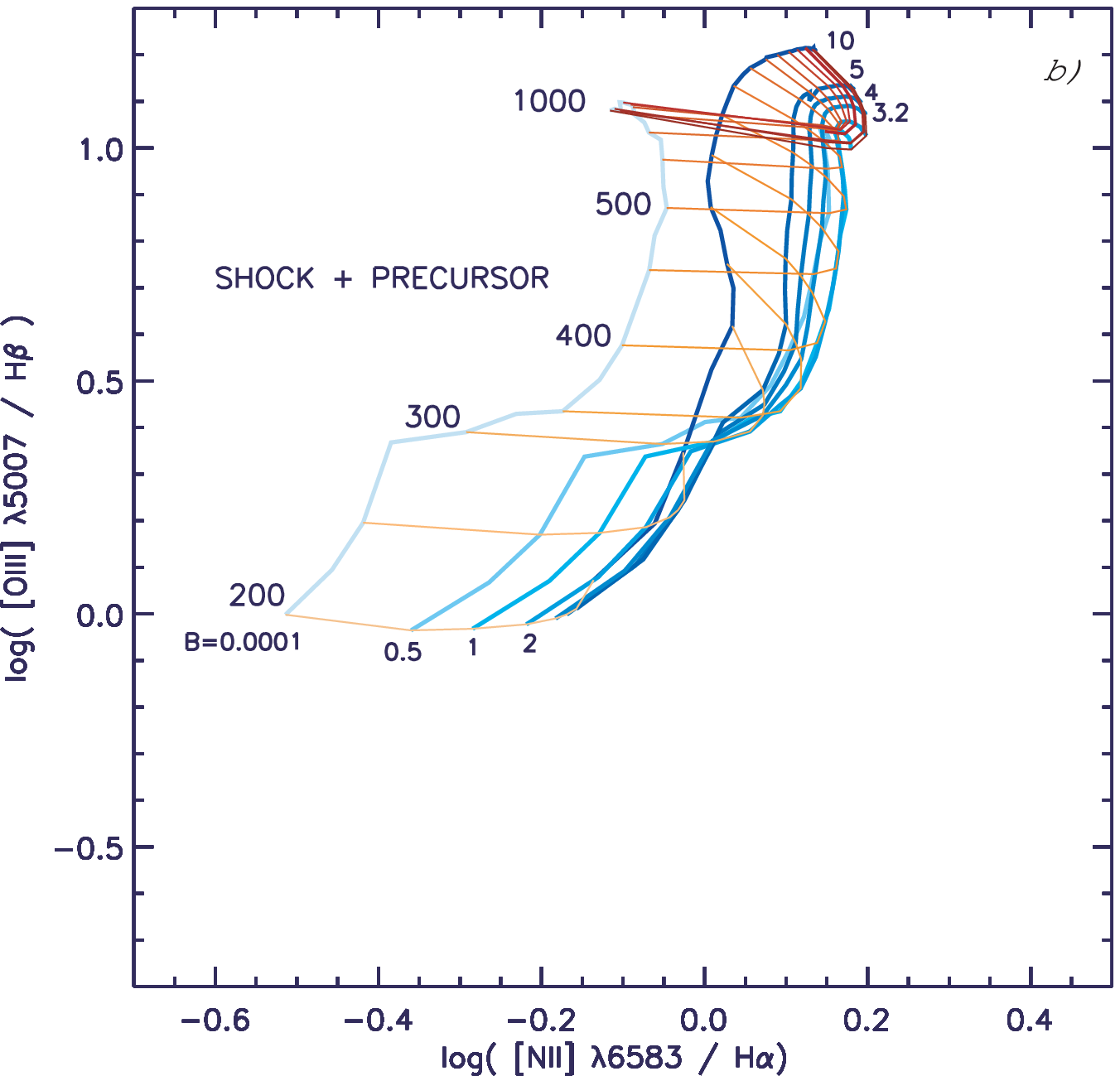}
\caption{ The classic \cite{veilleux1987} [\ion{O}{3}]$\lambda$5007/H$\beta$
versus [\ion{N}{2}]$\lambda$6583/H$\alpha$ diagnostic diagram for the solar
abundance n$=1$ cm$^{-3}$ models.  The left figure shows the shock-only
models, and the right figure shows the shock+precursor models. The grid is
comprised of lines of constant magnetic parameter shown with a bold linestyle,
and lines of constant shock velocity shown as the thin lines. The lines of
constant shock velocity are shown at 50 km~s$^{-1}$ intervals (and are color
coded with increasing red intensity for higher shock velocities in the
electronic edition). The velocity sequences are shown for all values of the
magnetic parameter (and are color coded with increasing blue intensity for
higher magnetic parameters in the electronic edition). Both the shock-only and
shock+precursor grids are shown for the shock velocities of 200--1000
km~s$^{-1}$, for which all the models have a fully ionized precursor.  See the
electronic edition of the Journal for a color version of this figure.
\label{lrat1}}
\end{figure*}                

The hydrogen ionizing radiation flux, $L_{\rm UV}$, integrated for all
energies $h\nu > 13.6$\ eV, and over 2$\pi$\ sr is listed in
Table~\ref{ion_prop_table}.  It is found to scale almost exactly as the
available enthalpy as:

\begin{equation}
\label{Luv_eq}
{\rm L}_{\rm UV} = 2.44 \times 10^{-4} \left(\frac{v_S}{100\ {\rm km}\
{\rm s}^{-1}}\right)^{3.02} \times \left(\frac{\rm n}{{\rm
cm}^{-3}}\right) {\rm ergs}\ {\rm cm}^{-2}\ {\rm s}^{-1} 
\end{equation}

The ionizing fields may also be characterized in terms of their ionization
parameter in the pre-shock gas.  As in DS96 we give ionization parameters,
$Q$, defined as the mean number of photons passing through unit area divided
by the total pre-shock particle density; $Q = N_{photons}/n_T$ (cm\
s$^{-1}$). Table~\ref{ion_prop_table} lists $Q$ in the \ion{H}{1}-ionizing
13.6$< h\nu <$ 24 eV band, $Q({\rm H})$, the \ion{He}{1}-ionizing 24$< h\nu <$
54 eV band, $Q({\rm He I})$, and the \ion{He}{2}-ionizing band $h\nu >$ 54 eV,
$Q({\rm HeII)}$ as a function of shock velocity. We also provide the
corresponding $R-$Type ionization front velocity $v_{ion}$\ and the
equilibrium electron temperatures, $T_e$ just ahead of the shock front.

\section{Shock and Precursor Structures}\label{sec:struct}

We now consider the ionization structures and the physical scales of
both the shock and precursor components. Figure \ref{Hstruct_eg} shows
the ionized and neutral hydrogen structures (as labelled) of the shock
and precursor components of the 500 km\ s$^{-1}$ solar abundance
model, along with the hydrogen density and electron temperature
profiles. Note that the hydrogen density is not plotted within the
precursor region as it is set at the constant value of n$_{\rm H}=1.0$
cm$^{-3}$ . The profiles are plotted in the frame of the shock front,
considered with respect to the time axis. This axis represents time
since (or prior to) the passage of the shock front, and can be
converted into a physical distance from the shock front via the shock
velocity.

In Figure~\ref{Hstruct_eg} (and subsequent ionization structure
Figures~\ref{Hstruct_V}--\ref{Sstruct_fig}), the structure of the precursor
is shown in the left panel and the shock structure is shown in the right
panel. The shock front is located at time $t=0$. The time taken to reach
equipartition in the electrons, ions, and un-ionized gas is assumed to be
negligible, so that the shock is therefore unresolved and the density and
electron temperature jump discontinuously at the shock front. Note also that
the time axes are shown in units of 10$^{12}$s, but the scales and ranges are
necessarily different for the shock and precursor panels of these diagrams, as
well as between the diagrams with different parameters.

As described in DS96 the a number of zones can be identified in the shock
structures. These are indicated in Figure~\ref{Hstruct_eg} and are,
respectively:

\begin{enumerate}
\item \emph{The ionization region.} This region, just after the shock
  front, is in an ionization state below that appropriate for the
  electron temperature assuming collisional ionization equilibrium
  (CIE), due to discontinuous temperature jump at the front.  After
  the passing of the front, the gas rapidly (barely resolved in the
  figures) adjusts from the pre-shock ionization state to the
  appropriate CIE state for the post-shock temperature. The rate at
  which it adjusts depends upon the collisional ionization rates of
  these species. As discussed in DS96, this region has strong line
  emission due to this state of high temperature and relatively
  low-ionization.

\item \emph{The high temperature radiative zone.} This is the zone in which
  most of the EUV and soft X-ray flux is emitted. The ionization state
  is approximately in coronal equilibrium for the temperature of the
  region. Even though the cooling rate of this region is low due to
  the high temperature, it is
  the dominant contributor to the radiation field of the shock,
  assisted by the high-ionization, optically-thin state of the gas.

\item \emph{The non-equilibrium cooling zone.} Once the temperature of
  the post shock gas drops below $\sim 10^6$ K, the cooling rate
  becomes very high and the recombination timescales of a large number
  of ions becomes longer than the local cooling timescale. As a
  result, the ionization state lags and the plasma is in a higher
  degree of ionization than collisional ionization equilibrium would
  suggest. The relatively high ionization state for the temperature
  means that the collisional
  line emission is weak for these intermediate ionization
  species, and that the gas remains optically thin to the diffuse
  radiation of the previous region.

\item \emph{The super-cooled zone} is the region where photoionization
  starts to become important in determining the ionization balance.
  Initially, however, the ionization state is still too high to
  efficiently absorb the ionizing photons. This leads to an
  over-cooling, and the temperature falls below the value given by
  photoionization equilibrium. The width of this region is dependent
  upon both the recombination and photoionization timescales of the ions.

\item \emph{The photoabsorption and recombination zone.} This is essentially
  an equilibrium plane-parallel \ion{H}{2} region illuminated by the
  downstream EUV photon field. The density in this region is much higher than
  the pre-ionized region ahead of the shock, so that the effective ionization
  parameter in this zone is much lower, and the \ion{H}{2} region is very much
  thinner. Eventually the EUV photons are absorbed, and hydrogen finally
  recombines. However, there is a cool, partially ionized region behind the
  main recombination front. This has a temperature of a few thousand degrees,
  and is ionized by the hard penetrating X-rays which undergo Auger processes
  and these lead to the emission of fast electrons which in turn produce
  secondary ionization cascades.

\end{enumerate}

\begin{figure}[htb]
\includegraphics[scale=0.6]{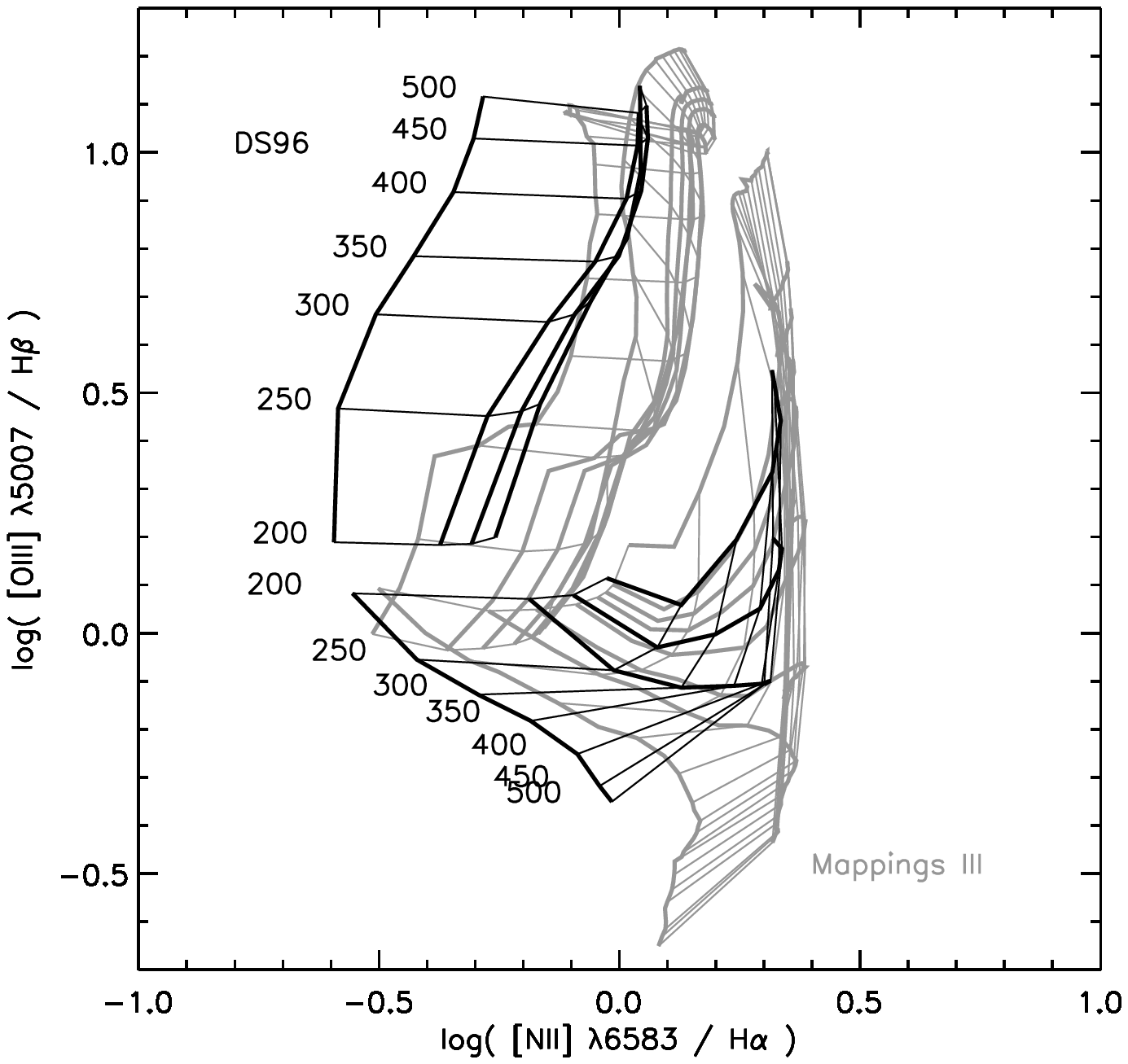} 
\caption{ Comparison of the new \mapiii\ models with the DS96 on the
[\ion{O}{3}]$\lambda$5007/H$\beta$ versus [\ion{N}{2}]$\lambda$6583/H$\alpha$
diagram.  The DS96 models are shown in black, they are labelled with the shock
velocity in km~s$^{-1}$, and include velocity sequences for magnetic
parameters of 0, 1, 2 and 4 $\mu$G\ cm$^{3/2}$ which increase from left to right
in both the shock-only (lower) and shock+precursor (upper) grids. The new
\mapiii\ models are shown in greyscale and cover the same range in shock
velocity and magnetic field as shown in Figure~\ref{lrat1}.  Note that the new
diagrams show systematic changes with respect to the earlier ones, and that
the new models show that these ratios become insensitive to shock velocity for
$v_{\rm s} > 500$ km~s$^{-1}$.
\label{lrat4}}
\end{figure}

\begin{figure*}[htb]              
\includegraphics[scale=0.6]{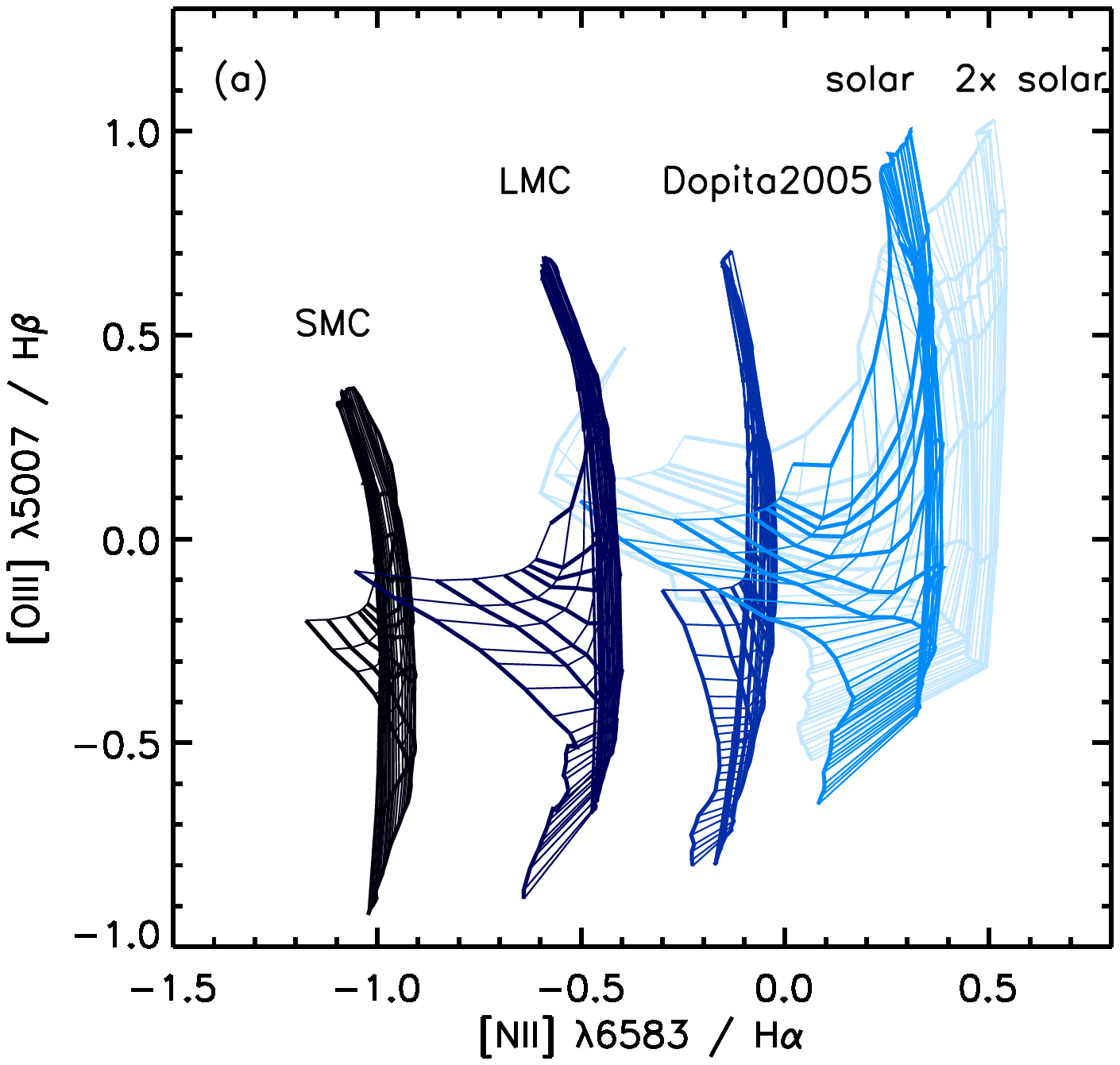} 
\includegraphics[scale=0.6]{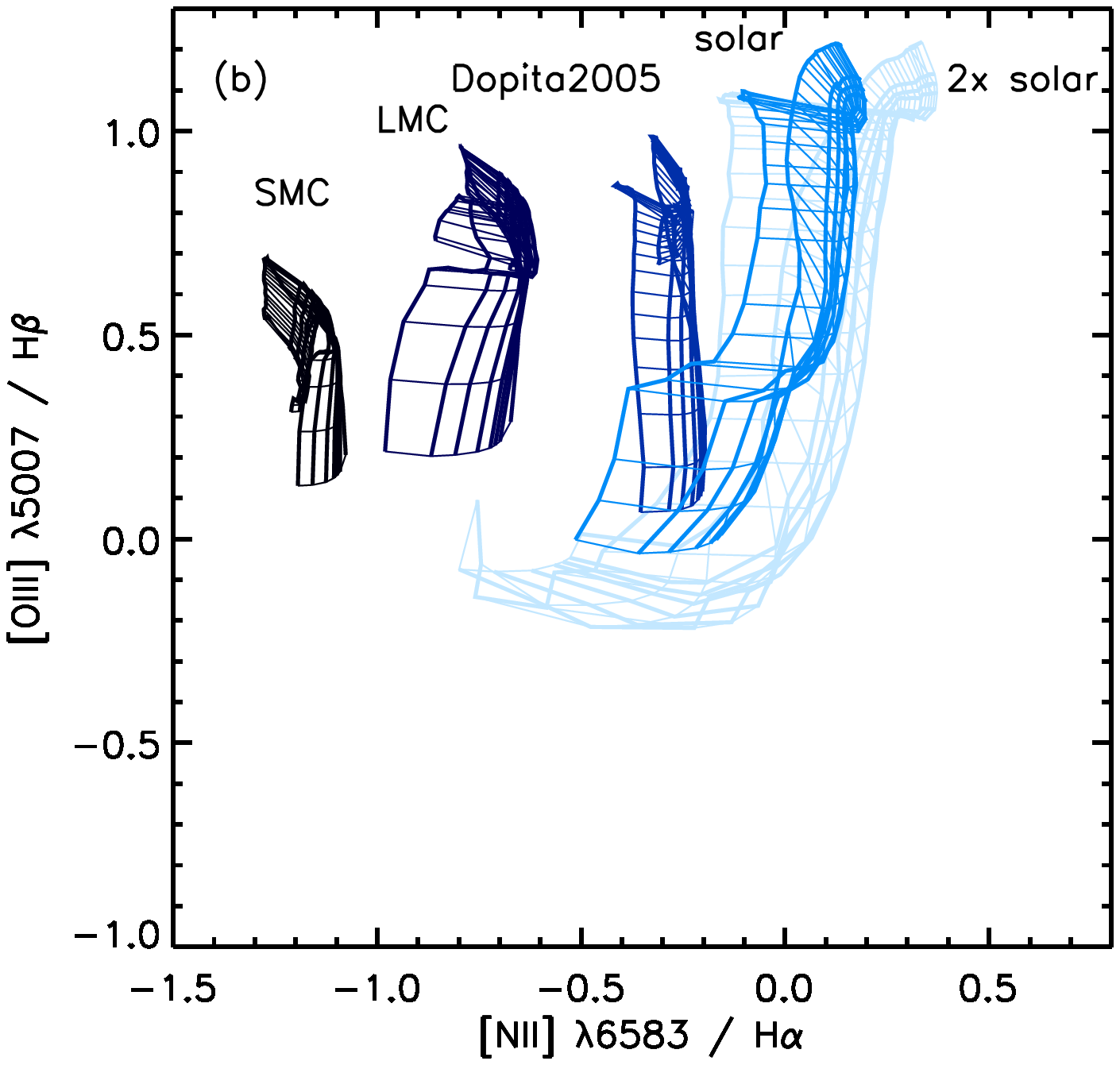} 
\caption{ Shock and shock+precursor model grids for the five different atomic
abundance sets used in the shock model library for the [\ion{O}{3}]/H$\beta$
versus [\ion{N}{2}]/H$\alpha$ diagram. The shock-only models are shown in (a),
and the shock+precursor models are shown in (b). All model grids are shown for
$v_{\rm s}$ 200--1000 km~s$^{-1}$, with lines of constant shock velocity drawn
at intervals of 50 km~s$^{-1}$.  Each grid is labelled with the abundance set
used, and generally moves from left to right as the total metallicity
increases.  See the electronic edition of the Journal for a color version of
this figure.
\label{lrat2}}
\end{figure*}                

\subsection{Hydrogen ionization structures}

Figures \ref{Hstruct_V} - \ref{Hstruct_abund} show, respectively, the
effects of changing velocity, magnetic field, pre-shock density and abundance
on the structures of the hydrogen ionization fraction, hydrogen number
density, n$_{\rm H}$, and the electron temperature, $T_{\rm e}$, in the shock
and precursor regions. Note that in each diagram the time scale changes due to
the effects of the changing parameters. In each diagram the parameters
of the model (abundance set, velocity, density, and magnetic field
respectively)  are listed in the top right corner for clarity.

\begin{deluxetable*}{lrrrrrr}
\tablecolumns{7}
\tablewidth{0pc}
\tablecaption{Shock Ionizing Properties \label{ion_prop_table}}
\tablehead{
\colhead{v$_{s}$}    &
\colhead{${\rm L}_{\rm UV}$}    &
\colhead{Q(H \sc{I})}    &
\colhead{Q(He \sc{I})}    &
\colhead{Q(H \sc{II})}    &
\colhead{v$_{\rm ion}$}         &
\colhead{T$_e$}             \\
\colhead{(km\ s$^{-1}$)}         &
\colhead{(ergs\ cm$^{-2}$s$^{-1}$)}    &
\colhead{(km\ s$^{-1}$})         &
\colhead{(km\ s$^{-1}$})         &
\colhead{(km\ s$^{-1}$})         &
\colhead{(km\ s$^{-1}$})         &
\colhead{(K)}             } 
\startdata
       200  & 8.041E-4  &       209.1   &      43.0    &    4.9     &    214.2    &   7033   \\
       300  & 3.519E-3  &       430.8   &      312.1   &    83.1    &    688.3    &   10288   \\
       400  & 9.106E-3  &       664.3   &      598.7   &    513.2   &    1480.0   &   12145  \\ 
       500  & 1.751E-2  &       985.2   &      1195.0  &    602.2   &    2318.3   &   15435  \\
       600  & 2.968E-2  &       1158.2  &      1628.2  &    800.7   &    2989.2   &   18486  \\
       700  & 4.631E-2  &       1559.2  &      1986.0  &    1094.0  &    3865.8   &   20516  \\
       800  & 6.815E-2  &       1958.8  &      2362.3  &    1542.7  &    4885.8   &   21934  \\
       900  & 9.645E-2  &       2375.0  &      2920.4  &    2206.6  &    6251.7   &   23389  \\
       1000 & 1.318E-1  &       2740.6  &      3262.7  &    2786.8  &    7325.0   &   24760  
\enddata
\end{deluxetable*}

The dominant parameter of the shock (and precursor) structure and emission is
the velocity, as is seen in Figure~\ref{Hstruct_V}. This figure shows our
fiducial n$=1.0$ cm$^{-3}$, B$/\sqrt{\rm n}=3.23~\mu$G cm$^{3/2}$, solar
abundance model with two velocities; $v_{\rm s}=200$ and 1000 km\,s$^{-1}$
(three including the $v_{\rm s}=500$ km\,s$^{-1}$ shown in Figure
\ref{Hstruct_eg}). There is a clear increase of both the extent and
temperature of the radiative zone as the velocity increases due to the greater
shock strength, leading to the changes in the ionizing radiation field shown in
Figures \ref{specfig1} and \ref{specfig2}. This increase of both the hardness
and luminosity of the radiation field leads to the greater extent and
ionization in the precursor region, visible through the increased temperature
in this figure and the higher ionization species seen in Figures
\ref{Hestruct_fig}--\ref{Sstruct_fig}. The increased radiation field also
affects the ionization of the post-shock recombination region. This is
mollified somewhat however by the density increase associated with the faster
shock, which leads to a smaller increase in the ionization parameter, though
the effects of the harder radiation are still visible.

\begin{figure*}             
\includegraphics[scale=0.6]{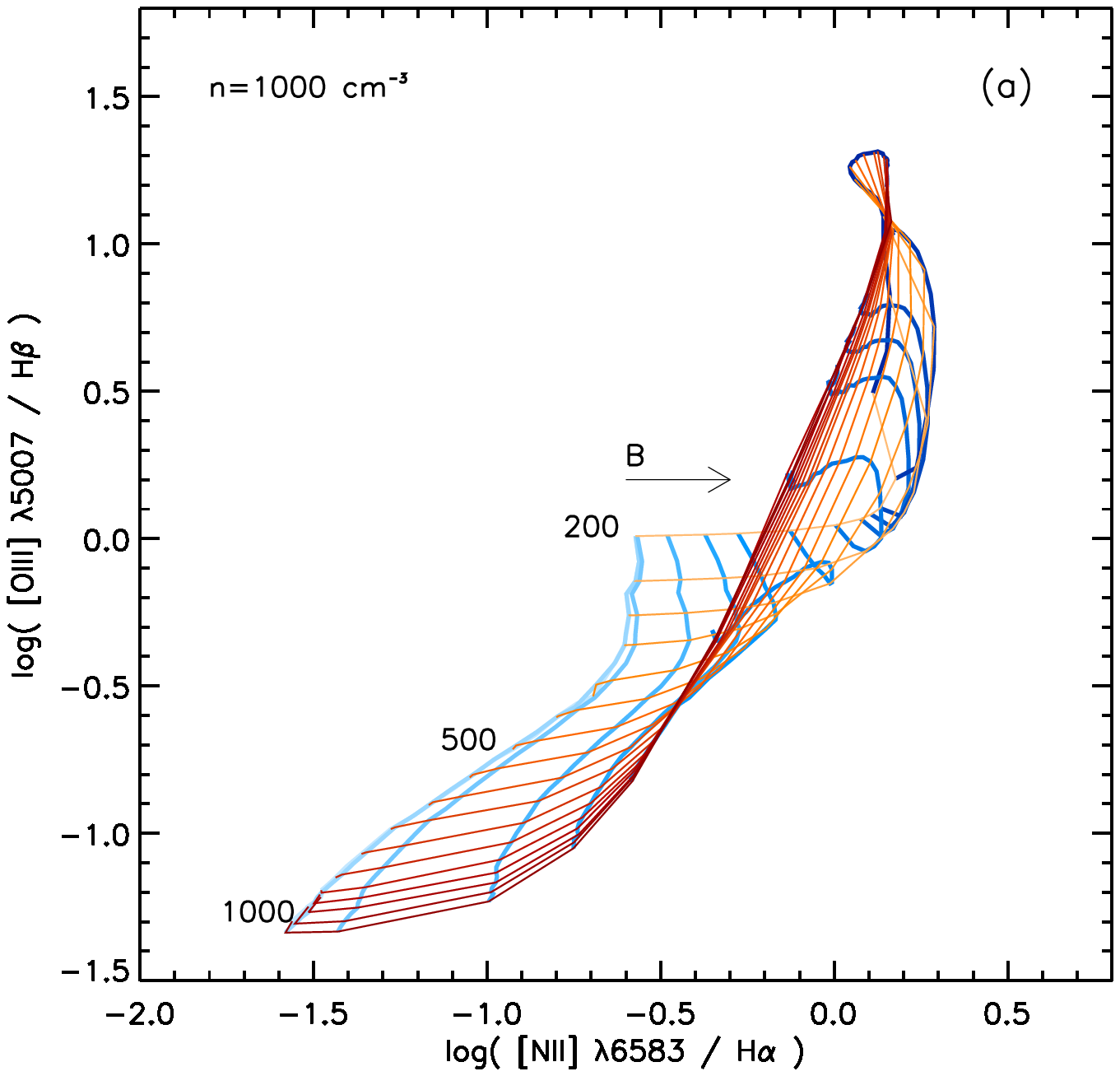}
\includegraphics[scale=0.6]{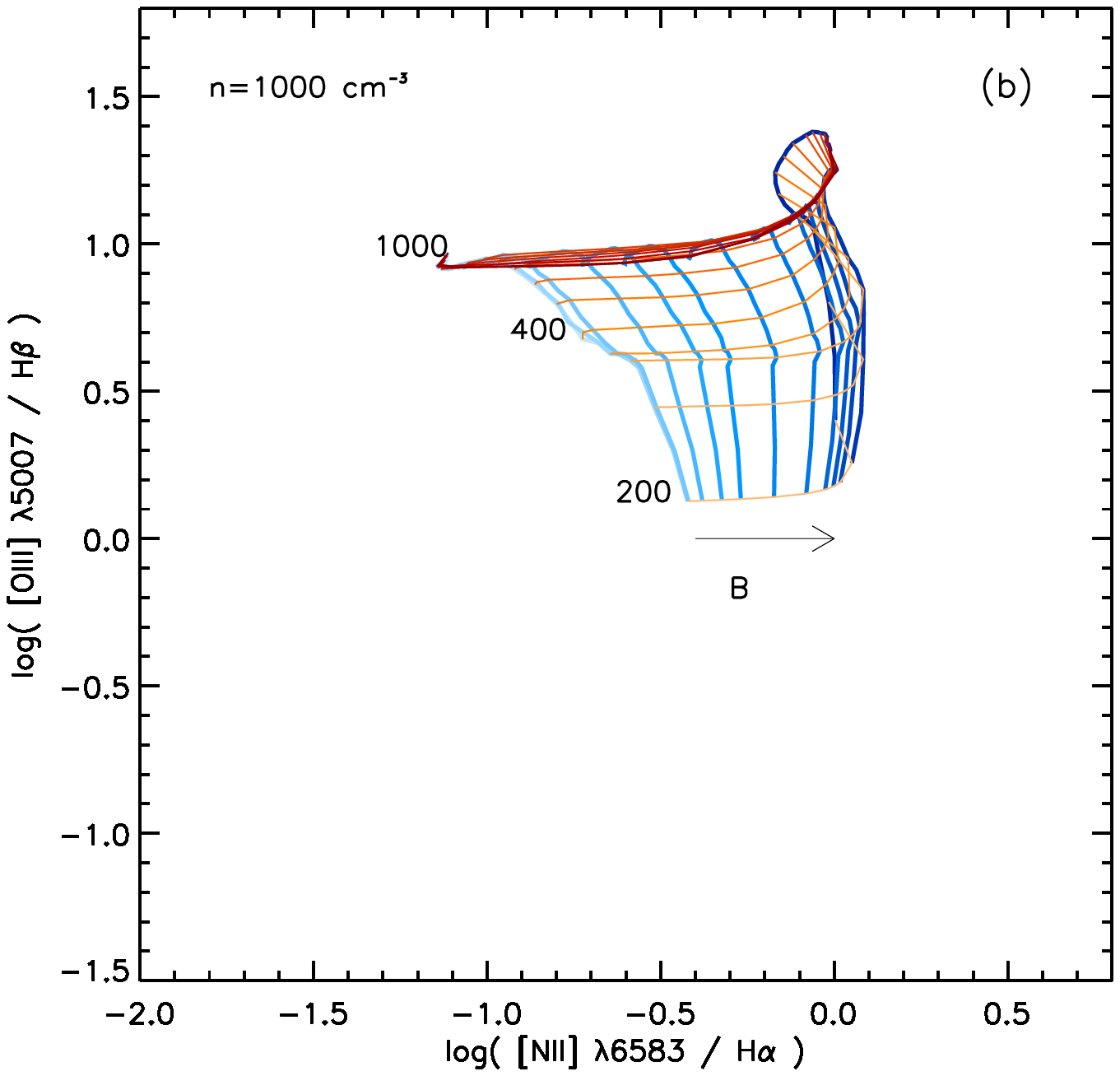} \\
\includegraphics[scale=0.6]{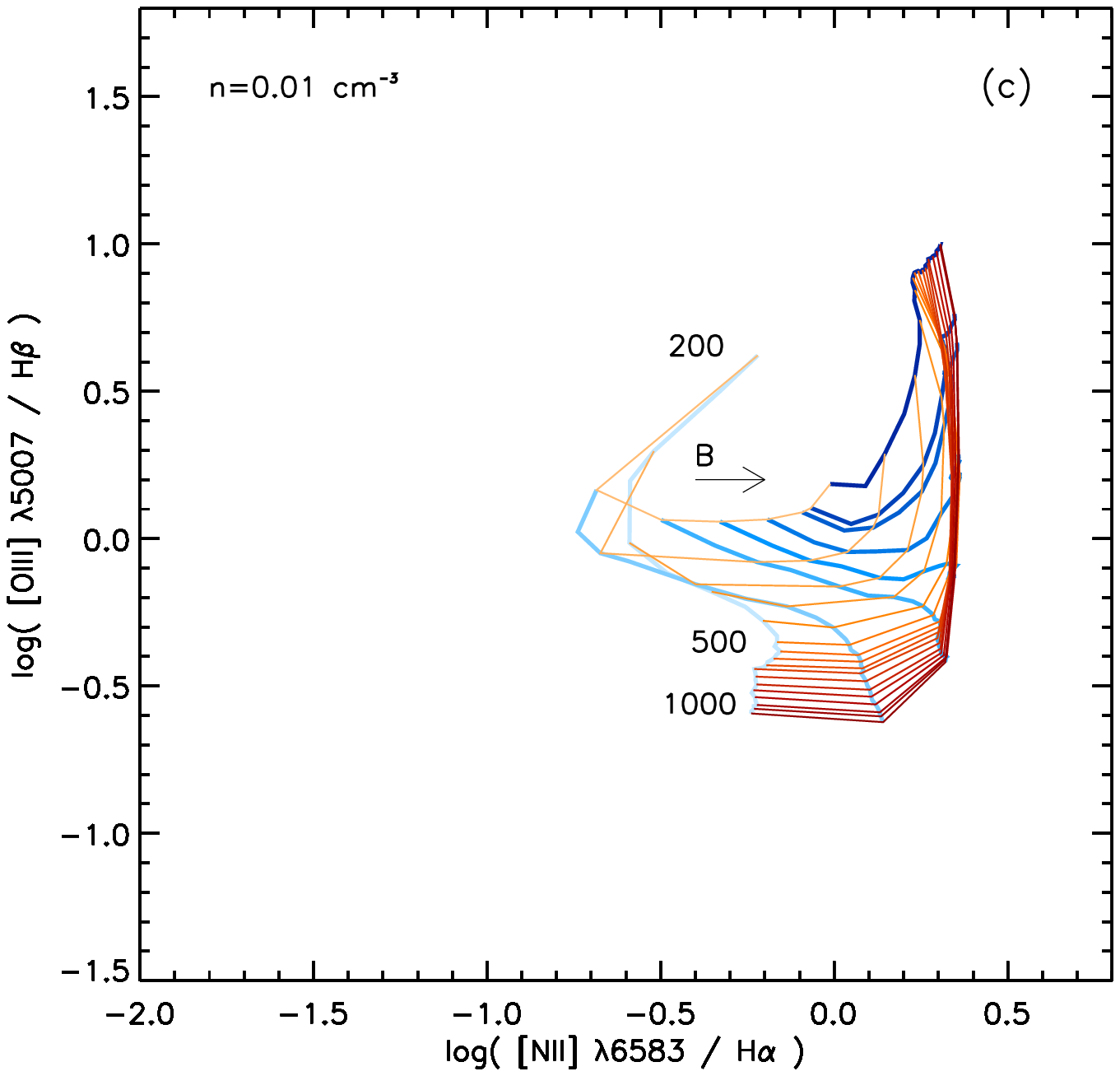}
\includegraphics[scale=0.6]{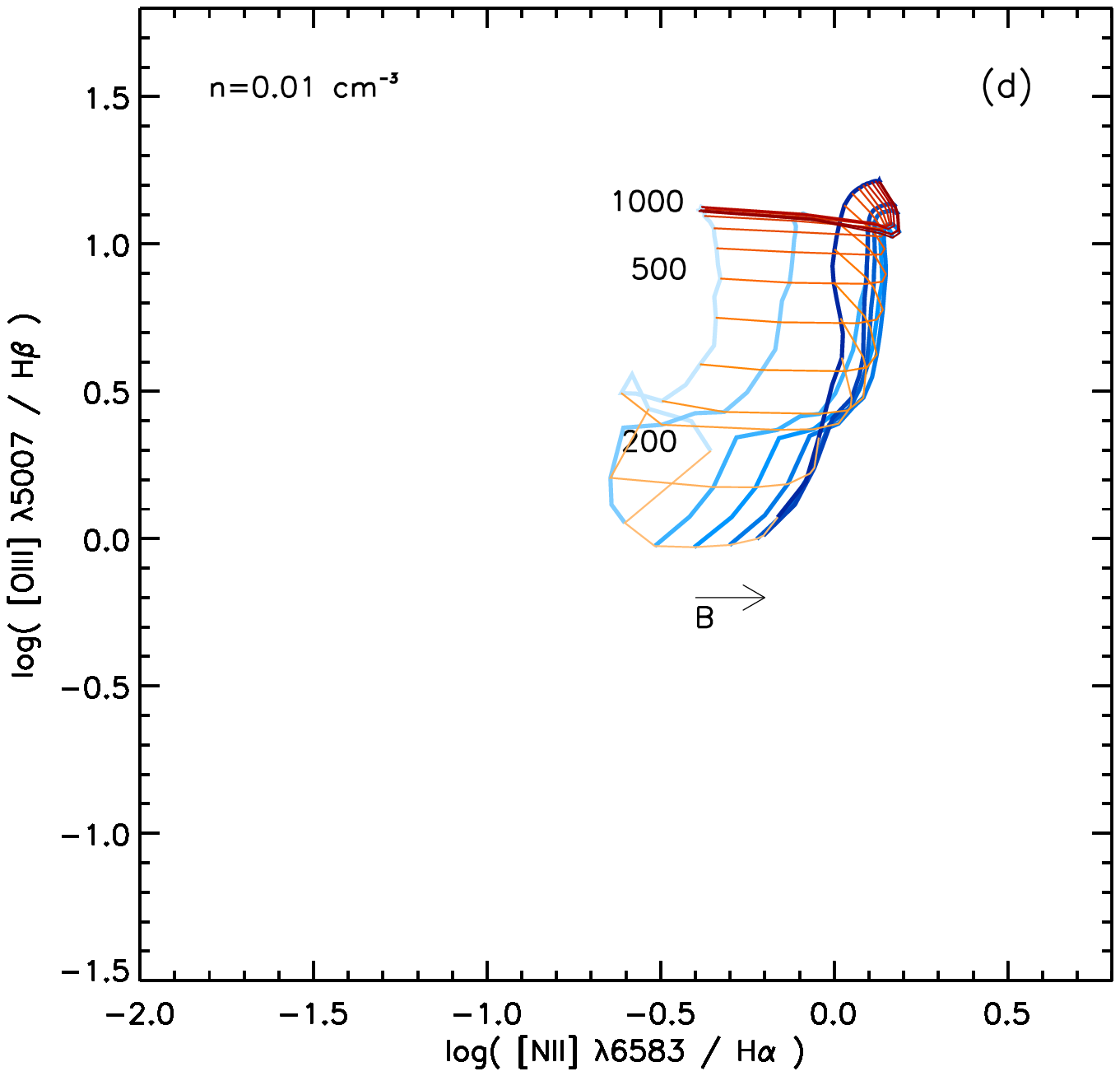}
\caption{Shock and shock+precursor model grids of solar abundance models for
two different pre-shock densities on the [\ion{O}{3}]$\lambda$5007/H$\beta$
versus [\ion{N}{2}]$\lambda$6583/H$\alpha$ diagram.  The top figures show the
high density case and the lower figures the low density case, with a) The
n$=1000$ cm$^{-3}$ shock model grid, b) the n$=1000$ cm$^{-3}$ shock+precursor
models, c) The n$=0.01$ cm$^{-3}$ shock model grid, and d) the n$=0.01$
cm$^{-3}$ shock+precursor models. The grids show shock velocities over the
range $v_{\rm s}$=200-1000 km~s$^{-1}$ with lines of constant magnetic
parameter shown as thick lines (and colored with increasing blue intensity for
higher magnetic parameter in the electronic edition). The thin lines represent
constant shock velocity and these are plotted for the full range of magnetic
parameters for each density as listed in Table~2 (and are shown colored with
increasing red intensity for higher shock velocities in the electronic
edition). See the electronic edition of the Journal for a color version of
this figure.
 \label{lrat3}}
\end{figure*}

Figure~\ref{Hstruct_B} shows models with n=1.0 cm$^{-3}$, $v_{\rm s}=500$
km\,s$^{-1}$, and solar abundance for differing magnetic fields. The top panel
shows a model with the magnetic parameter set to B$/\sqrt{\rm n}=0.0001 \mu$G
cm$^{3/2}$, and the bottom panel has B$/\sqrt{\rm n}=10.0 \mu$G cm$^{3/2}$,
which can be compared with the equipartition model with B$/\sqrt{\rm n}=3.23
\mu$G cm$^{3/2}$ model in Figure \ref{Hstruct_eg}.  In all three cases the
precursor structure is almost unchanged, as the radiation field and luminosity
arising from the shock are little affected by changes in B (except at
extremely high values), as discussed in sections \S \ref{sec:ionrad} and \S
\ref{sec:lum}.  The insensitivity of the radiation field to B can be seen in
this figure through the similarity of the radiative zone between the
models. This similarity, a result of the insensitivity of the temperature jump
to the magnetic field, means that the ionizing bremsstrahlung radiation that
dominates the energetics is practically the same. However, what is sensitive
to the transverse magnetic field is the density in the recombination zone
behind the shock front. As discussed in DS96, the maximum compression in this
post-shock region is driven by the ratio of the post- to pre-shock magnetic
fields. Thus a higher initial magnetic field results in a lower density
post-shock gas, and therefore, given the similarity in radiation field, a
higher effective post-shock ionization parameter. This results in a
significantly different ionization and emission structure in the post-shock
gas, leading to the strong diagnostics seen in DS95 and in this work (\S
\ref{sec:specsig}).

Figure \ref{Hstruct_dens} shows the effect of changing the pre-shock
density. We plot our fiducial solar abundance models, with $v_{\rm s}=500$
km\,s$^{-1}$ and B$/\sqrt{\rm n}=3.23~\mu$G cm$^{3/2}$, for each of the
densities n=0.1, 10, 
and 1000 cm$^{-3}$, from top to bottom (with the n=1.0 cm$^{-3}$ case shown in
Figure \ref{Hstruct_eg}). Note that the magnetic field, B, is also varied in
order to keep the magnetic parameter constant for the models shown.
This figure demonstrates that at a given shock 
velocity, the region influenced by the shock scales with density (to the
extent that collision de-excitation effects are unimportant), proving
explicitly that to first order the product n$_0t$ is an invariant with
changing density, as asserted by DS96.

\begin{figure*}[htb]              
\includegraphics[scale=0.6]{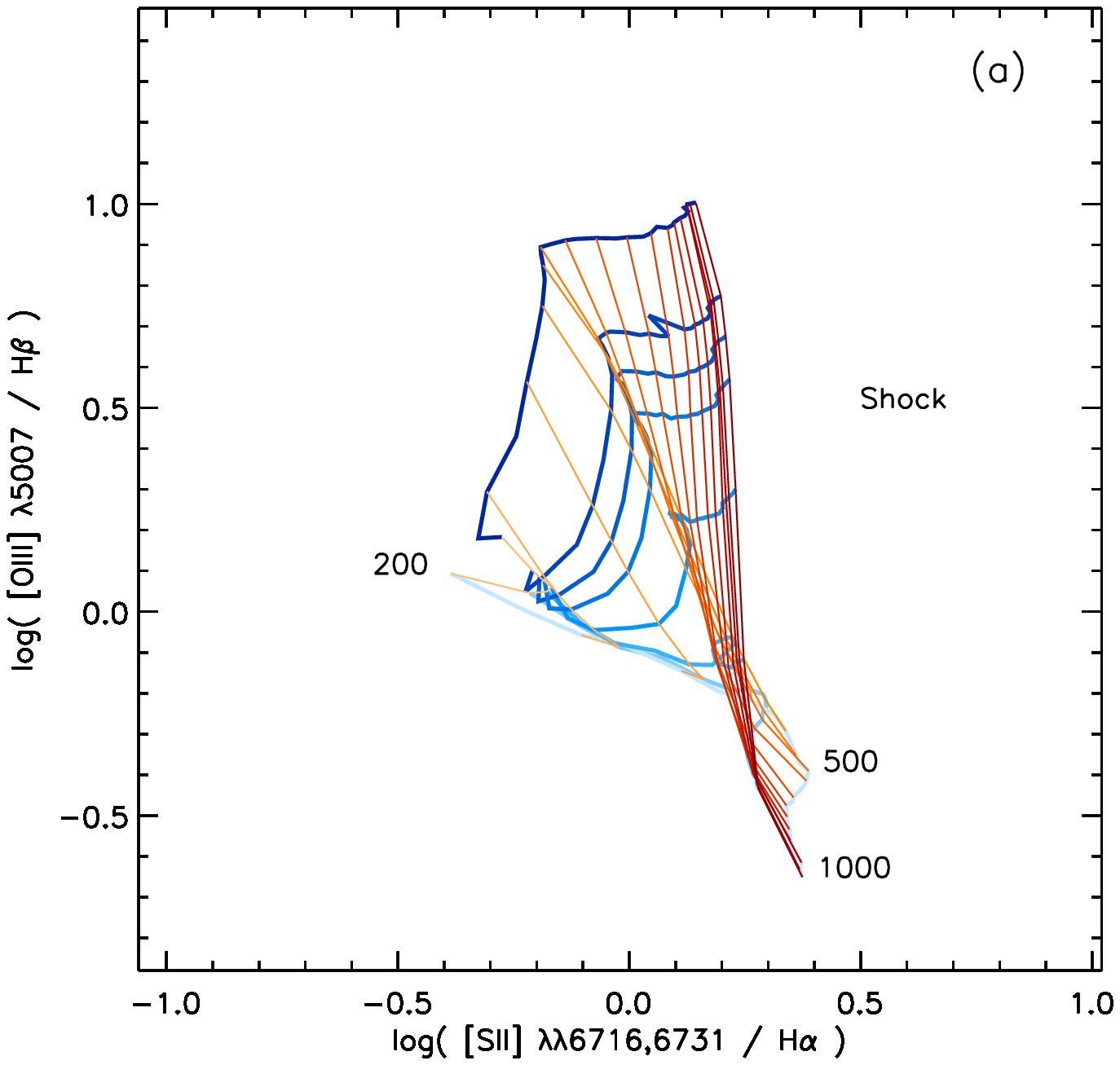} 
\includegraphics[scale=0.6]{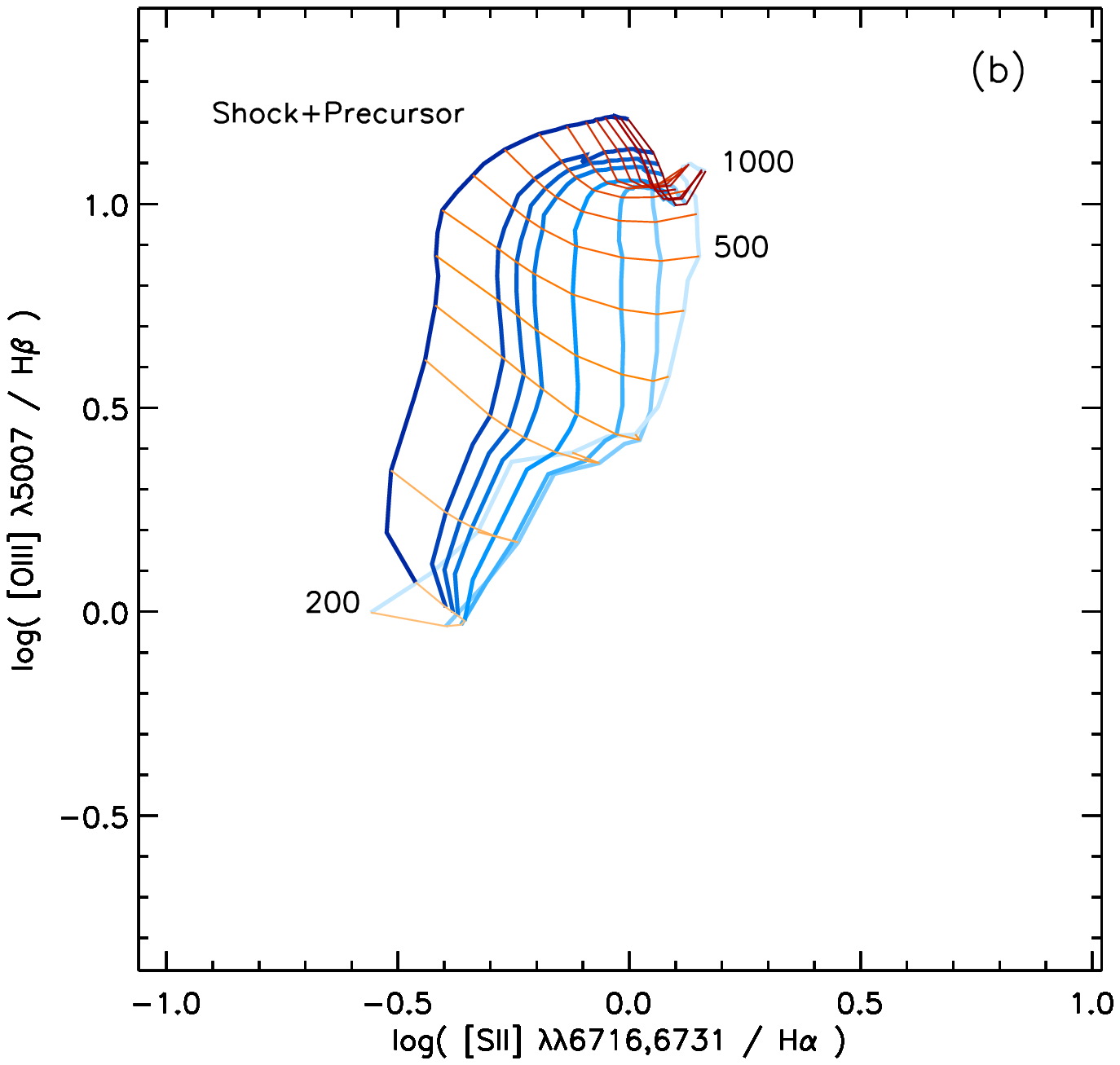} 
\caption{ The \cite{veilleux1987} plot of [\ion{O}{3}]$\lambda$5007/H$\beta$
versus [\ion{S}{2}]$\lambda\lambda$6716,31/H$\alpha$ diagram for a) the shock
and b) the shock + precursor models with n$=1$\ cm$^{-3}$ and solar abundance.
The range and step size of the shock velocity and magnetic parameter for these
grids are the same as in Figure~\ref{lrat1}.  See the electronic edition of
the Journal for a color version of this figure.
\label{lrat5}}
\end{figure*}                

\begin{figure*}[htb]              
\includegraphics[scale=0.6]{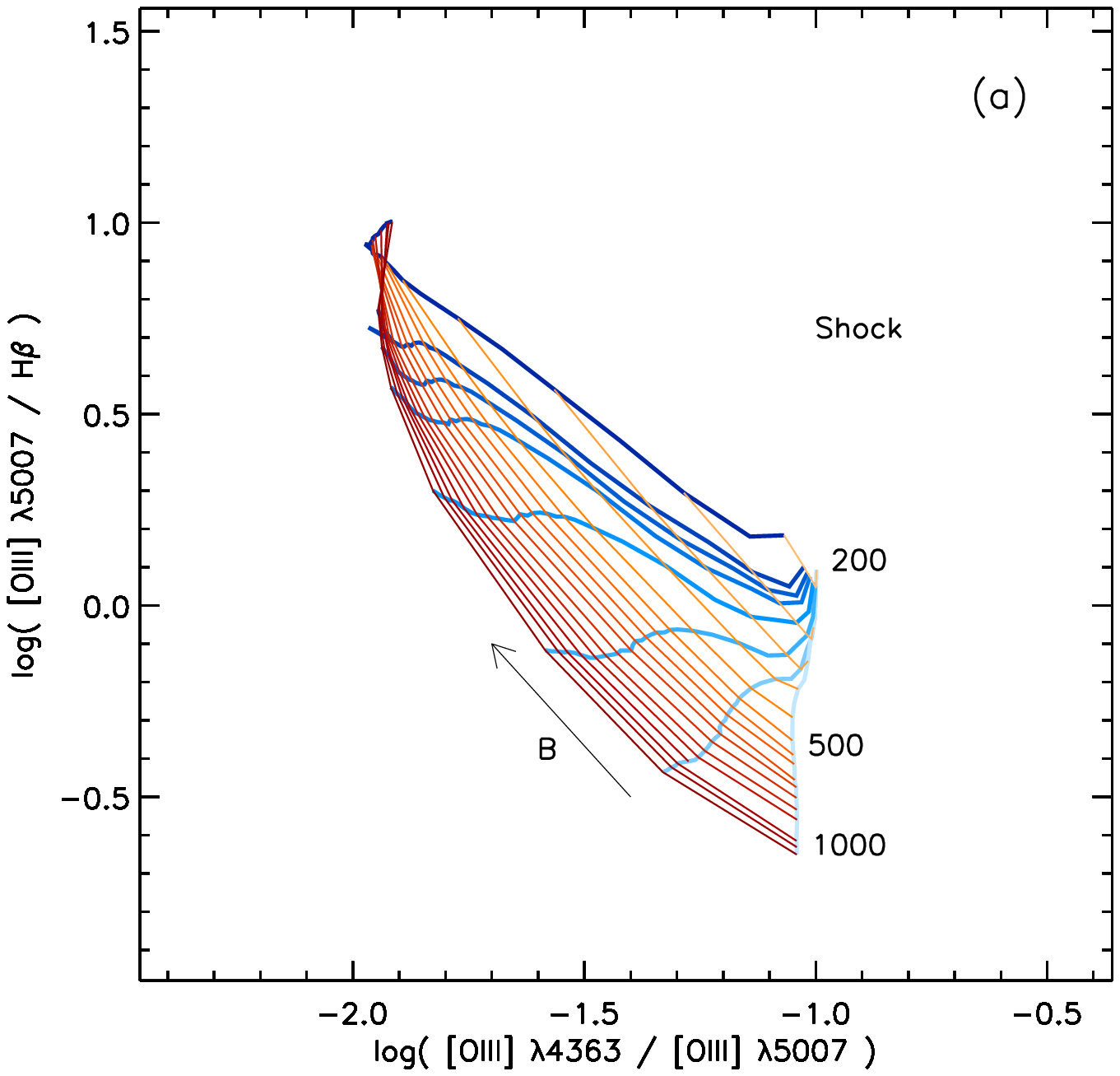} 
\includegraphics[scale=0.6]{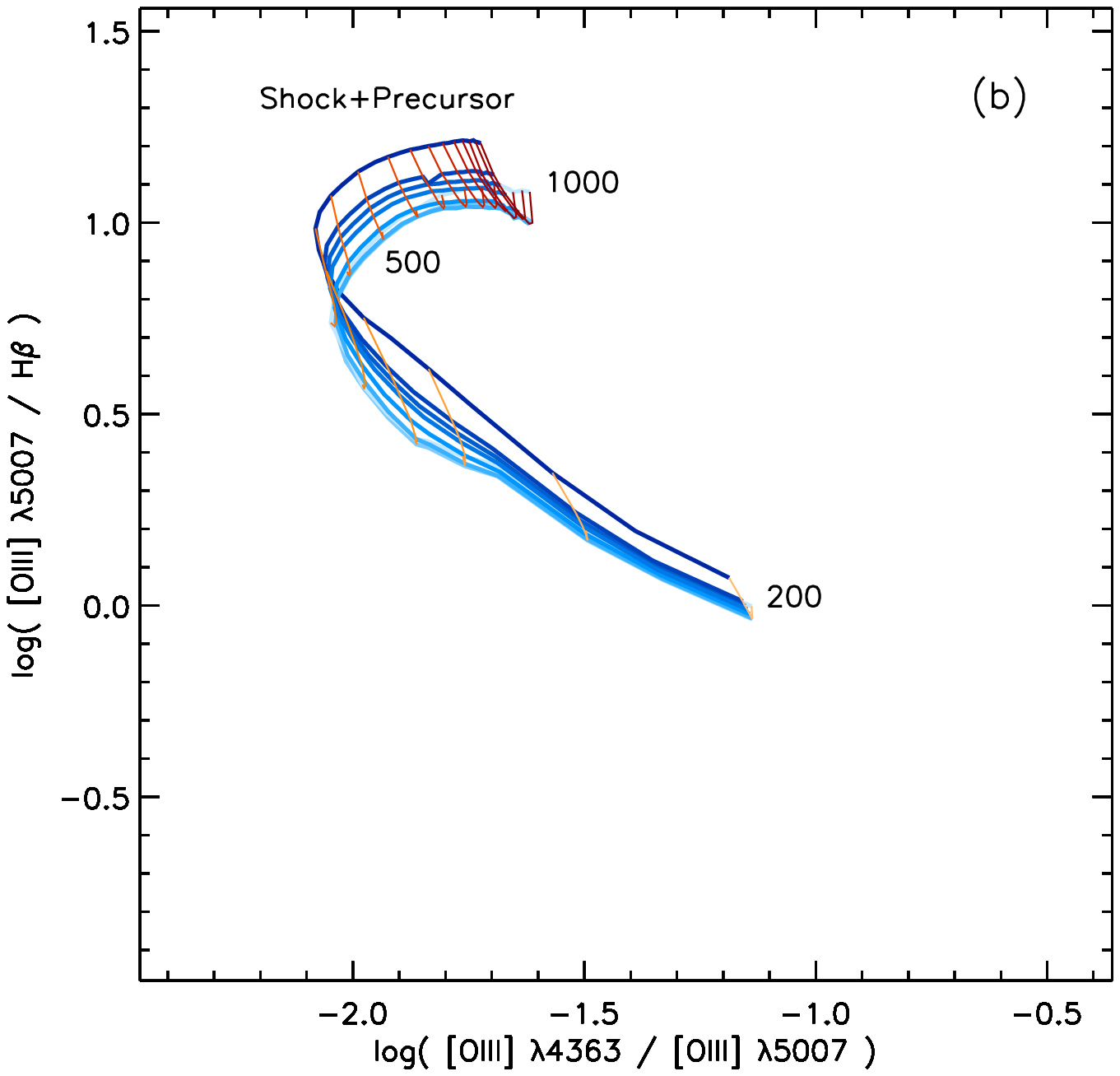} 
\caption{The temperature sensitive [\ion{O}{3}]$\lambda$4363/
[\ion{O}{3}]$\lambda$5007 ratio ($R_{\rm OIII}$) versus the
[\ion{O}{3}]$\lambda$5007/H$\beta$ line ratio diagram for the solar abundance
n$=1$\ cm$^{-3}$ models.  Panel (a) shows the shock grid, and panel (b) shows
the shock+precursor grid. The range and step size of the shock velocity and
magnetic parameter for these grids are the same as in Figure~\ref{lrat1}. See
the electronic edition of the Journal for a color version of this figure.
\label{lrat6}}
\end{figure*}                

Changes in the atomic abundances directly affect the cooling rate.
Figure~\ref{Hstruct_abund} shows the equipartition case, B/$\sqrt{\rm n}$ =
3.23~$\mu$G cm$^{3/2}$, with n=1.0 cm$^{-3}$ and $v_{\rm s}=500$
km\,s$^{-1}$ for models 
using the 2$\times$Solar, dopita2005, and SMC abundances. All three models
differ in structure, with the expected clear progression in cooling efficiency
as we increase in metallicity from the SMC abundances to the twice Solar. This
difference in efficiency leads to the changes visible in the spatial and
temporal extent of the models, in both the precursor and shock regions.  The
changes in metallicity also affect the ionizing radiation from the shock due
to the weakening of the heavy element free-bound features. However, as
discussed in section \ref{sec:ionrad}, the effect is minimal.
 
\subsection{Ionization Structures of Abundant Elements}

The shock model library includes the full fractional ionization structure of
all elements listed in Table~\ref{abund_table}.  In Figures
\ref{Hestruct_fig}--\ref{Sstruct_fig} we show the ionization structures of
helium, carbon, nitrogen, oxygen and sulfur for three velocities (200, 500 and
1000 km\ s$^{-1}$) of the solar abundance, n$=1.0$ cm$^{-3}$ models with the
equipartition magnetic parameter. The associated hydrogen structures are shown
in Figures \ref{Hstruct_eg} and \ref{Hstruct_V}. In each figure, the
ionization fraction, $X_{\rm ion}$, is shown at the bottom (with values $\le
1$), with each ionization state labelled and the odd and even ionization
states shown as the solid and dot-dashed curves respectively for clarity. Also
included in each diagram is the hydrogen density, n$_{\rm H}$, and electron
temperature structure, T$_{\rm e}$, to trace the different shock and precursor
zones.

Together, these figures elucidate further what has been discussed in the
previous sections; faster shocks lead to a higher temperature, and therefore
higher ionization, radiative zone. This in turn leads to a more luminous and
harder ionizing spectrum as can be traced within the precursor zone. Similarly
this figure allows one to trace the more complex ionization structure of
the post-shock recombination region, affected both by the increased ionizing
spectrum and increased density that results from faster shocks.

\begin{figure*}[h]              
\centering
\includegraphics[scale=0.55]{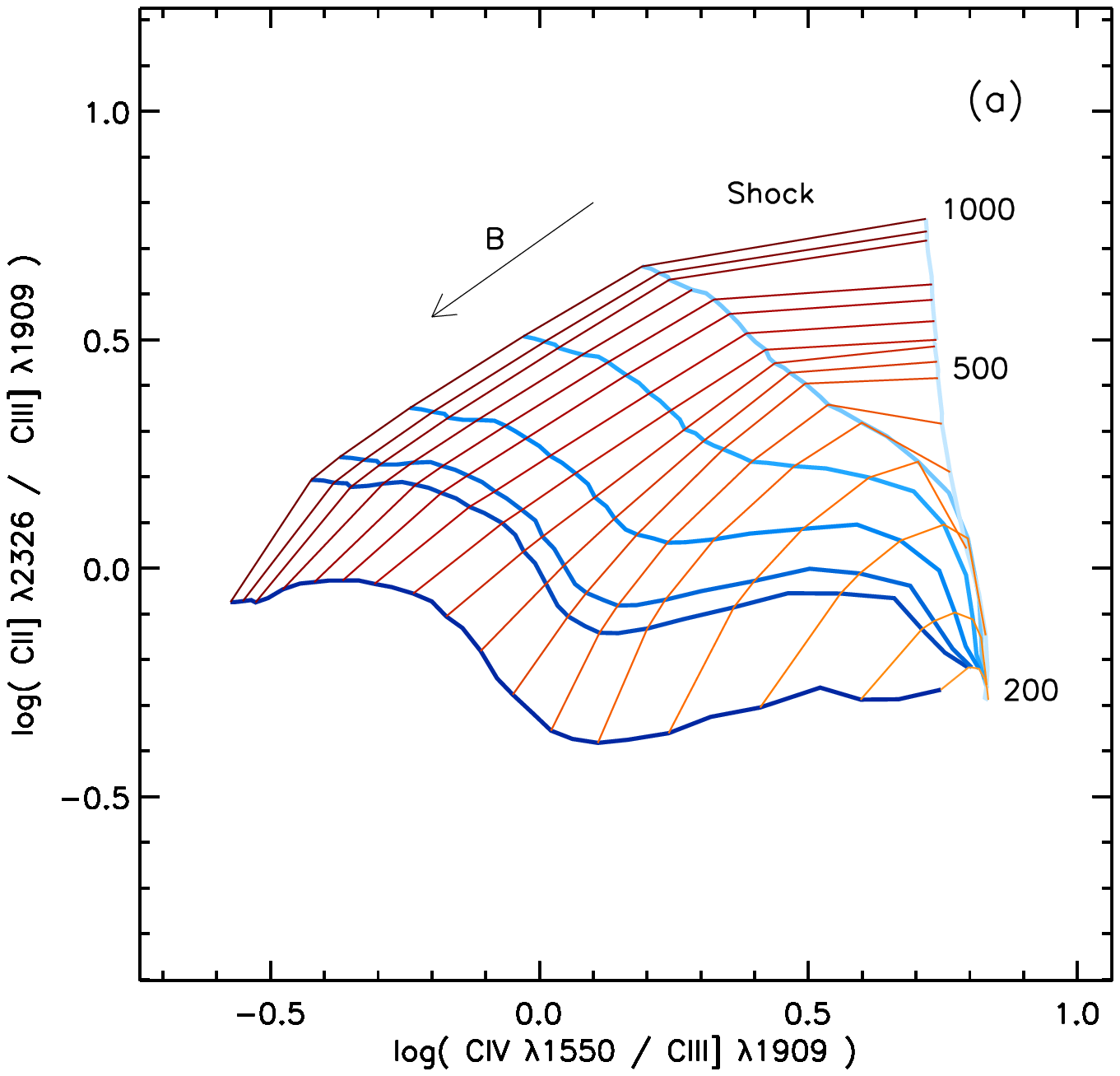} 
\includegraphics[scale=0.55]{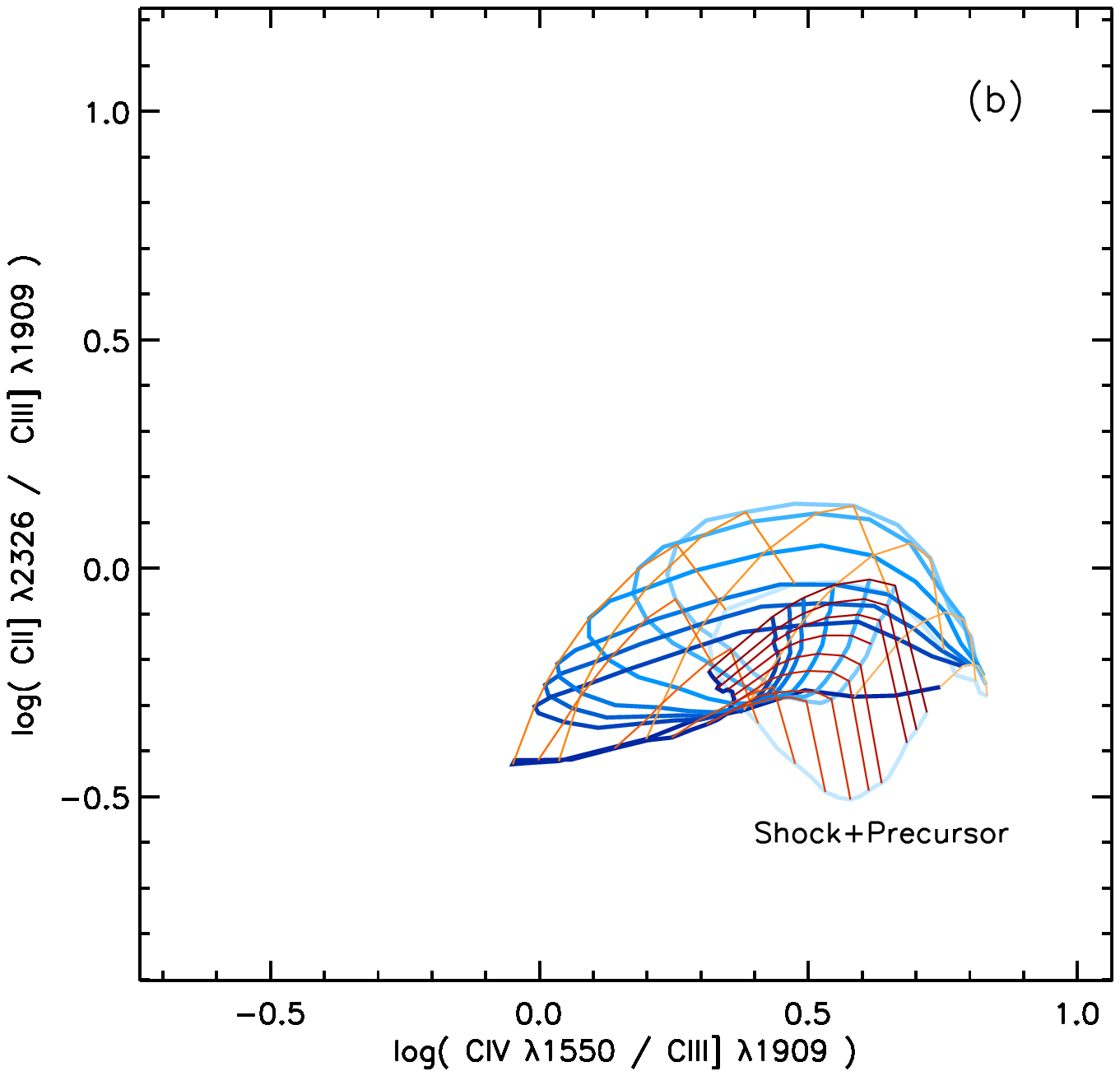} 
\caption{The ultraviolet carbon line ratio diagnostic plot for the solar
abundance n$=1$\ cm$^{-3}$ shock (a) and shock+precursor models (b).  The
shock only grid enables both the shock velocity and the magnetic parameter to
be independently determined, but the shock + precursor grid is multi-valued at
many positions.  The range and step size of the shock velocity and magnetic
parameter for these grids are the same as in Figure~\ref{lrat1}.  See the
electronic edition of the Journal for a color version of this figure.
\label{lrat7}}
\end{figure*}                

However the true strength of these figures, and the library in general, is
their diagnostic ability, allowing the reader to trace exactly where each
ionization state arises, which is especially interesting given the X-ray
ionized regions that occur far from the ionization front in both the precursor
and post-shock regions.

\subsection{Column Densities}

The column densities of the shock and precursor components of the models
inevitably increase with shock velocity, and can be as large as
$N$(H)=4$\times10^{21}$~cm$^{-2}$ in the shock itself for the highest velocity
solar models. The precursor column densities range from between 3 and 20 times
larger than the corresponding shock component, with this ratio decreasing for
the higher velocity shocks.

Figure \ref{coldens_fig} displays the integrated \ion{H}{2} column density of
the shock as a function of shock velocity. This is shown for the solar
abundance n$=1.0$ cm$^{-3}$ models with equipartition magnetic parameter, as
well as the for maximum and minimum magnetic parameters of 0.0001 and
10.0. The dashed line shows the scaling relation given by DS96, and the dotted
line shows the extrapolation of this beyond shock velocities of 500
km~s$^{-1}$. The diagram shows a weak dependence of the column density on the
magnetic parameter, and a decrease in the slope beyond shock velocities of 700
km~s$^{-1}$.

The column density of the precursor does not depend on the magnetic
parameter. All solar abundance, n=1.0 cm$^{-3}$ models fall on the same
curve, marked `PRECURSOR' in Figure~\ref{coldens_fig}. There is however a
systematic difference in the precursor column density compared to the DS96
models, with the newer models having a $\sim$10\% decrease in column density,
consistent with the slightly lower fraction of ionizing flux compared with
total flux produced in the cooling plasma in the newer models.

Tables of the column densities for all species for every library model are
available on-line (see section~\ref{online}). These columns are a function of
both the hydrogen column density and the ionization structures seen in the
previous figures. Example tables giving the column densities of the shock and
precursor components of the n$=1.0$ cm$^{-3}$ solar abundance models at
velocities of 200, 500 and 1000 km~s$^{-1}$ models are presented in
Tables~\ref{coldens_tab3} and 5.

\begin{figure*}[h]        
\centering      
\includegraphics[scale=0.6]{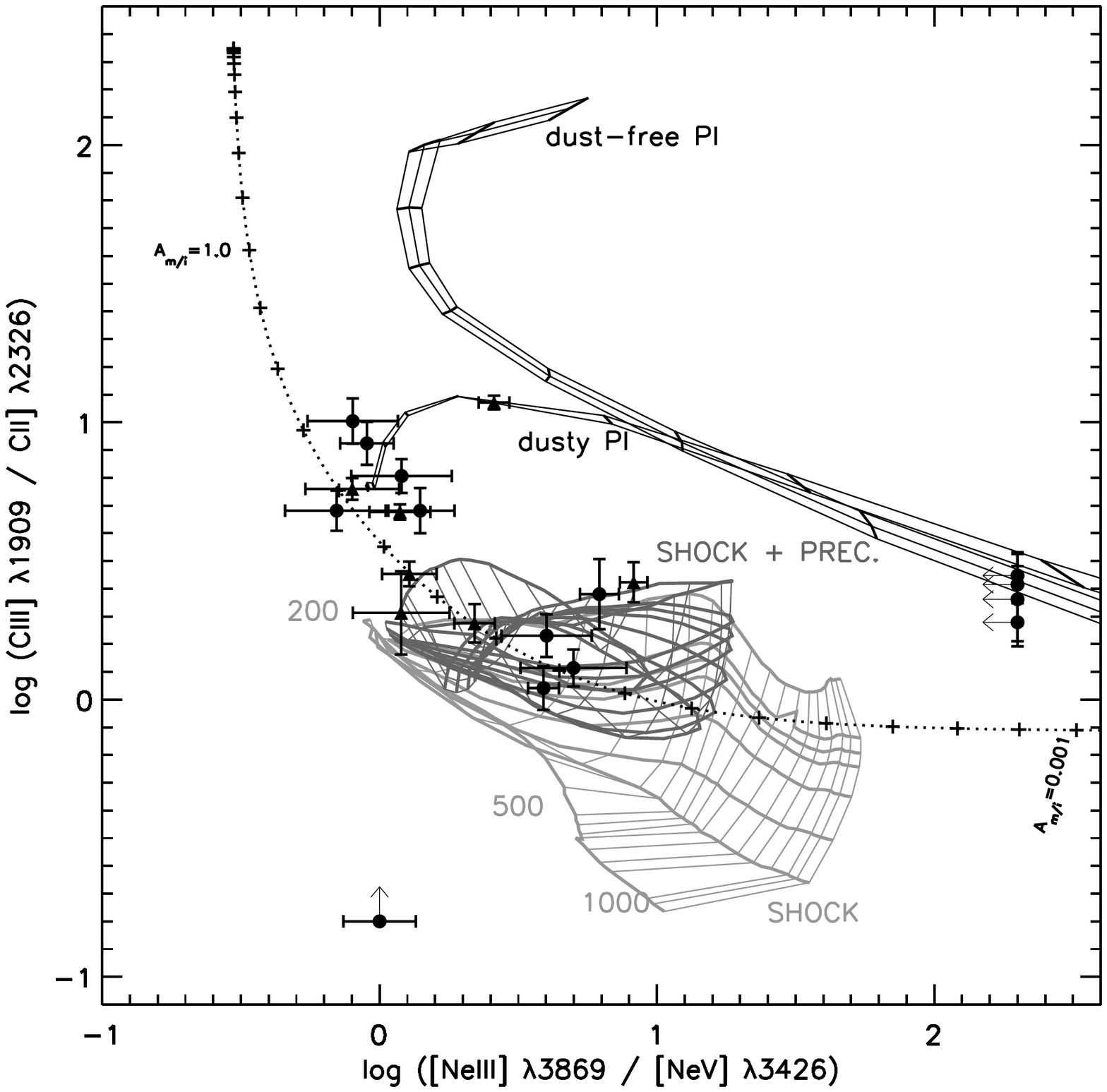} 
\caption{ The \ion{C}{3}]$\lambda$1909/\ion{C}{2}]$\lambda2326$ with
\ion{Ne}{3}$\lambda$3869/\ion{Ne}{5}]$\lambda$3426 diagram. The shock model
grid is shown in light gray (cyan in the electronic edition) and is labelled
with the shock velocity. The shock+precursor grid is shown in dark grey (green
in the electronic edition), and displays a twisted shape on these axes, and
also overlaps the shock grid. The dusty and dust-free models of
\cite{groves2004b} are also shown with the ionization parameter increasing to
the left of the diagram before the turn-over in these curves. The A$_{\rm
M/I}$ models from \cite{bws96} are shown as a dotted line with the + symbols
indicating increments of 0.25 dex. The data points are those from
\cite{best2000} and \cite{Inskip02} as indicated in the text.
\label{UV_C_Ne_diag}}
\end{figure*}                

\begin{figure*}             
\includegraphics[scale=0.6]{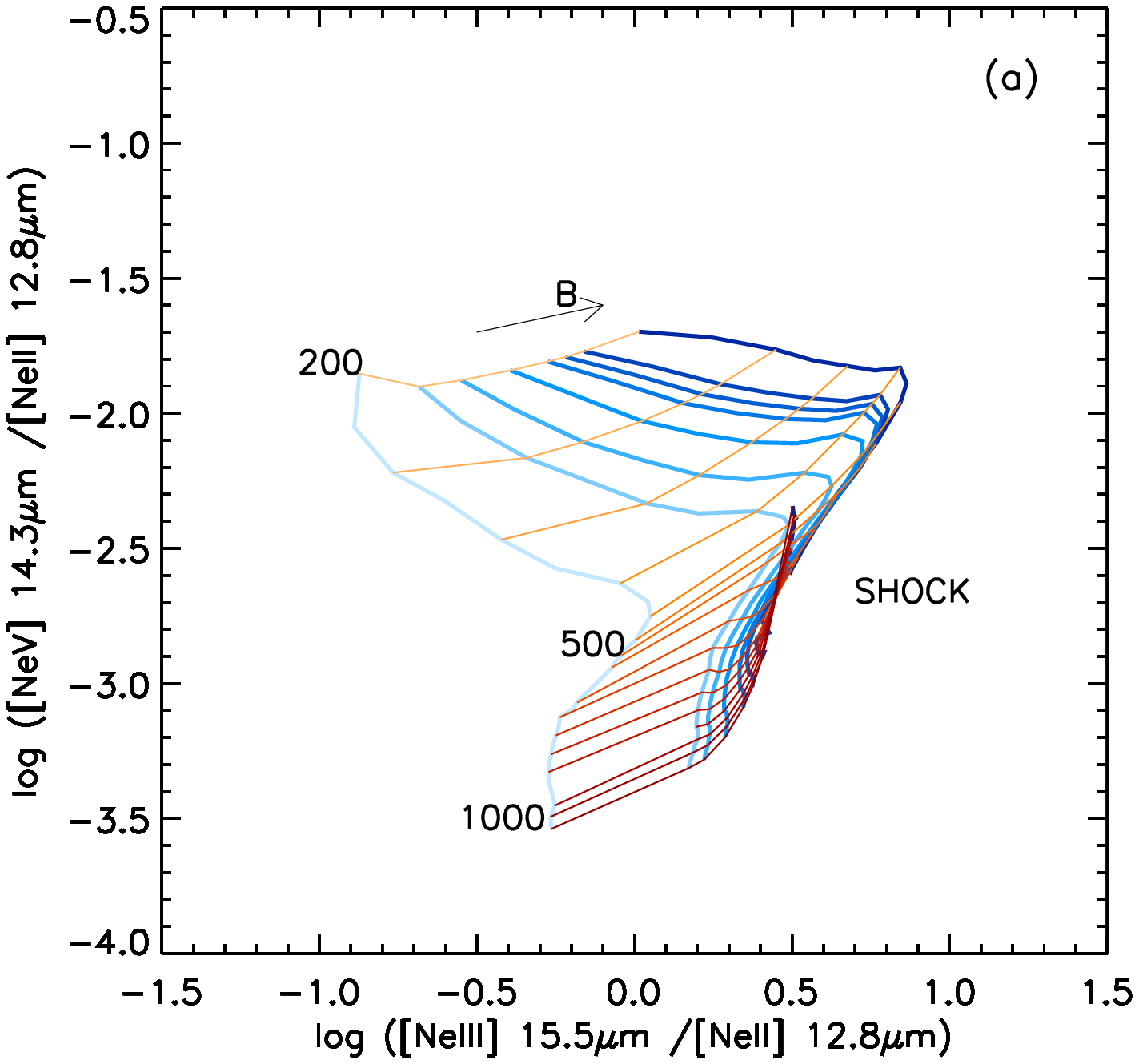} 
\includegraphics[scale=0.6]{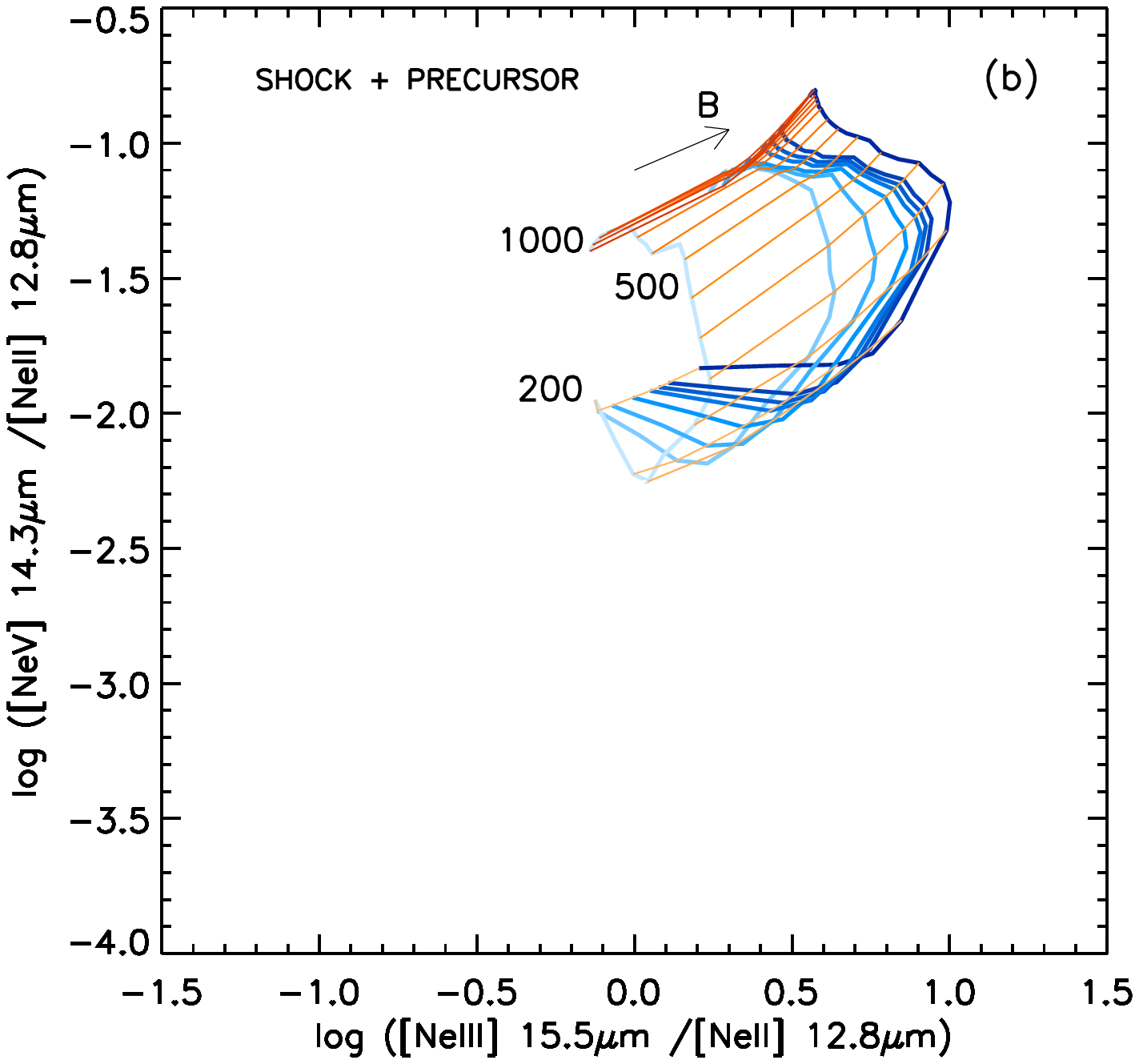} 
\caption{ IR diagnostic diagram of
[\ion{Ne}{5}]14.3$\mu$m/[\ion{Ne}{2}]12.8$\mu$m versus
[\ion{Ne}{3}]15.5$\mu$m/[\ion{Ne}{2}]12.8$\mu$m. Panels (a) and (b) show
respectively the shock and shock+precursor model grids for the solar abundance
n$=1$\ cm$^{-3}$ models.  The range and step size of the shock velocity and
magnetic parameter for these grids are the same as in Figure~\ref{lrat1}.  See
the electronic edition of the Journal for a color version of this figure.
\label{Ne_IR_diag}}
\end{figure*}

\subsection{Radiative fluxes of Shock and Precursor Components}\label{sec:lum}

A fully-radiative shock will, by definition, radiatively dissipate all of the
energy flux through the shock. Thus the total radiative flux, $L_{\rm Tot}$,
is equal to $0.5 \rho v_{s}^3$. DS96 provided scaling relations for the total
luminosity of a shock, and showed a similar scaling exists for the total
luminosity of the precursor emission.  This follows because, in the precursor,
all of the radiation emitted in the pre-shock direction is eventually
processed.

\begin{figure*}[htb]              
\includegraphics[scale=0.6]{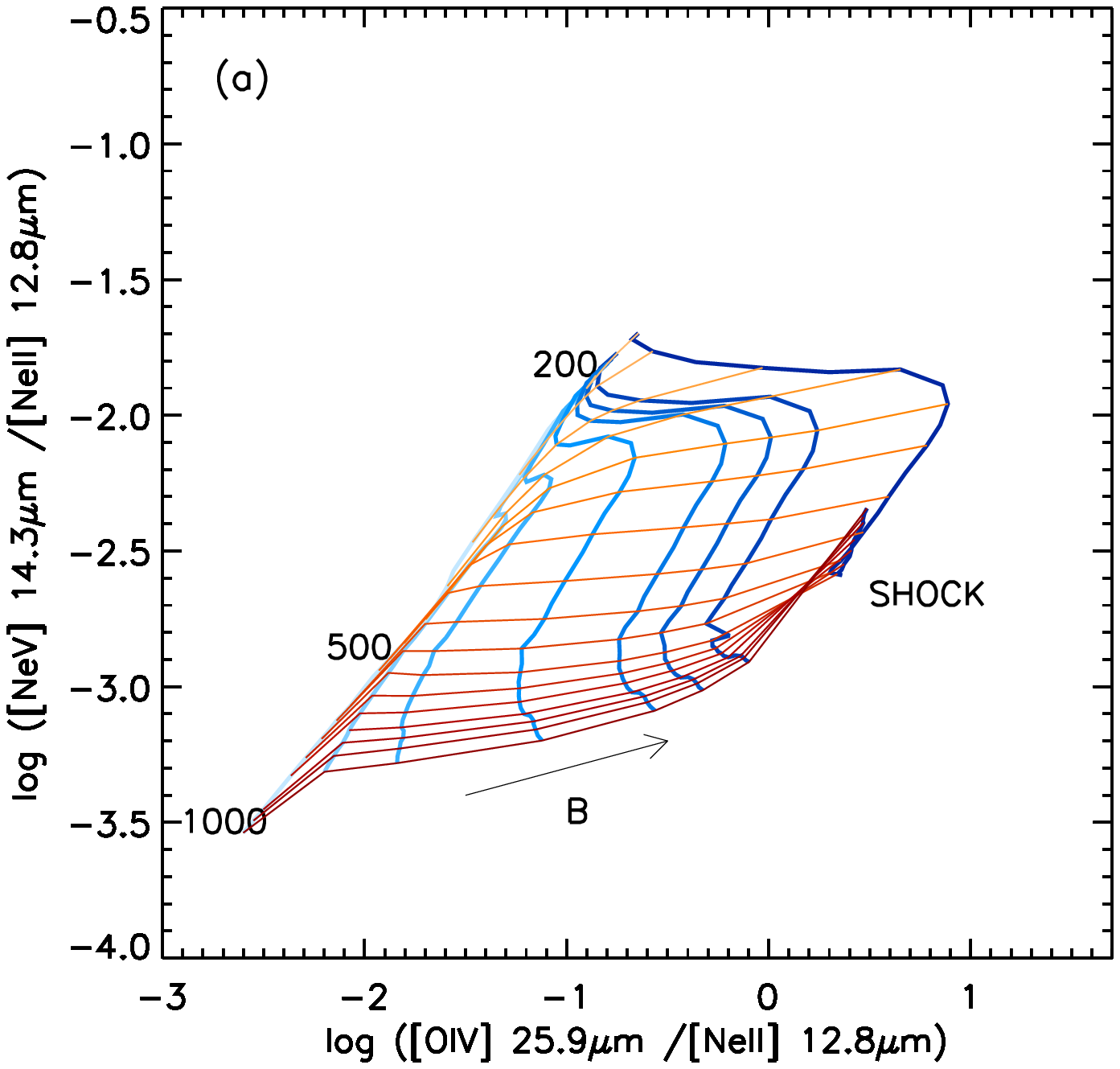} 
\includegraphics[scale=0.6]{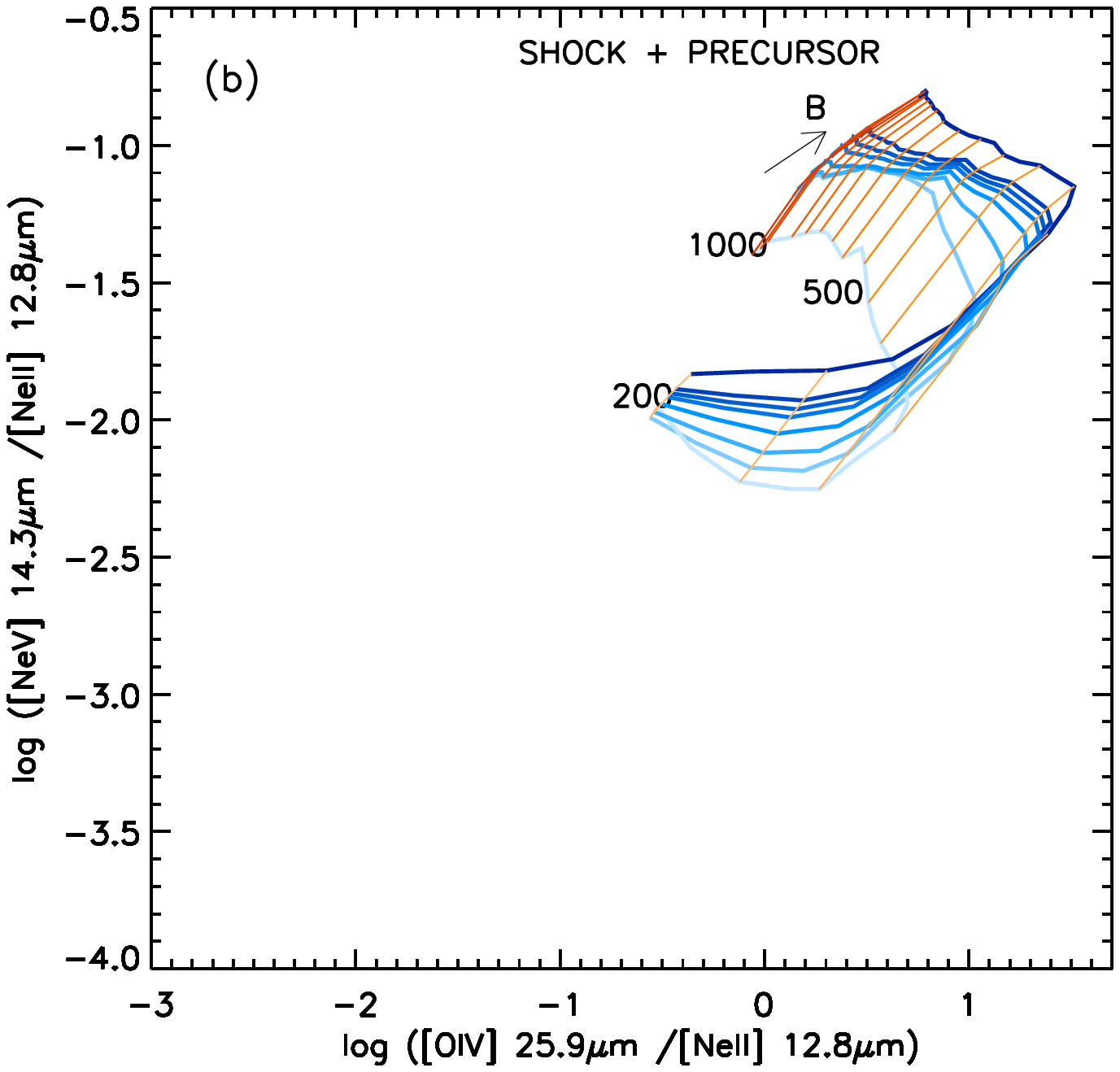} 
\caption{ IR diagnostic diagram of
[\ion{Ne}{5}]14.3$\mu$m/[\ion{Ne}{2}]12.8$\mu$m versus
[\ion{O}{4}]25.9$\mu$m/[\ion{Ne}{2}]12.8$\mu$m.  Panels (a) and (b) show
respectively the shock and shock+precursor model grids for the solar abundance
n$=1$\ cm$^{-3}$ models.  The range and step size of the shock velocity and
magnetic parameter for these grids are the same as in Figure~\ref{lrat1}.  See
the electronic edition of the Journal for a color version of this figure.
\label{Ne_O_IR_diag}}
\end{figure*}

In Figures~\ref{lum_compare1_fig} and \ref{lum_compare2_fig} we 
show the total radiative fluxes generated in the shock and in the precursor
respectively, in the \mapiii\ models with solar abundance and
equipartition magnetic fields. Here we cover the full range of pre-shock
density ($0.01-1000$ cm$^{-3}$). To illustrate the relatively minor
adjustments to the total radiated flux due to different abundances we also
show the luminosities for the models with 2$\times$Solar, Dopita2005, SMC and
LMC abundance sets for n$=1.0$ cm$^{-3}$. Changing the magnetic field from the
equipartition value produces too small an effect on the total radiative flux
to be visible on the scale of these plots.
 
\section{Spectral Signatures of the Shock and Precursor Gas}\label{sec:specsig}

While the total luminosities of the shock effectively depends only upon the
density and the shock velocity, the detailed emission line spectra depend
strongly on the physical and ionization structure of the shock. This is
determined primarily by the shock velocity and the magnetic
parameter. However, at the higher pre-shock densities (and at lower values of
the magnetic parameter) the density close to the photoionized tail and the
recombination zone of the shock becomes sufficiently high for collisional
de-excitation of forbidden lines to become important. In these circumstances,
the emission line spectrum of the shock becomes dependent upon the density as
well.

\begin{figure}[h]              
\includegraphics[scale=0.6]{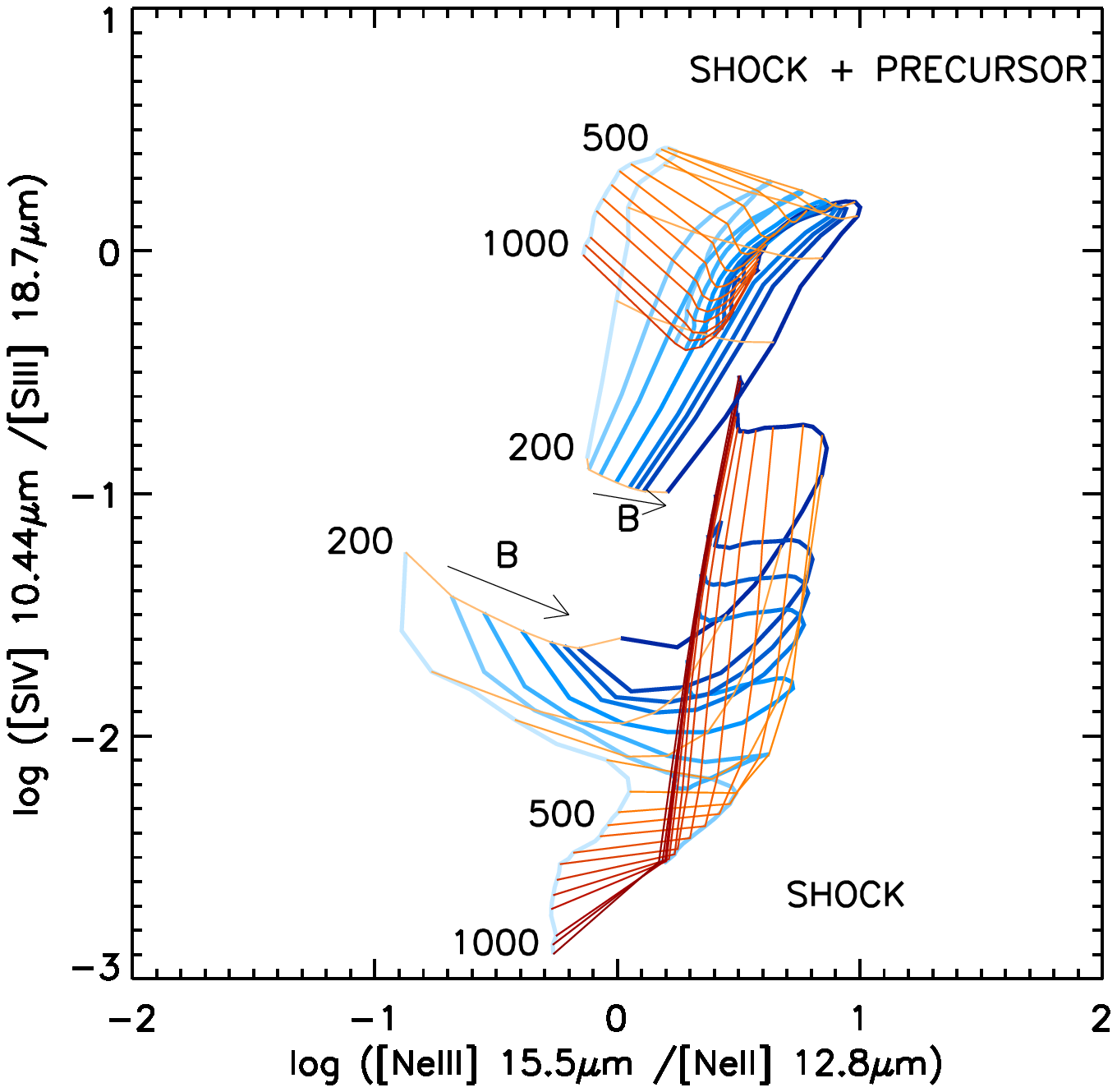} 
\caption{ IR diagram of [\ion{S}{4}]10.44$\mu$m/[\ion{S}{3}]18.7$\mu$m versus
[\ion{Ne}{3}]15.5$\mu$m/[\ion{Ne}{2}]12.8$\mu$m showing the shock and shock+precursor
grids of the solar abundance n$=1$\ cm$^{-3}$ models, as labelled. The range
and step size of the shock velocity and magnetic parameter for these grids are
the same as in Figure~\ref{lrat1}.  See the electronic edition of the Journal
for a color version of this figure. \label{S_Ne}}
\end{figure}                

Shocks are characterized by regions of high electron temperature and
ionization state. As a consequence, they display a rich spectrum of
collisionally excited UV lines. The shock velocity controls the shape of the
ionizing spectrum produced by the shock, and the magnetic parameter controls
the effective ionization parameter in the photoionized tail of the
shock. Higher ionization parameters yield a higher mean ionization in the
plasma, so therefore faster shocks give an (optical) spectrum of somewhat
higher ionization.

The emission line ratios (with respect to H$\beta$=1) of a selection of lines
for the shock, precursor and shock+precursor components of the solar
abundance, n=1.0 cm$^{-3}$ models are shown separately in
Figures~\ref{M_n1_be_s_llf}-\ref{M_n1_be_sp_llf} as a function of shock
velocity.

As discussed before, the precursor emission can be considered as an \ion{H}{2}
region with high ionization parameter. In this region strong cooling
lines, like [\ion{O}{3}]$\lambda$5007, are
strong and generally increase in strength with shock velocity
as the spectrum produced by the cooling plasma becomes harder, and its flux
increases. Eventually, as the shock velocity is increased, the average state
of ionization in the precursor region becomes high and the relative
intensity of the lines produced by lower ionization turn over,
flattening and even decreasing at the higher velocities. For example,
in Figure \ref{M_n1_be_p_llf} [\ion{O}{3}]$\lambda$5007 can be seen to
flatten out around 600 km~s$^{-1}$, and
even decrease at 1000 km~s$^{-1}$.

In addition the emission line ratios of 40 strong lines are tabulated for
shock, precursor and shock+precursor in Tables 6, 7, and 8 respectively. Complete tables of
emission line ratios of all lines, for all models are available electronically
( \emph{see} section~\ref{online}).

\subsection{Line Ratio Diagrams}

Diagnostic emission line ratio diagrams are well established as a powerful and
practical way to investigate the physics of emission line gas. At a basic
level they can be used empirically to identify the key excitation mechanism, and
divide active galaxies from their star-formation dominated counterparts. At
a deeper level, grids of theoretical models can be used to determine chemical
abundances and derive physical parameters such as electron temperature,
density, ionization parameter or shock velocity. Various sets of diagnostic
diagrams have been proposed and used - \cite{baldwin1981} (optical),
\cite{veilleux1987} (optical-IR), \cite{allen1998} (near-UV) and
\cite{groves2004b} (UV, optical, IR).  These 2-D diagrams remain useful
alongside multi-dimensional methods for comparing data to models because the
ratios can be chosen to be independent of reddening, and tuned to be sensitive
to particular physical properties of interest. For example, \cite{Dopita06}
have provided theoretical strong emission line diagnostics for ensembles of
\ion{H}{2} regions to enable more accurate measurement of chemical abundances
in unresolved starburst galaxies, and \cite{kewley2006} combined standard line
ratio diagrams with velocity dispersion measurement to help refine the
classification of active galaxies.

As part of the shock model library available online, we include interactive
plotting tools to plot diagnostic diagrams, such as the examples
presented in the following sections. These tools enable a
much clearer picture of these diagnostics than possible within the
paper, especially with those grids that 
twist and fold upon themselves in the 2-D diagrams. 
In addition, many other diagnostic ratios than those presented here
can be formed to gain insights into what diagnostics can be used when
only a limited sample of lines are available.

\subsection{Optical Diagnostics}

As a first example, we plot the new shock model grids on a number of the
standard optical diagnostic diagrams to reveal the general shape of the model
grids, and the sensitivity of the diagnostics to the parameters of shock
velocity, magnetic parameter and chemical abundance set.  Figure~\ref{lrat1}
shows the perhaps the most frequently used of the diagnostic diagrams, the
\cite{veilleux1987} plot of [\ion{O}{3}]$\lambda$5007/H$\beta$ versus
[\ion{N}{2}]$\lambda$6583/H$\alpha$, also commonly known as the BPT
diagram \citep{baldwin1981}. For clarity, the shock and shock+precursor grids for
the solar abundance n$=1$ cm$^{-3}$ models are plotted separately but
on the same scale, as is the case in most of the following diagrams,
due to the overlap of the two grids.  These two
grids represent the physical extremes of having little or no gas ahead of the
shock (shock-only), and having an extensive, radiation bounded, precursor
region ahead of the shock. As in DS95, the two grids show the range in
shock velocity and magnetic parameter, as labelled in the diagram. The
lines of constant velocity are plotted at 50 km\ s$^{-1}$ intervals as
the thin lines.
The lines of constant magnetic parameter are plotted as the
thick lines in the grid, with the value of the magnetic parameter as
marked. For a clearer picture of this and all other diagnostic
diagrams in the paper we provide color figures in the online journal.

\begin{figure}
\includegraphics[scale=0.6]{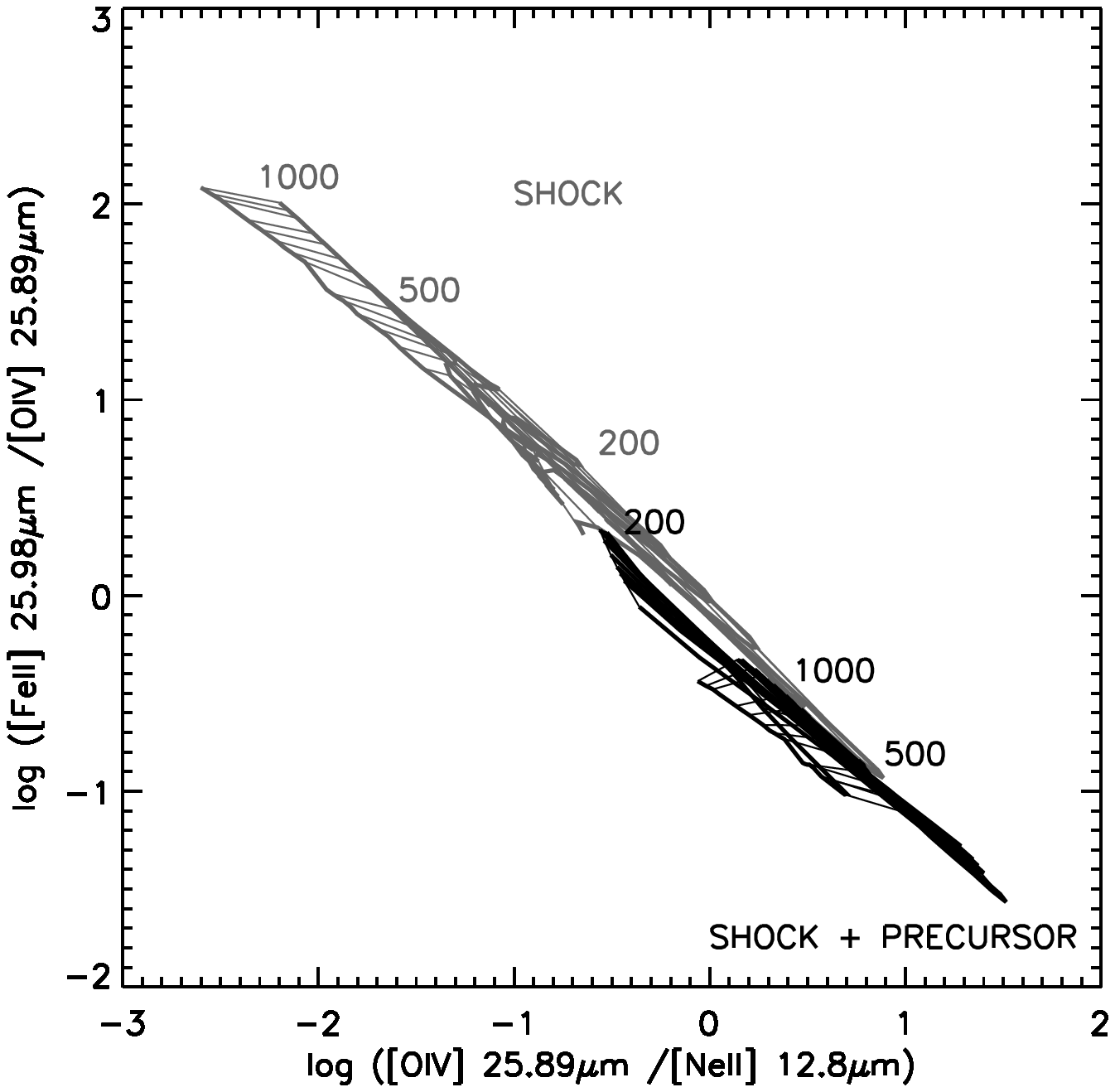}
\caption{ IR diagram [\ion{Fe}{2}]26.0$\mu$m/[\ion{O}{4}]25.9$\mu$m versus
[\ion{O}{4}]25.9$\mu$m/[\ion{Ne}{2}]12.8$\mu$m as introduced by
\cite{lutz2003}, overlaid with the shock and shock+precursor model
grids. \label{lutz_fig}}
\end{figure}

This figure reveals a similar picture to that in DS95 (their Fig.~2b),
with the shock-only and shock+precursor grids occupying different
regions of the diagram, a result of the different processes discussed
in the previous sections. Both grids shows broad steps in velocity
up to a velocity of $\sim550$\kms, at which the grids begin to
compress and turn upon themselves, becoming degenerate. However, for
the shock-only grid an increase in velocity causes an increase in the
[\ion{N}{2}]/H$\alpha$ ratio, while for the shock+precursor models in
results in an increase of the [\ion{O}{3}]/H$\beta$ ratio. The
situation is reversed for an increase of the magnetic parameter,
causing the spread observed in the grids, though the shock+precursor
models tend to be degenerate in this parameter at high values. The
spread of the grids and their separation indicate the diagnostic power
of these figures, allowing the determination of the physical
parameters in known shock-ionized plasmas.

\begin{figure}[h]              
\includegraphics[scale=0.6]{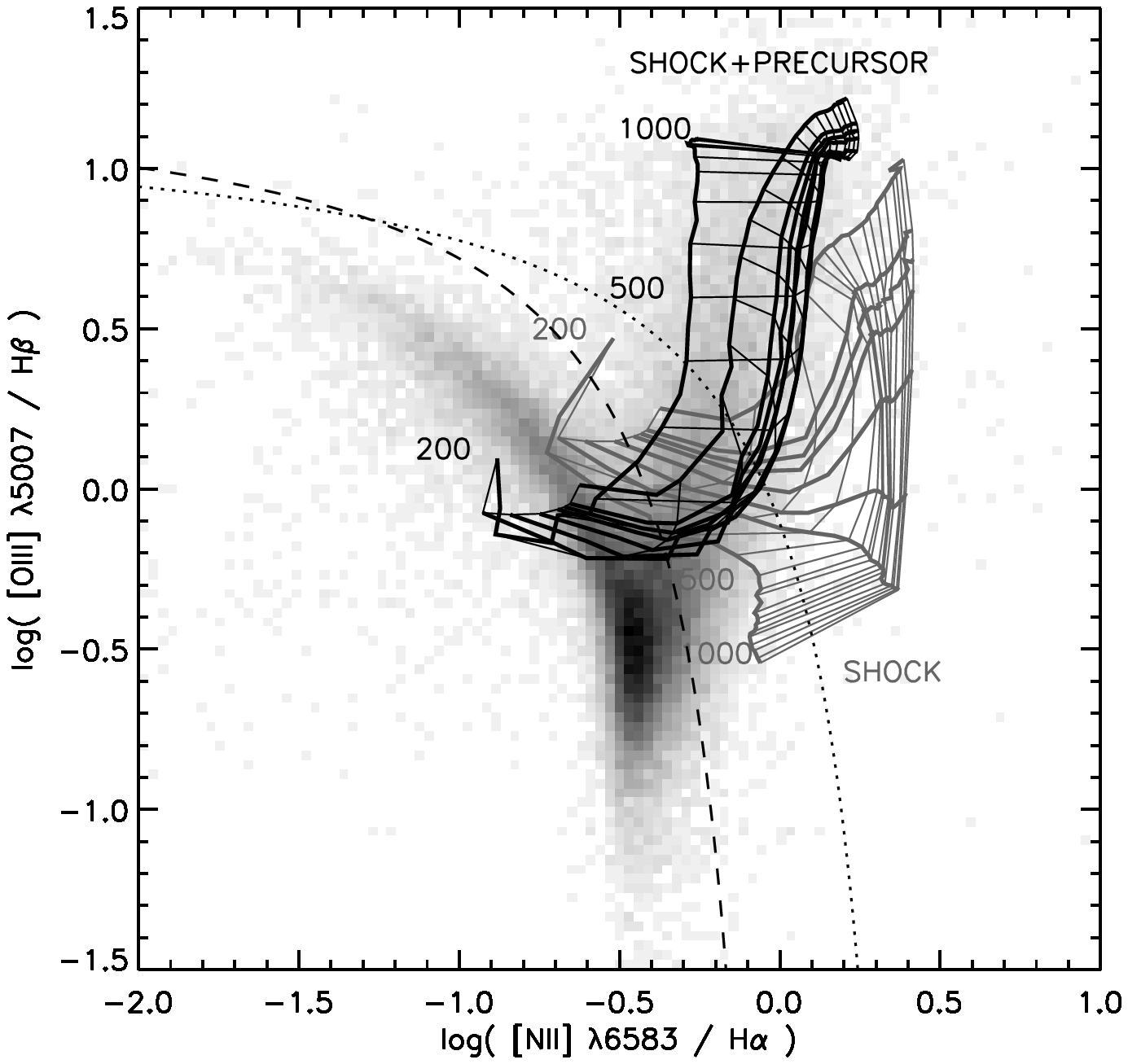} 
\caption{Comparison of the shock and shock+precursor models to SDSS line
ratios on the [\ion{O}{3}]$\lambda$5007/H$\beta$ versus
[\ion{N}{2}]$\lambda$6583/H$\alpha$ diagram. The model grids are those with
2$\times$solar abundance. The dotted and dashed curved lines represent the
\cite{kewley2001} and \cite{kauffmann2003} classification lines respectively
which divide \ion{H}{2} regions (lower left) from the region occupied by AGN.
\label{sdss_compare1}}
\end{figure}                

Improvements to the \mapiii\ code compared to the \mapii\ code used by DS96
means that there are some significant and systematic differences compared to
the DS95 grids. Figure \ref{lrat4} shows a direct comparison of the DS95 shock
and shock+precursor model grids overlaid on our new solar abundance shock and
shock+precursor models for the same diagnostic diagram.  The improvement of
our new models in terms of parameter space is clear, with DS95,96 models
limited to 500 km\ s$^{-1}$ in velocity space, and only covering magnetic
parameters of 0--4 $\mu$G cm$^{3/2}$. The shock component of the models agree
reasonably well over the common parameter ranges, although appears somewhat
offset. However the shock+precursor grid shows a difference of up to 0.5 dex
in the [\ion{O}{3}]$\lambda$5007/H$\beta$ and
[\ion{N}{2}]$\lambda$6583/H$\alpha$ ratios. We note again that on this diagram
the new shock+precursor models turn over at high shock velocity. One
consequence of this is that some active galaxies such as high-redshift radio
galaxies, which have been confirmed as being excited by shocks
and which were interpreted as matching 400-500 km\ s$^{-1}$
shock+precursor models, may in fact be characterized by higher shock
velocities. This would help resolve the mis-match between radial velocity
dispersions and shock velocities inferred by line ratios as inferred, for
example, by \citet{Reuland07}.

Figure \ref{lrat2} shows the same diagram for the different chemical
abundance sets which we have used. Each grid 
covers the same range in shock velocity and magnetic parameter. In
both shock and shock+precursor models the
[\ion{O}{3}]$\lambda$5007/H$\beta$ ratio is only weakly affected by
abundance, as the increased temperatures within the models from the
reduced cooling counteracts somewhat the effects of the drop in oxygen
abundance on this ratio (see Figure \ref{Hstruct_abund}). The
[\ion{N}{2}]$\lambda$6583/H$\alpha$ ratio however changes significantly 
with abundance, predominantly as a consequence of the larger relative
abundance differences in nitrogen due to its secondary nucleosynthesis
component.
In the shock models, the [\ion{O}{3}]/H$\beta$ does not change
significantly in range either, but the [\ion{N}{2}]/H$\alpha$ ratio
becomes compressed, meaning the diagrams are totally degenerate in
velocity at low metallicity, predominantly due to the changes
 photoionized and recombination regions of the shock with metallicity.
The case is similar for the shock+precursor, with the grids becoming
generally more ``compressed'' with decreasing metallicity, and
decreasing the diagnostic power of the diagrams. Note that for each
abundance set the shock-only and shock+precursor models still tend to
be discrete grids, and do not significantly overlap. 

In Figure~\ref{lrat3} we investigate the effect of changing the pre-shock
densities on this \cite{veilleux1987} diagnostic. For the solar abundance
grid, we display the results for the two extreme values of density n$=0.01$
cm$^{-3}$ and n$=1000$ cm$^{-3}$ for both the shock and shock+precursor
models.  Note the change in scale between this figure and Figure~\ref{lrat1},
necessary due to the changes in the grids.

As discussed in section \ref{sec:input}, the models with varying density were
computed not only with the standard magnetic parameter set, but also a range
of magnetic fields to allow for the comparison of both B and B$/\sqrt{\rm n}$
between the model grids. The range of both B and B$/\sqrt{\rm n}$ for these
models are given in Table \ref{param_table}. We include all these models in
Figure~\ref{lrat3} for completeness, meaning that the grids sample more finely
and extend further both the low and high values of the magnetic parameter than
seen in Figures \ref{lrat1}--\ref{lrat2}.

The low-density model grids are in most respects exactly the same as the n$=1$
cm$^{-3}$ grids seen in Figure \ref{lrat1}, with the only difference arising
due to the extended B$/\sqrt{\rm n}$ range sampled.  This similarity is not
surprising, as while the luminosity decreases linearly with the decreased
density, as seen in Equation \ref{Luv_eq}, the ionization parameter is
inversely proportional to density and therefore remains approximately the same
in both the shock recombination region and the precursor region. One
interesting thing to note is how degenerate the grids become upon the
introduction of the very high magnetic parameters.

However, at high pre-shock densities there are clear changes in both the shock
and shock+precursor grids. In the shock-only grid the low magnetic parameter
models extend to lower values in the both [\ion{O}{3}]/H$\beta$ and
[\ion{N}{2}]/H$\alpha$ ratios by up to 1 dex, while in the shock+precursor
grid only the [\ion{N}{2}]/H$\alpha$ ratio is reduced.  This is caused by
collisional de-excitation of the forbidden lines in the photoionized and
recombination regions of the shock, where the electron densities are very
high. The shock+precursor does not change significantly in
[\ion{O}{3}]/H$\beta$ as most of the emission of the [\ion{O}{3}] line arises
from the precursor, which is not high enough density to collisionally
de-excite this transition.  There is less of a difference between the grids at
high magnetic parameters and at low to intermediate velocities, because lower
velocities lead to smaller compression factors, and because magnetic pressure
support helps to lessen the degree of gas compression in the tail end of the
shock, meaning that the high density needed to collisionally de-excite these
transitions is not reached.

\begin{figure}[h]              
\includegraphics[scale=0.6]{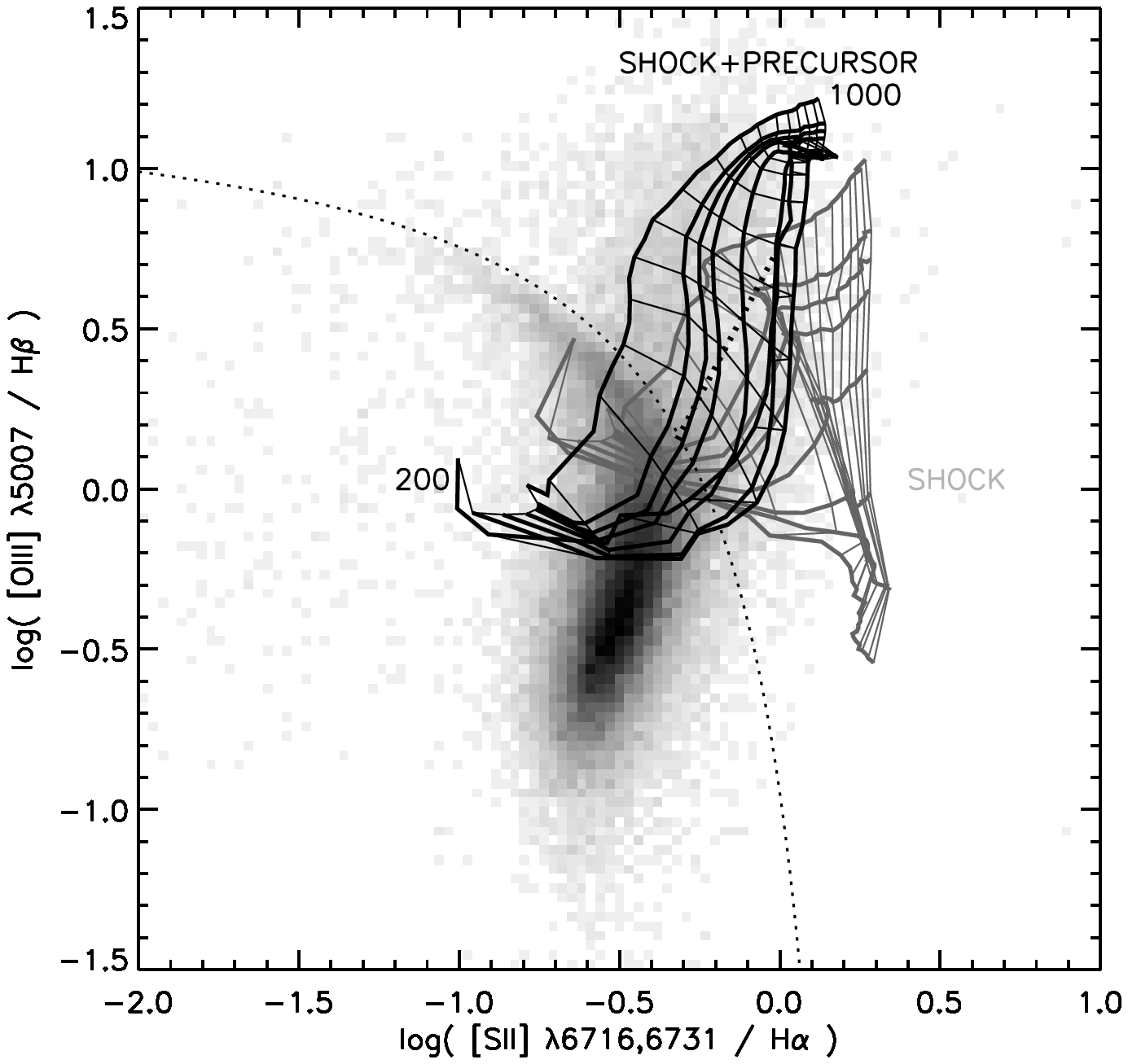} 
\caption{Comparison to SDSS line ratios to the 2$\times$solar abundance model
grids.  The thin dotted line represents the \cite{kewley2001} starburst/AGN
classification line, while the thick straight dotted line represents the
Seyfert-LINER division described by \cite{kewley2006}.
\label{sdss_compare2}}
\end{figure}                

Another commonly used \cite{veilleux1987} diagnostic is the
[\ion{O}{3}]$\lambda$5007/H$\beta$ versus
[\ion{S}{2}]$\lambda\lambda$6716,6731/H$\alpha$ diagram.  In
Figure~\ref{lrat5} we plot the new grids for this diagram, with the shock and
shock+precursor models plotted separately for clarity, as the overlap between
the grids makes distinguishing them difficult.  As with the previous
diagnostic diagram, both the shock and shock+precursor models turn over at
high velocities, becoming twisted and with little separation between the
velocities. Similarly the magnetic parameters show a similar range in spread
in both models, and become more spread with increasing velocity in the
shock-only model.  The shape of the shock+precursor grid on this diagram is
again rather different compared to DS95. Like in Figure \ref{lrat4}, the new
shock+precursor model covers a wider range in [\ion{O}{3}]/H$\beta$, with a
lower value at $v_{\rm s}=200$\kms, while it appears somewhat offset to higher
values of [\ion{S}{2}]/H$\alpha$, though not significantly. It also does not
appear to show the degenerate nature see in DS95 (their Figure 2a). The shock models
are harder to compare due to the wider parameter range in the model, but in
general show the same features of increased [\ion{S}{2}]/H$\alpha$ with
velocity, with the turn over around $\sim$500 \kms. The greater range in
magnetic parameter of the new models also means that the spread in
[\ion{O}{3}]/H$\beta$ is much greater at higher velocities as well.

Figure \ref{lrat6} is a plot of the well-known temperature-sensitive ratio
[\ion{O}{3}]$\lambda$4363/[\ion{O}{3}]$\lambda$5007 (also known as $R_{\rm
OIII}$) versus the [\ion{O}{3}]$\lambda$5007/H$\beta$ ratio, with again the
shock-only and shock+precursor spectra plotted separately.  Considering first
the shock-only grid, at low magnetic parameter the electron temperature in the
O$^{2+}$ zone is consistently high, with only the
[\ion{O}{3}]$\lambda$5007/H$\beta$ ratio decreasing with increasing shock
velocity.  However, as the magnetic parameter rises, the decreased compression
in the photoionized tail of the shock leads to a higher ionization parameter
in this zone, and greater dominance of [\ion{O}{3}] in the photoionized zone,
which has an electron temperature close to 10000K. As a consequence, the
electron temperature decreases and the [\ion{O}{3}]$\lambda$5007/H$\beta$
ratio increases with increasing shock velocity. Together this leads to the fan
shape observed in the diagram.  The shock+precursor grid is dominated by the
bright [\ion{O}{3}] emission of the precursor region, and as the precursor is not
sensitive to the magnetic parameter, this grid covers a reduced area of the parameter
space in the diagram. In terms of velocity, the curve increases
in [\ion{O}{3}]$\lambda$5007/H$\beta$ and decreases in $R_{\rm OIII}$ up to a
velocity of $\sim500$\kms, then turns over and begins to increase in
temperature again.  The shock+precursor grid at velocities less than $\sim$450
\kms\ and the high magnetic parameter shock-only grid overlap, meaning this
diagnostic is degenerate in this range. This diagram is very similar to the
DS95, with only the extension to higher velocities and magnetic parameter
range differentiating it. However, this similarity also indicates that the
well-known ``Temperature Problem'', discussed in DS95 and many other papers,
still exists. A comparison of our models with the observations used in DS95 or
larger datasets like SDSS show that the $R_{\rm OIII}$ ratio is too low, and
even different abundance sets are unable to fully solve this issue.

\subsection{UV Diagnostics}

In \cite{allen1998} we emphasized the use of UV line ratio diagrams for the
discrimination between shocks and photoionization models for the NLR.  These
diagrams use relatively bright emission lines and reddening-insensitive
ratios. One of the most useful diagrams involves the various ionization stages
of carbon \ion{C}{2}]$\lambda2326$/ \ion{C}{3}] $\lambda$1909
vs.~\ion{C}{4} $\lambda$1550 / \ion{C}{3}] $\lambda$1909. This diagram
is shown in 
Figure~\ref{lrat7} with the new solar abundance n$=1$ cm$^{-3}$ shock models.
There is significantly more overlap between the new shock and shock+precursor
grids than with the DS96 models because the higher magnetic parameter shock
models form a fan-like grid on these axes, which almost completely encompasses
the region covered by the shock+precursor models. The shock+precursor grid is
twisted on these axes making it multi-valued at most positions, hence not good
for assessing shock parameters. This diagram does however remain of great use
for separating shock and photoionization models.

Another useful diagram combines the UV carbon ration of \ion{C}{3}]
$\lambda$1909 / \ion{C}{2}]~$\lambda2326$\ with [\ion{Ne}{3}]~$\lambda$3869 /
[\ion{Ne}{5}]~$\lambda$3426. This has been used by \cite{best2000} and
\cite{Inskip02} to identify shocks and photoionization in the emission line
gas of 3CR and 6C radio galaxies. They show that the ionization state of the
gas varies with radio size such that large radio sources ($>120$ kpc) are
consistent with AGN photoionization, while smaller sources are consistent with
shocks associated with the expanding radio source. Furthermore the extreme gas
kinematics in the smaller radio sources, and detailed consideration of the
energetics and observability of shocks in \cite{Inskip02} supports the
interpretation of shock excitation.

\begin{figure}[h]   
\includegraphics[scale=0.6]{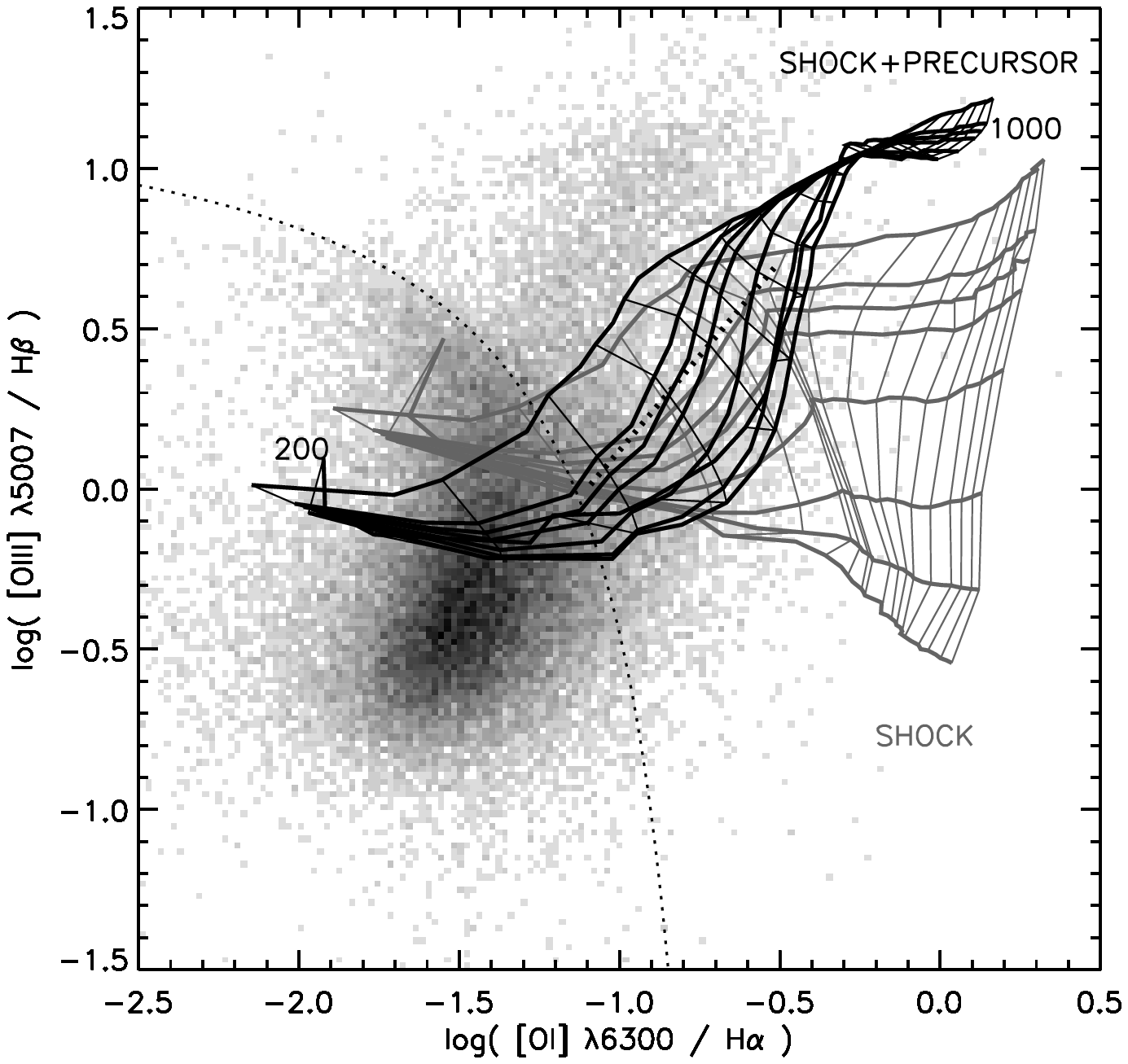} 
\caption{Comparison to SDSS line ratios to the 2$\times$ solar abundance model
grids.  The thin dotted line represents the \cite{kewley2001} starburst/AGN
classification line, while the thick straight dotted line represents the
Seyfert-LINER division described by \cite{kewley2006}.
\label{sdss_compare3}}
\end{figure}

As described in \cite{Inskip02}, there is a continuous sequence of objects
which fall in between the shock dominated and photoionization dominated
groups. In these, both mechanisms are likely to play a role and the balance
between shocks and photoionization is may be linked to the radio source
size. As the definition of the shock and photoionization groups relies on the
coverage of the respective model grids, it is important that these grids are
complete. In Figure~\ref{UV_C_Ne_diag} we reconstruct the same \ion{C}{3}]~
$\lambda$1909 / \ion{C}{2}]~$\lambda2326$ versus
[\ion{Ne}{3}]~$\lambda$3869/ [\ion{Ne}{5}]~$\lambda$3426 diagram as used by
\cite{best2000} and \cite{Inskip02} using the data tabulated in those
papers. We overlay the new shock model grids to demonstrate how these
relate to, and possible help explain, these previously-made
conclusions. For comparison, we also plot the set of new and updated
AGN photoionization models from \cite{groves2004b} as well as the
A$_{M/I}$ models from \cite{bws96} that combine matter- and
ionization-bounded clouds to create the observed sequence.  

The shock and shock+precursor models are those for new solar abundance n$=1$
cm$^{-3}$.  As in the previous diagram, the shock models define a fan shaped
grid, and the shock+precursor grid turns over in both ratios. With the
wider range in shock velocity and magnetic parameter the new models cover a
greater region of this diagram than the DS96 models used by \cite{best2000}
and \cite{Inskip02}. Indeed a number of 6C sources which have \ion{C}{3}] /
$\lambda$1909 \ion{C}{2}]$\lambda2326$\ ratios intermediate between the main
shock and photoionization groups, fall within the higher velocity
shock+precursor grid (6C~1017, 6C~1256, 6C~0943, 6C~1019).

The dust-free and dusty photoionization models
from \cite{groves2004b} demonstrate the other extreme of ionization,
where the emission arises totally from gas excited by the ionizing
radiation emitted by the accretion of gas onto the central black hole
of the AGN.  The dust-free models represent \mapiii\
photoionization models that have been calculated for a sequence of ionization
parameters and densities. The 1Z$_{\odot}$, $\alpha$=-1.4, n$_{\rm
H}$=10$^2$-10$^4$ models shown here behave in a similar way to the
\mapii\ models described in \cite{allen1998} (and used by
\cite{best2000} and \cite{Inskip02}).  The dusty models incorporate the effect
of radiation pressure and result in a stagnation of the ionization parameter
at high values, offering an explanation for the similarity of Seyfert NLR
spectra. Figure~\ref{UV_C_Ne_diag} shows that the dusty photoionization models
can produce line ratios as observed in the larger 3CR and 6C radio
sources. The dusty photoionization model plotted here is the 1Z$_{\odot}$,
$\alpha$=-1.4, n$_{\rm H}$=10$^2$ and n$_{\rm H}$=10$^4$ model, and a
discussion of these models in terms of the \cite{best2000} and
\cite{Inskip02} data, as well as a full
grid of such models on the same diagram can be found in
\cite{groves2004b}.
%, specifically section 3.2 and Figure~5.

\subsection{IR Diagnostics}

The mid- and far-IR emission from galaxies is dominated by the emission of
dust that is heated by UV radiation.  IR spectra contain PAH and dust features
and are also rich in atomic fine structure emission lines. These spectral
features, as observed by ISO and Spitzer offer a wealth of information for
studying the nature of the circumnuclear dust, and the contributions of the
AGN, shocks and starbursts to the total IR emission.

 \cite{genzel1998} introduced infrared diagnostic diagrams using ratios of 
[\ion{Ne}{5}]14.3$\mu$m / [\ion{Ne}{2}]12.8$\mu$m and
[\ion{O}{4}]25.9$\mu$m / [\ion{Ne}{2}]12.8$\mu$m to investigate starbursts,
ULIRGs and AGNs . They used these line ratios, combined with PAH strengths to
place limits on the percentage contribution of AGN and starburst contributions
to the IR emission of ULIRGs.  Mid-IR diagnostics for LINERs are described by
\cite{sturm2006}, where they use the high ionization [\ion{O}{4}]25.9$\mu$m
and [\ion{Ne}{5}]14.3$\mu$m lines to identify the presence of AGN in
$\sim90$\% of LINERs. They confirm the differences in properties between
IR-faint and IR-luminous LINERs, and identify the need to disentangle the
various stellar, \ion{H}{2} region and AGN processes at work in these objects.

\cite{groves2006} investigated the IR emission of the NLR distinct from the
emission of the torus, showing that the NLR emission can contribute up to
$\sim10$\% of the IRAS 25$\mu$m flux.  They emphasize the fact that high
ionization lines like [\ion{Ne}{5}]14.3$\mu$m arises only in the NLR, and
their diagnostic diagram which utilizes only IR lines of neon provides a very
useful indicator of AGN and starburst contributions.

Here we present a set of IR diagnostic diagrams drawn from these previous
works, overlaid with the solar abundance shock and shock + precursor model
grids. Figure~\ref{Ne_IR_diag} 
([\ion{Ne}{5}]14.3$\mu$m / [\ion{Ne}{2}]12.8$\mu$m versus
[\ion{Ne}{3}]15.5$\mu$m / [\ion{Ne}{2}]12.8$\mu$m) combines three different
ionization stages of neon removing any abundance dependence. Figure
\ref{Ne_O_IR_diag}, [\ion{Ne}{5}]14.3$\mu$m / [\ion{Ne}{2}]12.8$\mu$m versus
[\ion{O}{4}]25.9$\mu$m / [\ion{Ne}{2}]12.8$\mu$m, utilizes two high ionization
species of \ion{O}{4} and \ion{Ne}{5} whose ratios to \ion{Ne}{2} provide
strong discriminants of AGN versus starburst processes.  Figure ~\ref{S_Ne}
displays the model results for [\ion{S}{4}]10.44$\mu$m / [\ion{S}{3}]18.7$\mu$m
versus [\ion{Ne}{3}]15.5$\mu$m / [\ion{Ne}{2}]12.8$\mu$m.

\cite{lutz2003} used the ratios of
[\ion{Fe}{2}]26.0$\mu$m / [\ion{O}{4}]25.9$\mu$m versus
[\ion{O}{4}]25.9$\mu$m / [\ion{Ne}{2}]12.8$\mu$m to distinguish between gas
ionized by early-type stars, AGN and shocks.  \cite{sturm2006} showed how the
same diagram separates starburst galaxies, Seyfert galaxies and supernova
remnants.  Figure~\ref{lutz_fig} shows how the shock and shock + precursor
model grids form relatively tight shock velocity sequences in this
diagram. The slope of these sequences is similar to the distribution of
observed ratios in \cite{sturm2006} but the models predict systematically
higher ratios than observed, except for some LINER objects which fall in the
region of the lower velocity shock model grid.

\subsection{Comparison with SDSS Observations}
We now compare the new shock and shock+precursor models to emission line
ratios of AGN and star-forming galaxies observed in the Sloan Digital Sky
Survey (SDSS).  To do this we have adopted the sample of narrow emission line
galaxies compiled by \cite{hao2005}. The classification of the sample into
broad- and narrow-line AGN, and star-forming galaxies is considered in detail
in \cite{hao2005}. Here we choose to use the complete narrow emission line
galaxies sample of $\sim 42000$ sources, and use the AGN/star-forming galaxy
separation schemes of \cite{kewley2001}, \cite{kauffmann2003}, and
\cite{kewley2006} in order to
emphasize where our model grids lie with respect to these different classes of
objects.

\begin{figure}[h]   
\includegraphics[scale=0.6]{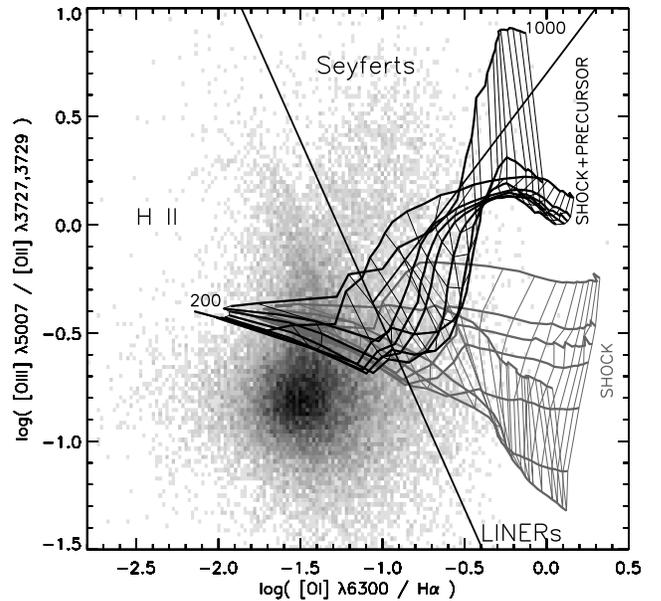} \\
\caption{Comparison to SDSS line ratios to an oxygen excitation diagnostic in
which the regions identified as being excited by \ion{H}{2} regions, Seyferts
and LINERs are labelled. Note that shock-only spectra may characterize some
LINERs, but that the Seyfert galaxies are not well fit by shock+precursor
models on this particular diagram. 
\label{sdss_compare4}}
\end{figure}                

Figure \ref{sdss_compare1} shows the first of the familiar
\citet{veilleux1987} diagnostics; [\ion{O}{3}]$\lambda$5007/H$\beta$
versus [\ion{N}{2}]$\lambda$6583/H$\alpha$. We have overlaid as a
density plot the line ratios observed in the SDSS the narrow emission line
sample.

The distribution of emission line galaxies shows two main branches.  The star
forming galaxy branch sweeps in a curve showing relatively small scatter from
upper left to lower right.  This is largely an abundance sequence, with
abundances increasing towards the lower right \citep{dopita2000,kewley2001b,
Dopita06}. The AGN are mostly distributed in an arm which extends from the
base of the star-forming sequence up towards the upper right of the
diagram. AGN generally have higher values of
[\ion{N}{2}]$\lambda$6583/H$\alpha$, and the distribution extends to higher
values of [\ion{O}{3}]$\lambda$5007/H$\beta$. The dotted and dashed curved
lines represent the \cite{kewley2001} and \cite{kauffmann2003} classification
lines respectively. The \cite{kewley2001} classification is based on the
theoretical maximum line ratios possible by pure stellar photoionization.
Galaxies above this line are most likely dominated by AGN.  The
\cite{kauffmann2003} line is a purely empirical dividing line between pure
star-forming galaxies, and Seyfert-H{\rm II} composite objects.

For comparison, we show in Figure \ref{sdss_compare1} the shock and
shock+precursor grids for the $2\times$solar metallicity models. This choice
of metallicity is driven by the work of \cite{groves2004} and
\cite{groves2006}, who find that super-solar metallicity photoionization
models best reproduce the observed narrow-line ratios in AGN, and the work of
\cite{kauffmann2003}, who find that AGN are typically hosted in galaxies more
massive than $10^{10}$ M$_{\odot}$, and therefore likely to contain high
metallicity gas \citep[e.g.][]{tremonti04}

Figure \ref{sdss_compare1} reveals that both the shock and shock+precursor
models are generally located in the AGN region of this diagram, lying mostly
above the \cite{kewley2001} classification curve.  The shock+precursor grid
overlaps well with the strong AGN or Seyfert branch, lying above the
\cite{kewley2006} LINER/Seyfert dividing line, and extends along the branch
with increasing shock velocity.  The high shock velocity portion of the grid
extends roughly to the limit of the observed distribution before folding over
on itself at the highest velocities. At lower velocities, the shock+precursor
models extend into the ``composites'' region and even into the starformation
or H{\sc II} region below the \cite{kauffmann2003} curve at the lowest
velocities and magnetic parameters.

The lower velocity shock-only models also overlap the AGN branch, but are
predominantly located in the LINER region, simultaneously extending out and
spreading out (with magnetic parameter) at the higher shock velocities.

The emission-line galaxies form similar distributions on the other two
\citet{veilleux1987} diagnostic diagrams; [\ion{O}{3}]$\lambda$5007/H$\beta$
versus [\ion{S}{2}]$\lambda\lambda$6716,6731/H$\alpha$, and the plot of
[\ion{O}{3}]$\lambda$5007/H$\beta$ versus [\ion{O}{1}]$\lambda$6300/H$\alpha$
(see Figures \ref{sdss_compare2} and \ref{sdss_compare3}). The
\cite{kewley2001} division between AGN and \ion{H}{2} region excitation is
shown as dotted lines on these figures.

In these figures, the AGN branches show a bifurcation, that has been used by
\cite{kewley2006} to distinguish between Seyfert and LINER galaxies, and so
provide a general classification scheme for AGN host galaxies. The
Seyfert-LINER dividing line is shown in Figures \ref{sdss_compare2} and
\ref{sdss_compare3} as a thick dotted line.  The shock and shock+precursor
grids overlap the AGN distributions, but tend to fall in the region of the
Seyfert-LINER division. The shock-only models mostly fall in the LINER region
of these diagnostic plots, but tend to extend to extend to higher values of
[\ion{S}{2}]$\lambda\lambda$6716,6731/H$\alpha$, and
[\ion{O}{1}]$\lambda$6300/H$\alpha$ than is observed. There is also more
overlap with the shock+precursor models than in seen in
Figure~\ref{sdss_compare1}.

Figure~\ref{sdss_compare4} shows the
[\ion{O}{3}]$\lambda$5007/[\ion{O}{2}]$\lambda\lambda$3726,3729 versus
[\ion{O}{1}]$\lambda$6300/H$\alpha$ diagnostic diagram over-plotted with the
classification scheme from \cite{kewley2006}.  This diagram provides the
cleanest separation between the two AGN branches. In photoionized plasmas, the
[\ion{O}{3}]$\lambda$5007/[\ion{O}{2}]$\lambda\lambda$3726,3729 is sensitive
to the specific intensity of the radiation field, and the
[\ion{O}{1}]$\lambda$6300/H$\alpha$ ratio to the hardness or spectral index of
the radiation field. As pointed out by \cite{kewley2006} this diagram provides
a simple method for classification, but is more sensitive to reddening
correction of the
[\ion{O}{3}]$\lambda$5007/[\ion{O}{2}]$\lambda\lambda$3726,3729 ratio.  Once
again, the shock+precursor grids fall predominantly close to the Seyfert-LINER
dividing line, and so describe neither the Seyferts or the LINERs particularly
well. The pure shock models are a much better fit to the LINER sequence,
provided that the shock velocities are not too great. The highest shock
velocities have too strong [\ion{O}{1}]$\lambda$6300/H$\alpha$ ratios.

In conclusion, Seyfert galaxies are not well described by the shock +
precursor models. For these, the radiation-pressure dominated photoionization
models \citep{dopita2002, groves2004a, groves2004b} provide a much better
description of the spectra. The LINERs, on the other hand, fit better to shock
only models, and it is likely that at least some of these objects are in fact
shock-excited. A good example of a LINER which is known to be shock-excited is
the nuclear disk of M87 \citep{dopita1997}.

\section{The On-line Library}
\label{online}

The complete electronic files that comprise the \mapiii\ Library of Fast Shock
Models are available via the Shock Model Portal of the MAPPINGS online web
pages at \url{http://cdsweb.u-strasbg.fr/\~{}allen/shock.html}.
The original \mapiii\ output files are available for each model in the
library. We also provide tables of emission line ratios, and column
densities. These files are organized into the various model sets characterized
by a given chemical abundance and pre-shock density and then into velocity
sequences for a given magnetic field. The files are listed on the online pages
using the model names shown in Table \ref{param_table}.

The emission line ratio tables are wavelength ordered lists of the flux ratios
of all the emission lines calculated in \mapiii, and are given with respect to
H$\beta$=1. Each table contains the ratios for a velocity sequence of models,
100-1000 km\ s$^{-1}$, for a given abundance set, density and magnetic
field. Shock and precursor components are tabulated separately and we also
provide tables of the emission line ratios for the combination of
shock+precursor. Each emission line ratio table also includes the absolute
luminosity of the H$\beta$ line in units of Log$_{10}$(erg
cm$^{-2}$~s$^{-1}$).

The column density tables contain the integrated model column densities for
all of the ionic stages of each of the elements in the corresponding abundance
list.  Each table includes the column densities for a velocity sequence of
models in units of cm$^{-2}$. Column densities for the shock and precursor
components are tabulated separately.

In addition to the tabulated model data, we also provide programs for
accessing the library of models, and for generating various plots.  These
programs are coded using IDL\footnote{\url{http://www.ittvis.com/idl/}} and
can be used as interactive graphical user interface widgets, and also via the
IDL command line. The SHOCKPLOT package allows plotting the models on 2-D line
ratio diagrams using any linear combination of line ratios. SHOCKPLOT enables
quick browsing through the many grids of models, and can also be used as a
procedure call from other IDL programs to over-plot model grids on observed
line ratio data.

\section{Summary}

We have presented an extensive library of radiative shock models covering a
wide range of shock velocities, magnetic fields, densities and abundances. The
shock model predictions for the ionizing radiation, temperatures, and
luminosities generated by shocks will be applicable in a wide range of
astrophysical situations. The solar abundance models supersede the models of
DS96, with model code improvements leading to some differences in the
predictions of nitrogen species, and also for models with very low magnetic
parameter. Differences in the input abundances result in the most significant
variations between the grids of models, and the range and sampling of the
shock velocity and magnetic field provides much a more detailed and complete
set of model predictions than previously available.  The extension to higher
shock velocities is important for the analysis of emission line regions of
active galaxies where shocks of this speed are expected in jet-cloud
interactions \citep{saxton2005}. At these shock velocities some of the
commonly used line ratios such as [\ion{O}{3}]$\lambda$5007/H$\beta$ turn over
and do not follow simple extrapolations from the DS96 model grids.

Included as a part of this library, the physical structure of the
shocks and their precursors give insight into how the physical parameters
of the shock lead to the resulting continuum and line emission. We
have included in this paper examples of these, exploring the range of
the parameters and demonstrating the effects of these of the density,
temperature and ionization structure.

We have presented the model grids on a set of standard UV, optical and IR line
ratio diagrams, and we have compared of the new models to the example data
sets of radio galaxies and the SDSS sample of narrow emission line
galaxies. The updated version of the \ion{C}{3}]$\lambda$1909 /
[\ion{C}{2}]$\lambda2326$ vs.~[\ion{Ne}{3}]$\lambda$3869 /
[\ion{Ne}{5}]$\lambda$3426 diagram used by \cite{best2000} and \cite{Inskip02}
for the analysis of emission line regions of radio galaxies, shows that some
of the 6C sources fall within the new higher velocity shock+precursor
grid. This supports the interpretation that the emission line regions in
smaller radio sources are excited by shocks.  The new library of models
presented here, and a more complete set of emission line observations of these
sources will allow a much more detailed analysis of the contributions of shock
and photoionization in these sources.

Large samples of emission line ratios such as now available from the SDSS
provide an extremely valuable resource for classifying and analyzing the
emission line excitation mechanisms in active and star-forming galaxies.  The
comparison of the narrow emission line galaxies sample (compiled by
\cite{hao2005}) to the shock models shows that shocks do predict line ratios
in the observed range, and that shocks may indeed provide the best explanation
for LINER spectra. This library of shock models, combined with detailed grids
of (AGN and star-formation) photoionization models now available should allow a
new statistical approach to analyzing the contributions of shocks,
star-formation and AGN photoionization to excitation of emission lines in
galaxies.

The complete set of electronic files that comprise the library are available
on-line, along with tools to assist in the comparison of observations to the
model predictions line ratio diagrams. Together this library provides
one of the largest databases of radiative shock models and a unique
tool in the interpretation and diagnosis of fast shocks.

\acknowledgments We thank the anonymous referee for comments which have improved the 
clarity of this paper. BG would like to thank the Observatoire Astronomique de
Strasbourg for their financial support and hospitality. MAD acknowledges the
support of both the Australian National University and of the Australian
Research Council (ARC) through his (2002-2006) ARC Australian Federation
Fellowship, and also under the ARC Discovery projects DP0208445 and
DP0342844. Both Dopita and Sutherland acknowledge support under the ARC
Discovery project DP0664434.

\clearpage

\LongTables
\begin{deluxetable}{lrr}
\tablecolumns{3}
\tablewidth{0pc}
\tabletypesize{\small}
\tablecaption{Model Parameters \label{param_table}}
\tablehead{
\colhead{Model name}    &
\colhead{B ($\mu$G)}    &
\colhead{B/$\sqrt{\rm n}$ ($\mu$G\ cm$^{3/2}$)}   }
\startdata
\cutinhead{Solar abundance, n=1.0, v=100,125,...1000}
M\_n1\_b0    & 1.0e-04 &   1.0e-04 \\
M\_n1\_b0.5  & 0.50    &   0.50   \\
M\_n1\_b1    & 1.00     &   1.00    \\
M\_n1\_b2    & 2.00     &   2.00    \\
M\_n1\_be    & 3.23     &   3.23    \\
M\_n1\_b4    & 4.00     &   4.00    \\
M\_n1\_b5    & 5.00     &   5.00    \\
M\_n1\_b10   & 10.0     &   10.0   \\
\cutinhead{Solar $\times$2 abundance, n=1.0, v=100,125,...1000}
R\_n1\_b0    &  1.0e-04 & 1.0e-04     \\
R\_n1\_b0.5  &  0.50    & 0.50   \\
R\_n1\_b1    &  1.00     & 1.00     \\
R\_n1\_b2    &  2.00     & 2.00     \\
R\_n1\_be    &  3.23     & 3.23     \\
R\_n1\_b4    &  4.00     & 4.00     \\
R\_n1\_b5    &  5.00     & 5.00     \\
R\_n1\_b10   &  10.0     & 10.0    \\
\cutinhead{Dopita 2005 abundance, n=1.0, v=100,125,...1000}
J\_n1\_b0    &  1.0e-04  & 1.0e-04 \\
J\_n1\_b0.5  &  0.50     & 0.50   \\
J\_n1\_b1    &  1.00     & 1.00   \\
J\_n1\_b2    &  2.00     & 2.00   \\
J\_n1\_be    &  3.23     & 3.23   \\
J\_n1\_b4    &  4.00     & 4.00   \\
J\_n1\_b5    &  5.00     & 5.00   \\
J\_n1\_b10   &  10.0     & 10.0   \\
\cutinhead{SMC abundance, n=1.0, v=100,125,...1000}
P\_n1\_b0    &  1.0e-04 & 1.0e-04 \\
P\_n1\_b0.5  &  0.50    & 0.50   \\
P\_n1\_b1    &  1.00    & 1.00   \\
P\_n1\_b2    &  2.00    & 2.00   \\
P\_n1\_be    &  3.23    & 3.23   \\
P\_n1\_b4    &  4.00    & 4.00   \\
P\_n1\_b5    &  5.00    & 5.00   \\
P\_n1\_b10   &  10.0    & 10.0   \\
\cutinhead{LMC abundance, n=1.0, v=100,125,...1000}
Q\_n1\_b0    &  1.0e-04 & 1.0e-04 \\
Q\_n1\_b0.5  &  0.50    & 0.50    \\
Q\_n1\_b1    &  1.00    & 1.00    \\
Q\_n1\_b2    &  2.00    & 2.00    \\
Q\_n1\_be    &  3.23    & 3.23    \\
Q\_n1\_b4    &  4.00    & 4.00    \\
Q\_n1\_b5    &  5.00    & 5.00    \\
Q\_n1\_b10   &  10.0    & 10.0    \\
\cutinhead{solar abundance, n=0.01, v=100,125,...1000}
T\_n0.01\_b0.001  &  0.001 &   0.010\\
T\_n0.01\_b0.01   &  0.010 &   0.10  \\
T\_n0.01\_b0.05   &  0.050 &   0.50 \\
T\_n0.01\_b0.1    &  0.10 &    1.00 \\
T\_n0.01\_b0.2    &  0.20 &    2.00 \\
T\_n0.01\_b0.4    &  0.40 &    4.00 \\
T\_n0.01\_b0.5    &  0.50 &    5.00 \\
T\_n0.01\_b1      &   1.0 &    10.0 \\
T\_n0.01\_b10     &   10.0 &   100 \\
\cutinhead{solar abundance, n=0.1, v=100,125,...1000}
U\_n0.1\_b0.0001   & 1.0e-04      &  0.000316 \\
U\_n0.1\_b0.001    & 0.001        &  0.00316  \\
U\_n0.1\_b0.01     & 0.01         &  0.0316   \\
U\_n0.1\_b0.05     & 0.05         &  0.158    \\
U\_n0.1\_b0.1      & 0.10         &  0.316    \\
U\_n0.1\_b0.2      & 0.20         &  0.632    \\
U\_n0.1\_b0.32     & 0.32         &  1.01     \\
U\_n0.1\_b0.4      & 0.40         &  1.26     \\
U\_n0.1\_b0.5      & 0.50         &  1.58     \\
U\_n0.1\_b0.632    & 0.63         &  2.00     \\
U\_n0.1\_b1.0      & 1.00         &  3.16     \\
U\_n0.1\_b1.26     & 1.26         &  4.00     \\
U\_n0.1\_b1.58     & 1.58         &  5.00     \\
U\_n0.1\_b2.0      & 2.00         &  6.32     \\
U\_n0.1\_b3.16     & 3.16         &  10.0     \\
U\_n0.1\_b4.0      & 4.00         &  12.6     \\
U\_n0.1\_b5.0      & 5.00         &  15.8     \\
U\_n0.1\_b10       & 10.0         &  31.6     \\
\cutinhead{solar abundance, n=10.0, v=100,125,...1000}
V\_n10\_b0.001     &  0.001       &  0.000316 \\
V\_n10\_b0.01      &  0.01        &  0.00316  \\
V\_n10\_b0.1       &  0.10        &  0.0316   \\
V\_n10\_b1         &  1.00        &  0.316    \\
V\_n10\_b1.58      &  1.58       &  0.500    \\
V\_n10\_b3.16      &  3.16       &  1.00    \\
V\_n10\_b5         & 5.00     &  1.58     \\
V\_n10\_b6.32      & 6.32     &  2.00     \\
V\_n10\_b10        & 10.0     &  3.16     \\
V\_n10\_b10.2      & 10.2     &  3.23     \\
V\_n10\_b12.65     & 12.6     &  4.00     \\
V\_n10\_b15.8      & 15.8     &  5.00     \\
V\_n10\_b20        & 20.0     &  6.32     \\
V\_n10\_b30        & 30.0     &  9.49     \\
V\_n10\_b40        & 40.0     &  12.6     \\
V\_n10\_b50        & 50.0     &  15.8     \\
V\_n10\_b100       & 100      &  31.6    \\
\cutinhead{solar abundance, n=100, v=100,125,...1000}
L\_n100\_b0.001    &  0.001  &  0.0001      \\
L\_n100\_b0.01     &  0.01   &  0.001       \\
L\_n100\_b0.1      &   0.10  &  0.010       \\
L\_n100\_b1        &  1.00   &  0.10        \\
L\_n100\_be        & 32.3    &  3.23    \\
L\_n100\_b5        &  5.0    &  0.50   \\
L\_n100\_b10       & 10.0    &  1.00   \\
L\_n100\_b20       & 20.0    &  2.00   \\
L\_n100\_b40       & 40.0    &  4.00   \\
L\_n100\_b50       & 50.0    &  5.00   \\
L\_n100\_b100      & 100     &  10.0   \\
\cutinhead{solar abundance, n=1000, v=100,125,...1000}
S\_n1000\_b0.01    &  0.01     &  0.000316 \\
S\_n1000\_b0.1     &   0.1     &  0.00316  \\
S\_n1000\_b1       &   1.0     &  0.0316   \\
S\_n1000\_b5       &  5.00     &  0.158   \\
S\_n1000\_b10      &  10.0     &  0.316   \\
S\_n1000\_b16      &  16.0     &  0.51   \\
S\_n1000\_b32      & 32.00     &  1.01    \\
S\_n1000\_b63      & 63.00     &  2.00    \\
S\_n1000\_b100     & 100.0     &  3.16    \\
S\_n1000\_b126     & 126.0     &  4.00    \\
S\_n1000\_b160     & 160.0     &  5.06    \\
S\_n1000\_b316     & 316.0     &  10.0    \\
S\_n1000\_b1000    & 1000      &  31.6    \\
\enddata 
\end{deluxetable}

\clearpage
\LongTables
\begin{landscape}
\include{tab4}
\end{landscape}

\clearpage
\LongTables
\begin{landscape}
\include{tab5}

\end{landscape}

\LongTables
\include{tab6}

\LongTables
\include{tab7}

\LongTables
\include{tab8}

\end{document}

%% file: tab4.tex
\begin{deluxetable}{lccccccccccccc}
\tablewidth{0pc}
\tabletypesize{\tiny}
\tablecaption{Integrated Column Densities for solar abundance models, n=1.0, B=3.23 \label{coldens_tab3}}
\tablehead{}
%\rotate
\startdata
  V=200 & H & He & C & N & O & Ne & Mg & Al & Si & S & Ar & Ca & Fe\\
  I & 4.771E+18 & 4.745E+17 & 9.812E+13 & 5.334E+14 & 4.286E+15 & 2.376E+14 & 9.506E+13 & 7.978E+11 & 3.265E+12 & 3.178E+11 & 3.083E+12 & 5.587E+12 & 7.278E+12\\
  II & 8.275E+18 & 5.560E+17 & 3.604E+15 & 6.686E+14 & 4.875E+15 & 9.445E+14 & 1.954E+14 & 2.849E+13 & 3.708E+14 & 1.361E+14 & 2.913E+13 & 5.947E+12 & 1.803E+14\\
  III & 0. & 2.444E+17 & 2.138E+14 & 9.468E+12 & 8.270E+13 & 1.640E+14 & 1.223E+14 & 1.836E+12 & 7.185E+12 & 3.884E+13 & 7.223E+12 & 2.812E+12 & 5.884E+13\\
  IV & 0. & 0. & 9.258E+12 & 3.141E+12 & 7.130E+13 & 3.353E+13 & 1.062E+13 & 1.090E+12 & 6.375E+11 & 5.825E+11 & 3.989E+11 & 3.556E+11 & 6.636E+12\\
  V & 0. & 0. & 8.096E+14 & 8.003E+12 & 9.384E+13 & 9.124E+13 & 2.490E+13 & 2.312E+12 & 1.639E+13 & 4.817E+11 & 4.554E+11 & 5.272E+11 & 5.019E+12\\
  VI & 0. & 0. & 1.466E+12 & 2.411E+14 & 2.154E+14 & 9.982E+13 & 3.827E+13 & 2.780E+12 & 5.026E+13 & 8.279E+11 & 5.534E+11 & 8.970E+11 & 1.850E+13\\
  VII & 0. & 0. & 2.691E+07 & 3.517E+09 & 1.479E+15 & 3.349E+13 & 8.895E+12 & 1.143E+12 & 1.346E+13 & 3.285E+13 & 7.710E+11 & 7.485E+11 & 2.034E+13\\
  VIII & 0. & 0. & 0. & 392. & 7.578E+07 & 6.668E+11 & 4.744E+11 & 5.668E+10 & 8.241E+11 & 1.546E+12 & 6.290E+11 & 2.732E+11 & 8.302E+12\\
  IX & 0. & 0. & 0. & 0. & 0. & 9.882E+10 & 4.098E+09 & 4.896E+08 & 5.974E+09 & 1.080E+10 & 5.118E+12 & 4.342E+10 & 5.829E+11\\
  X & 0. & 0. & 0. & 0. & 0. & 9.599E-03 & 1.339E+07 & 5.513E+05 & 6.303E+06 & 1.028E+07 & 2.164E+09 & 4.174E+09 & 1.049E+10\\
  XI & 0. & 0. & 0. & 0. & 0. & 0. & 1.731E+04 & 0.346 & 3.483E-10 & 458. & 1.245E+05 & 2.375E+09 & 1.145E+08\\
  XII & 0. & 0. & 0. & 0. & 0. & 0. & 0. & 0. & 0. & 0. & 0. & 7.354E+03 & 3.437E+05\\
  XIII & 0. & 0. & 0. & 0. & 0. & 0. & 0. & 0. & 0. & 0. & 0. & 0. & 691.\\
  XIV & 0. & 0. & 0. & 0. & 0. & 0. & 0. & 0. & 0. & 0. & 0. & 0. & 0.\\
\hline
  V=500 & H & He & C & N & O & Ne & Mg & Al & Si & S & Ar & Ca & Fe\\
  I & 1.691E+20 & 1.393E+19 & 1.645E+14 & 1.868E+16 & 1.447E+17 & 3.288E+15 & 5.336E+14 & 2.179E+12 & 1.214E+13 & 7.472E+11 & 1.272E+14 & 6.354E+12 & 4.911E+13\\
  II & 3.595E+20 & 5.446E+18 & 6.777E+16 & 4.060E+15 & 2.670E+16 & 1.283E+16 & 4.240E+15 & 5.956E+14 & 7.183E+15 & 1.690E+15 & 5.393E+14 & 4.291E+13 & 4.081E+15\\
  III & 0. & 3.228E+19 & 6.639E+15 & 3.179E+14 & 3.496E+15 & 9.174E+15 & 3.040E+15 & 5.101E+12 & 7.374E+13 & 1.639E+15 & 7.958E+13 & 2.074E+14 & 5.038E+14\\
  IV & 0. & 0. & 5.160E+13 & 8.445E+12 & 1.630E+14 & 2.742E+13 & 7.505E+12 & 3.520E+12 & 2.195E+13 & 3.829E+12 & 5.183E+11 & 1.360E+13 & 1.671E+14\\
  V & 0. & 0. & 1.582E+15 & 8.530E+12 & 3.537E+13 & 3.987E+13 & 1.218E+13 & 1.532E+12 & 8.400E+12 & 4.092E+11 & 2.159E+11 & 1.179E+12 & 2.069E+13\\
  VI & 0. & 0. & 7.674E+15 & 2.059E+15 & 3.511E+14 & 7.120E+13 & 4.045E+13 & 2.269E+12 & 3.645E+13 & 4.455E+11 & 1.948E+11 & 4.021E+11 & 1.167E+13\\
  VII & 0. & 0. & 1.080E+17 & 6.578E+15 & 5.399E+16 & 1.276E+14 & 7.112E+13 & 6.610E+12 & 6.678E+13 & 3.404E+13 & 2.149E+11 & 5.050E+11 & 3.001E+13\\
  VIII & 0. & 0. & 0. & 2.760E+16 & 1.009E+17 & 3.080E+14 & 9.916E+13 & 1.173E+13 & 1.598E+14 & 5.995E+13 & 3.221E+11 & 4.466E+11 & 7.183E+13\\
  IX & 0. & 0. & 0. & 0. & 1.196E+17 & 2.941E+16 & 9.537E+13 & 1.530E+13 & 3.196E+14 & 1.190E+14 & 3.318E+13 & 3.413E+11 & 8.945E+13\\
  X & 0. & 0. & 0. & 0. & 0. & 9.101E+15 & 4.634E+14 & 1.409E+13 & 7.834E+14 & 2.558E+14 & 4.313E+13 & 7.012E+11 & 1.012E+14\\
  XI & 0. & 0. & 0. & 0. & 0. & 6.612E+14 & 1.125E+16 & 6.150E+13 & 2.222E+15 & 3.424E+14 & 8.267E+13 & 3.582E+13 & 1.236E+14\\
  XII & 0. & 0. & 0. & 0. & 0. & 0. & 2.397E+14 & 8.369E+14 & 7.505E+14 & 4.629E+14 & 2.018E+14 & 5.329E+13 & 1.442E+14\\
  XIII & 0. & 0. & 0. & 0. & 0. & 0. & 7.814E+11 & 3.668E+12 & 7.111E+15 & 4.898E+14 & 2.720E+14 & 9.393E+13 & 2.019E+14\\
  XIV & 0. & 0. & 0. & 0. & 0. & 0. & 0. & 2.310E+09 & 6.598E+12 & 1.086E+15 & 2.349E+14 & 1.259E+14 & 2.027E+14\\
  XV & 0. & 0. & 0. & 0. & 0. & 0. & 0. & 0. & 7.767E+08 & 2.389E+15 & 1.205E+14 & 7.731E+13 & 3.087E+14\\
  XVI & 0. & 0. & 0. & 0. & 0. & 0. & 0. & 0. & 0. & 6.630E+10 & 1.146E+14 & 2.961E+13 & 6.709E+14\\
  XVII & 0. & 0. & 0. & 0. & 0. & 0. & 0. & 0. & 0. & 2.104E+05 & 6.890E+13 & 5.633E+12 & 4.895E+15\\
  XVIII & 0. & 0. & 0. & 0. & 0. & 0. & 0. & 0. & 0. & 0. & 5.618E+07 & 1.329E+12 & 6.709E+14\\
  XIX & 0. & 0. & 0. & 0. & 0. & 0. & 0. & 0. & 0. & 0. & 0. & 1.834E+11 & 4.623E+13\\
  XX & 0. & 0. & 0. & 0. & 0. & 0. & 0. & 0. & 0. & 0. & 0. & 2.735E+03 & 1.513E+12\\
  XXI & 0. & 0. & 0. & 0. & 0. & 0. & 0. & 0. & 0. & 0. & 0. & 0. & 1.668E+10\\
  XXII & 0. & 0. & 0. & 0. & 0. & 0. & 0. & 0. & 0. & 0. & 0. & 0. & 1.074E+08\\
  XXIII & 0. & 0. & 0. & 0. & 0. & 0. & 0. & 0. & 0. & 0. & 0. & 0. & 4.021E+05\\
  XXIV & 0. & 0. & 0. & 0. & 0. & 0. & 0. & 0. & 0. & 0. & 0. & 0. & 631.\\
  XXV & 0. & 0. & 0. & 0. & 0. & 0. & 0. & 0. & 0. & 0. & 0. & 0. & 0.748\\
\hline
  V=1000 & H & He & C & N & O & Ne & Mg & Al & Si & S & Ar & Ca & Fe\\
  I & 1.155E+21 & 1.037E+20 & 1.239E+15 & 1.294E+17 & 9.895E+17 & 1.790E+16 & 2.886E+15 & 7.454E+12 & 5.299E+13 & 7.066E+12 & 1.046E+15 & 2.790E+13 & 2.490E+14\\
  II & 5.182E+21 & 2.248E+19 & 4.561E+17 & 1.648E+16 & 1.113E+17 & 8.702E+16 & 3.268E+16 & 3.832E+15 & 4.600E+16 & 1.263E+16 & 3.563E+15 & 3.042E+14 & 2.670E+16\\
  III & 0. & 4.931E+20 & 1.707E+16 & 7.406E+14 & 1.104E+16 & 5.579E+16 & 1.411E+16 & 8.313E+12 & 1.966E+14 & 8.538E+15 & 1.363E+14 & 1.327E+15 & 3.314E+15\\
  IV & 0. & 0. & 1.306E+14 & 2.681E+13 & 6.054E+14 & 1.092E+14 & 2.608E+13 & 6.004E+12 & 1.209E+14 & 2.059E+13 & 9.671E+11 & 5.413E+13 & 3.245E+14\\
  V & 0. & 0. & 7.762E+14 & 8.459E+12 & 7.118E+13 & 3.062E+13 & 9.145E+12 & 3.753E+12 & 9.651E+12 & 1.603E+12 & 1.651E+11 & 9.494E+12 & 4.767E+13\\
  VI & 0. & 0. & 7.591E+15 & 9.491E+14 & 1.772E+14 & 4.771E+13 & 2.410E+13 & 1.630E+12 & 2.513E+13 & 9.129E+11 & 1.695E+11 & 5.678E+11 & 1.022E+13\\
  VII & 0. & 0. & 1.818E+18 & 7.068E+15 & 2.644E+16 & 5.922E+13 & 3.919E+13 & 3.621E+12 & 3.753E+13 & 1.955E+13 & 2.838E+11 & 3.288E+11 & 1.781E+13\\
  VIII & 0. & 0. & 0. & 5.564E+17 & 1.392E+17 & 2.570E+14 & 5.023E+13 & 5.999E+12 & 8.105E+13 & 3.215E+13 & 2.036E+11 & 2.867E+11 & 3.898E+13\\
  IX & 0. & 0. & 0. & 0. & 4.115E+18 & 3.012E+16 & 4.592E+13 & 7.334E+12 & 1.521E+14 & 5.775E+13 & 1.649E+13 & 1.944E+11 & 4.592E+13\\
  X & 0. & 0. & 0. & 0. & 0. & 8.210E+16 & 8.303E+14 & 7.369E+12 & 3.740E+14 & 1.172E+14 & 2.059E+13 & 3.205E+11 & 4.982E+13\\
  XI & 0. & 0. & 0. & 0. & 0. & 5.062E+17 & 3.787E+16 & 1.427E+14 & 1.990E+15 & 1.511E+14 & 3.660E+13 & 1.593E+13 & 5.858E+13\\
  XII & 0. & 0. & 0. & 0. & 0. & 0. & 5.653E+16 & 4.898E+15 & 2.956E+15 & 2.058E+14 & 8.579E+13 & 2.392E+13 & 6.610E+13\\
  XIII & 0. & 0. & 0. & 0. & 0. & 0. & 9.584E+16 & 5.155E+15 & 8.485E+16 & 3.104E+14 & 1.259E+14 & 4.419E+13 & 9.124E+13\\
  XIV & 0. & 0. & 0. & 0. & 0. & 0. & 0. & 4.623E+15 & 6.004E+16 & 3.959E+15 & 1.494E+14 & 8.034E+13 & 9.065E+13\\
  XV & 0. & 0. & 0. & 0. & 0. & 0. & 0. & 0. & 2.796E+16 & 6.014E+16 & 1.810E+14 & 9.811E+13 & 1.469E+14\\
  XVI & 0. & 0. & 0. & 0. & 0. & 0. & 0. & 0. & 0. & 1.489E+16 & 1.731E+15 & 1.292E+14 & 4.983E+14\\
  XVII & 0. & 0. & 0. & 0. & 0. & 0. & 0. & 0. & 0. & 1.691E+15 & 1.480E+16 & 1.803E+14 & 6.596E+15\\
  XVIII & 0. & 0. & 0. & 0. & 0. & 0. & 0. & 0. & 0. & 0. & 1.088E+15 & 1.033E+15 & 7.854E+15\\
  XIX & 0. & 0. & 0. & 0. & 0. & 0. & 0. & 0. & 0. & 0. & 2.805E+13 & 4.900E+15 & 1.071E+16\\
  XX & 0. & 0. & 0. & 0. & 0. & 0. & 0. & 0. & 0. & 0. & 0. & 1.231E+14 & 1.434E+16\\
  XXI & 0. & 0. & 0. & 0. & 0. & 0. & 0. & 0. & 0. & 0. & 0. & 7.223E+11 & 1.477E+16\\
  XXII & 0. & 0. & 0. & 0. & 0. & 0. & 0. & 0. & 0. & 0. & 0. & 0. & 1.458E+16\\
  XXIII & 0. & 0. & 0. & 0. & 0. & 0. & 0. & 0. & 0. & 0. & 0. & 0. & 1.996E+16\\
  XXIV & 0. & 0. & 0. & 0. & 0. & 0. & 0. & 0. & 0. & 0. & 0. & 0. & 1.466E+16\\
  XXV & 0. & 0. & 0. & 0. & 0. & 0. & 0. & 0. & 0. & 0. & 0. & 0. & 1.334E+16\\
  XXVI & 0. & 0. & 0. & 0. & 0. & 0. & 0. & 0. & 0. & 0. & 0. & 0. & 3.648E+12\\
  XXVII & 0. & 0. & 0. & 0. & 0. & 0. & 0. & 0. & 0. & 0. & 0. & 0. & 1.857E+08\\
\hline
\enddata
\end{deluxetable}

%% file: tab5.tex
\begin{deluxetable}{lccccccccccccc}
\tablewidth{0pc}
\tabletypesize{\tiny}
\tablecaption{Integrated Column Densities for the precursor component of solar abundance models, n=1.0, B=3.23 \label{precursor_coldens_tab1}}
\tablehead{}
%\rotate
\startdata
  V=200 & H & He & C & N & O & Ne & Mg & Al & Si & S & Ar & Ca & Fe\\
  I & 8.836E+19 & 8.699E+18 & 4.030E+15 & 9.792E+15 & 7.582E+16 & 8.448E+15 & 1.350E+14 & 2.057E+11 & 1.271E+12 & 1.935E+11 & 1.637E+14 & 3.158E+12 & 1.500E+12\\
  II & 1.198E+20 & 1.127E+19 & 4.675E+16 & 7.516E+15 & 7.372E+16 & 9.629E+15 & 4.418E+15 & 3.741E+14 & 5.617E+15 & 1.536E+15 & 2.199E+14 & 2.774E+13 & 2.343E+15\\
  III & 0. & 3.756E+17 & 2.471E+16 & 5.905E+15 & 2.685E+16 & 7.473E+15 & 3.305E+15 & 1.799E+14 & 1.247E+15 & 1.765E+15 & 3.684E+14 & 2.278E+14 & 1.594E+15\\
  IV & 0. & 0. & 1.052E+14 & 1.471E+14 & 8.204E+14 & 6.430E+13 & 5.709E+13 & 5.960E+13 & 5.124E+14 & 7.515E+13 & 4.014E+12 & 1.520E+13 & 9.311E+14\\
  V & 0. & 0. & 9.281E+11 & 4.191E+11 & 4.332E+12 & 2.582E+11 & 1.521E+11 & 6.680E+11 & 9.974E+12 & 4.006E+11 & 5.185E+09 & 5.757E+11 & 1.105E+13\\
  VI & 0. & 0. & 1.429E+05 & 9.819E+08 & 3.015E+09 & 1.112E+08 & 1.138E+08 & 5.264E+08 & 1.494E+10 & 1.158E+10 & 1.928E+07 & 1.834E+09 & 6.478E+10\\
  VII & 0. & 0. & 0. & 0. & 0. & 0. & 0. & 2.240E+04 & 3.348E+05 & 8.522E+05 & 2.597E+04 & 6.880E+05 & 1.057E+08\\
\hline
  V=500 & H & He & C & N & O & Ne & Mg & Al & Si & S & Ar & Ca & Fe\\
  I & 3.923E+21 & 3.676E+20 & 9.843E+15 & 4.300E+17 & 3.322E+18 & 2.010E+17 & 3.353E+14 & 3.900E+11 & 2.609E+12 & 4.559E+11 & 6.127E+15 & 8.244E+11 & 1.177E+13\\
  II & 2.134E+21 & 1.688E+20 & 1.490E+18 & 4.651E+16 & 2.496E+17 & 2.103E+17 & 1.361E+17 & 1.280E+16 & 1.510E+17 & 4.937E+16 & 9.052E+15 & 4.467E+13 & 9.166E+16\\
  III & 0. & 5.563E+19 & 3.977E+17 & 8.208E+16 & 1.088E+18 & 2.747E+17 & 5.874E+16 & 1.473E+14 & 1.658E+15 & 2.348E+16 & 2.249E+15 & 6.091E+15 & 8.385E+15\\
  IV & 0. & 0. & 1.280E+17 & 9.142E+16 & 3.732E+17 & 4.007E+16 & 1.562E+16 & 1.884E+15 & 1.431E+16 & 1.683E+16 & 3.971E+15 & 1.158E+15 & 3.125E+16\\
  V & 0. & 0. & 1.596E+17 & 1.415E+16 & 8.908E+16 & 1.707E+16 & 1.177E+16 & 1.954E+15 & 2.236E+16 & 4.867E+15 & 5.156E+14 & 5.959E+14 & 8.587E+15\\
  VI & 0. & 0. & 1.412E+16 & 1.475E+16 & 3.122E+16 & 2.027E+15 & 5.601E+15 & 9.322E+14 & 1.997E+16 & 3.629E+15 & 7.041E+13 & 8.656E+13 & 1.861E+15\\
  VII & 0. & 0. & 2.303E+14 & 7.781E+14 & 2.508E+15 & 9.202E+12 & 1.897E+15 & 1.379E+14 & 5.318E+15 & 4.522E+13 & 8.022E+12 & 7.946E+12 & 2.311E+14\\
  VIII & 0. & 0. & 0. & 3.292E+12 & 2.088E+13 & 2.861E+11 & 2.077E+14 & 2.284E+13 & 3.325E+14 & 1.316E+13 & 9.636E+10 & 5.233E+11 & 1.628E+13\\
  IX & 0. & 0. & 0. & 0. & 1.810E+10 & 6.851E+09 & 4.352E+12 & 8.903E+11 & 1.887E+13 & 8.379E+11 & 7.297E+08 & 1.572E+10 & 6.101E+11\\
  X & 0. & 0. & 0. & 0. & 0. & 0. & 5.528E+10 & 8.765E+09 & 3.715E+11 & 1.191E+10 & 5.742E+07 & 8.657E+07 & 2.423E+10\\
  XI & 0. & 0. & 0. & 0. & 0. & 0. & 2.444E+08 & 4.969E+07 & 1.768E+09 & 2.439E+08 & 1.302E+06 & 1.864E+05 & 9.224E+07\\
  XII & 0. & 0. & 0. & 0. & 0. & 0. & 0. & 1.431E+05 & 5.939E+06 & 2.380E+06 & 8.874E+03 & 4.899E+03 & 4.744E+04\\
  XIII & 0. & 0. & 0. & 0. & 0. & 0. & 0. & 0. & 1.078E+04 & 5.912E+03 & 69.6 & 23.8 & 78.0\\
  XIV & 0. & 0. & 0. & 0. & 0. & 0. & 0. & 0. & 0. & 11.5 & 0.132 & 2.108E-02 & 9.953E-02\\
  XV & 0. & 0. & 0. & 0. & 0. & 0. & 0. & 0. & 0. & 0. & 0. & 2.409E-05 & 0.\\
\hline
  V=1000 & H & He & C & N & O & Ne & Mg & Al & Si & S & Ar & Ca & Fe\\
  I & 3.711E+22 & 3.599E+21 & 5.026E+15 & 4.214E+18 & 3.187E+19 & 2.568E+18 & 1.684E+14 & 2.034E+11 & 1.355E+12 & 1.640E+11 & 7.177E+16 & 5.669E+09 & 6.059E+12\\
  II & 1.165E+22 & 6.939E+20 & 1.415E+19 & 2.041E+17 & 1.591E+18 & 1.583E+18 & 1.272E+18 & 1.167E+17 & 1.394E+18 & 5.163E+17 & 7.005E+16 & 9.875E+12 & 8.638E+17\\
  III & 0. & 4.722E+20 & 6.107E+17 & 8.576E+16 & 2.672E+18 & 1.105E+18 & 2.865E+17 & 1.564E+14 & 2.784E+15 & 1.231E+17 & 3.955E+15 & 5.201E+16 & 5.999E+16\\
  IV & 0. & 0. & 3.272E+17 & 2.090E+17 & 1.041E+18 & 1.308E+17 & 4.236E+16 & 3.643E+15 & 1.659E+16 & 4.578E+16 & 1.681E+16 & 4.763E+15 & 1.064E+17\\
  V & 0. & 0. & 1.055E+18 & 8.139E+16 & 1.134E+18 & 2.714E+17 & 3.561E+16 & 5.219E+15 & 5.125E+16 & 1.984E+16 & 2.512E+15 & 1.945E+15 & 1.720E+16\\
  VI & 0. & 0. & 1.284E+18 & 3.894E+17 & 1.525E+18 & 3.198E+17 & 4.021E+16 & 4.439E+15 & 7.136E+16 & 2.582E+16 & 3.146E+15 & 1.468E+15 & 1.826E+16\\
  VII & 0. & 0. & 2.683E+17 & 2.619E+17 & 1.260E+18 & 1.640E+16 & 7.340E+16 & 3.461E+15 & 7.833E+16 & 4.853E+15 & 7.444E+15 & 1.755E+15 & 3.025E+16\\
  VIII & 0. & 0. & 0. & 2.588E+16 & 3.918E+17 & 4.008E+15 & 7.667E+16 & 5.970E+15 & 5.138E+16 & 2.269E+16 & 9.964E+14 & 1.680E+15 & 2.729E+16\\
  IX & 0. & 0. & 0. & 0. & 1.916E+16 & 1.325E+15 & 2.119E+16 & 3.668E+15 & 4.620E+16 & 2.376E+16 & 1.095E+14 & 6.088E+14 & 1.274E+16\\
  X & 0. & 0. & 0. & 0. & 0. & 3.607E+13 & 5.216E+15 & 6.005E+14 & 1.639E+16 & 6.145E+15 & 1.663E+14 & 3.904E+13 & 6.620E+15\\
  XI & 0. & 0. & 0. & 0. & 0. & 1.792E+11 & 4.759E+14 & 7.096E+13 & 1.380E+15 & 2.123E+15 & 7.028E+13 & 1.185E+12 & 4.243E+14\\
  XII & 0. & 0. & 0. & 0. & 0. & 0. & 2.139E+12 & 3.989E+12 & 9.250E+13 & 3.101E+14 & 8.348E+12 & 8.244E+11 & 4.262E+12\\
  XIII & 0. & 0. & 0. & 0. & 0. & 0. & 1.609E+09 & 4.506E+09 & 3.189E+12 & 1.289E+13 & 1.452E+12 & 1.142E+11 & 1.425E+11\\
  XIV & 0. & 0. & 0. & 0. & 0. & 0. & 0. & 0. & 1.494E+09 & 6.780E+11 & 8.089E+10 & 4.113E+09 & 4.040E+09\\
  XV & 0. & 0. & 0. & 0. & 0. & 0. & 0. & 0. & 0. & 1.078E+10 & 1.054E+09 & 1.893E+08 & 6.014E+07\\
  XVI & 0. & 0. & 0. & 0. & 0. & 0. & 0. & 0. & 0. & 0. & 1.465E+07 & 3.442E+06 & 1.810E+04\\
  XVII & 0. & 0. & 0. & 0. & 0. & 0. & 0. & 0. & 0. & 0. & 7.387E+04 & 1.705E+04 & 28.9\\
  XVIII & 0. & 0. & 0. & 0. & 0. & 0. & 0. & 0. & 0. & 0. & 0. & 58.0 & 0.290\\
  XIX & 0. & 0. & 0. & 0. & 0. & 0. & 0. & 0. & 0. & 0. & 0. & 5.533E-02 & 7.180E-04\\
  XX & 0. & 0. & 0. & 0. & 0. & 0. & 0. & 0. & 0. & 0. & 0. & 0. & 5.351E-07\\
  XXI & 0. & 0. & 0. & 0. & 0. & 0. & 0. & 0. & 0. & 0. & 0. & 0. & 5.694E-10\\
\hline
\enddata
\end{deluxetable}

%% file: tab6.tex
\begin{deluxetable}{lrrrrrrrrrr}
\tablewidth{0pc}
\tabletypesize{\scriptsize}
\tablecaption{Line Ratios for shock components of solar abundance
models, n=1.0, B=3.23\label{lr_M_n1_be_s}}  
\tablehead{
\colhead{Line} & 
\multicolumn{10}{c}{Shock Velocity (km\ s$^{-1}$) } \\ \cline{2-11}
\colhead{}     &
\colhead{100 } &
\colhead{200 } &
\colhead{300 } &
\colhead{400 } &
\colhead{500 } &
\colhead{600 } &
\colhead{700 } &
\colhead{800 } &
\colhead{900 } &
\colhead{1000}}
%\rotate
\startdata 
C {\sc iii} $\lambda$977                         &       11.300     & 0.987      & 0.463      & 0.314      & 0.244      & 0.194      & 0.161      & 0.139      & 0.121      & 0.107         \\
N {\sc iii} $\lambda$991                         &       2.919      & 0.620      & 0.348      & 0.225      & 0.174      & 0.140      & 0.116      & 0.099      & 0.085      & 0.074         \\
O {\sc vi} $\lambda\lambda$1032, 1037            &       0.000      & 51.100     & 28.835     & 22.357     & 18.281     & 14.626     & 11.860     & 10.072     & 8.643      & 7.506         \\
Ly\ $\alpha$\ $\lambda$1215                      &       43.570     & 28.720     & 28.820     & 33.040     & 34.360     & 37.500     & 42.800     & 49.360     & 55.710     & 60.000        \\
N {\sc v} $\lambda$1240                          &       0.012      & 1.349      & 0.777      & 0.522      & 0.382      & 0.298      & 0.242      & 0.207      & 0.177      & 0.154         \\
Si {\sc iv} + O {\sc iv} $\lambda$1400           &       4.681      & 2.411      & 1.298      & 0.924      & 0.703      & 0.574      & 0.477      & 0.416      & 0.370      & 0.332         \\
N {\sc iv} $\lambda$1486                         &       0.208      & 0.122      & 0.064      & 0.048      & 0.035      & 0.028      & 0.022      & 0.020      & 0.017      & 0.016         \\
C {\sc iv} $\lambda$1550                         &       15.032     & 4.593      & 2.262      & 1.398      & 1.044      & 0.825      & 0.677      & 0.578      & 0.498      & 0.436         \\
He {\sc ii} $\lambda$1640                        &       0.469      & 0.479      & 1.010      & 2.921      & 2.204      & 1.774      & 1.647      & 1.678      & 1.812      & 1.822         \\
O {\sc iii}$]$ $\lambda$1664                     &       1.735      & 0.711      & 0.425      & 0.304      & 0.252      & 0.214      & 0.182      & 0.162      & 0.149      & 0.137         \\
N {\sc iii} $\lambda$1750                        &       0.400      & 0.084      & 0.051      & 0.038      & 0.032      & 0.027      & 0.024      & 0.022      & 0.021      & 0.021         \\
C {\sc iii}$]$ $\lambda$1909                     &       5.621      & 0.722      & 0.574      & 0.816      & 0.854      & 0.814      & 0.820      & 0.875      & 0.985      & 1.027         \\
C {\sc ii}$]$ $\lambda$2326                      &       2.031      & 0.438      & 0.559      & 0.696      & 0.754      & 0.955      & 1.207      & 1.464      & 1.661      & 1.797         \\
Mg {\sc ii} $\lambda$2800                        &       0.584      & 0.563      & 1.343      & 1.744      & 1.952      & 2.492      & 2.926      & 3.463      & 3.846      & 4.326         \\
$[$Ne {\sc v}$]$ $\lambda$3426                   &       2.074e-5   & 0.288      & 0.156      & 0.099      & 0.077      & 0.064      & 0.055      & 0.048      & 0.043      & 0.038         \\
$[$Ne {\sc iii}$]$ $\lambda$3869                 &       0.612      & 0.309      & 0.694      & 1.253      & 1.312      & 1.414      & 1.577      & 1.767      & 1.956      & 2.058         \\
$[$O {\sc ii}$]$ $\lambda\lambda$3727, 3729      &       10.652     & 5.964      & 11.045     & 14.156     & 14.601     & 14.141     & 14.230     & 14.770     & 15.428     & 15.446        \\
$[$S {\sc ii}$]$ $\lambda\lambda$4067, 4076      &       0.066      & 0.057      & 0.109      & 0.138      & 0.133      & 0.144      & 0.160      & 0.177      & 0.189      & 0.200         \\
$[$O {\sc iii}$]$ $\lambda$4363                  &       0.290      & 0.115      & 0.073      & 0.061      & 0.055      & 0.050      & 0.045      & 0.044      & 0.045      & 0.045         \\
He {\sc ii} $\lambda$4686                        &       0.039      & 0.060      & 0.154      & 0.452      & 0.337      & 0.269      & 0.249      & 0.253      & 0.273      & 0.274         \\
H\ $\beta$\ $\lambda$4861                        &       1.000      & 1.000      & 1.000      & 1.000      & 1.000      & 1.000      & 1.000      & 1.000      & 1.000      & 1.000         \\
$[$O {\sc iii}$]$ $\lambda$5007                  &       3.190      & 1.188      & 1.253      & 2.688      & 3.045      & 3.060      & 3.039      & 3.167      & 3.490      & 3.714         \\
$[$N {\sc i}$]$ $\lambda$5200                    &       0.109      & 0.080      & 0.242      & 0.292      & 0.462      & 0.968      & 1.428      & 1.793      & 2.005      & 2.206         \\
$[$Fe {\sc vii}$]$ $\lambda$6085                 &       0.000      & 0.013      & 0.015      & 0.010      & 0.008      & 0.006      & 0.005      & 0.004      & 0.004      & 0.003         \\
$[$O {\sc i}$]$ $\lambda$6300                    &       0.145      & 0.242      & 0.896      & 1.225      & 1.531      & 2.467      & 3.365      & 4.112      & 4.599      & 4.998         \\
H\ $\alpha$\ $\lambda$6563                       &       3.214      & 3.010      & 2.938      & 2.936      & 2.947      & 2.968      & 2.999      & 3.029      & 3.060      & 3.082         \\
$[$N {\sc ii}$]$ $\lambda$6583                   &       1.641      & 2.024      & 4.155      & 5.103      & 5.056      & 4.871      & 4.870      & 5.040      & 5.265      & 5.269         \\
$[$S {\sc ii}$]$ $\lambda\lambda$6716, 6731      &       1.445      & 1.975      & 2.919      & 3.200      & 3.050      & 3.504      & 4.066      & 4.543      & 4.801      & 5.074         \\
$[$O {\sc ii}$]$ $\lambda\lambda$7318, 7324      &       0.449      & 0.170      & 0.195      & 0.247      & 0.264      & 0.261      & 0.272      & 0.292      & 0.318      & 0.326         \\
$[$S {\sc iii}$]$ $\lambda\lambda$9069, 9532     &       0.751      & 0.840      & 2.421      & 3.227      & 3.896      & 4.871      & 5.672      & 6.311      & 6.763      & 7.046         \\
Br\ $\alpha$\ $\lambda$4.051$\micron$            &       0.112      & 0.096      & 0.087      & 0.084      & 0.085      & 0.086      & 0.087      & 0.087      & 0.086      & 0.086         \\
$[$Ar {\sc vi}$]$ $\lambda$4.530$\micron$        &       0.000      & 0.000      & 0.000      & 0.000      & 9.454e-5   & 7.785e-5   & 6.537e-5   & 5.679e-5   & 4.948e-5   & 4.345e-5      \\
$[$Ar {\sc ii}$]$ $\lambda$6.983$\micron$        &       0.089      & 0.092      & 0.080      & 0.080      & 0.107      & 0.171      & 0.222      & 0.252      & 0.267      & 0.280         \\
$[$Ar {\sc iii}$]$ $\lambda$6.983$\micron$       &       0.038      & 0.114      & 0.300      & 0.339      & 0.334      & 0.302      & 0.273      & 0.256      & 0.248      & 0.239         \\
$[$Ne {\sc vi}$]$ $\lambda$7.652$\micron$        &       0.000      & 0.014      & 0.011      & 0.007      & 0.005      & 0.004      & 0.004      & 0.003      & 0.003      & 0.002         \\
$[$S {\sc iv}$]$ $\lambda$10.44$\micron$         &       0.022      & 0.011      & 0.016      & 0.041      & 0.053      & 0.064      & 0.077      & 0.093      & 0.112      & 0.129         \\
$[$Ne {\sc ii}$]$ $\lambda$12.8$\micron$         &       0.628      & 0.840      & 0.741      & 0.569      & 0.907      & 1.543      & 2.028      & 2.224      & 2.293      & 2.390         \\
$[$Ne {\sc v}$]$ $\lambda$14.5$\micron$          &       0.000      & 0.013      & 0.007      & 0.004      & 0.004      & 0.003      & 0.003      & 0.002      & 0.002      & 0.002         \\
$[$Ne {\sc iii}$]$ $\lambda$15.5$\micron$        &       0.310      & 0.446      & 2.230      & 3.229      & 3.435      & 3.936      & 4.427      & 4.799      & 5.088      & 5.291         \\
$[$S {\sc iii}$]$ $\lambda$18.7$\micron$         &       0.269      & 0.440      & 1.052      & 1.298      & 1.647      & 2.094      & 2.402      & 2.594      & 2.712      & 2.783         \\
$[$O {\sc iv}$]$ $\lambda$25.9$\micron$          &       0.083      & 0.134      & 0.097      & 0.347      & 0.344      & 0.348      & 0.370      & 0.433      & 0.556      & 0.655         \\ \hline
log$_{10}$\ H$\beta$ (erg\ cm$^{-2}$\ s$^{-1}$)     &       -5.317     & -4.444     & -3.996     & -3.675     & -3.469     & -3.303     & -3.155     & -3.032     & -2.918     & -2.814        \\
\enddata 
\end{deluxetable}

%% file: tab7.tex
\begin{deluxetable}{lrrrrrrrrr}
\tablewidth{0pc}
\tabletypesize{\scriptsize}
\tablecaption{Line Ratios for precursor components of solar abundance models, n=1.0, B=3.23\label{lr_M_n1_be_p}}
\tablehead{
\colhead{Line} & 
\multicolumn{9}{c}{Shock Velocity (km\ s$^{-1}$) } \\ \cline{2-10}
\colhead{}     &
\colhead{200 } &
\colhead{300 } &
\colhead{400 } &
\colhead{500 } &
\colhead{600 } &
\colhead{700 } &
\colhead{800 } &
\colhead{900 } &
\colhead{1000}}
%\rotate
\startdata 
C {\sc iii} $\lambda$977                       & 0.000      & 7.075e-5   & 0.001      & 0.009      & 0.075      & 0.200      & 0.275      & 0.288      & 0.285           \\
N {\sc iii} $\lambda$991                       & 0.000      & 1.013e-5   & 0.000      & 0.002      & 0.015      & 0.038      & 0.053      & 0.050      & 0.050           \\
O {\sc vi} $\lambda\lambda$1032, 1037          & 0.000      & 0.000      & 0.000      & 0.011      & 0.653      & 4.650      & 12.084     & 22.156     & 30.910          \\
Ly\ $\alpha$\ $\lambda$1215                    & 19.540     & 19.610     & 20.190     & 22.220     & 26.280     & 30.870     & 34.090     & 37.440     & 41.390          \\
N {\sc v} $\lambda$1240                        & 0.000      & 0.000      & 6.491e-5   & 0.014      & 0.217      & 0.673      & 1.055      & 1.333      & 1.468           \\
Si {\sc iv} + O {\sc iv} $\lambda$1400         & 1.099e-5   & 0.002      & 0.012      & 0.117      & 0.689      & 1.396      & 1.831      & 2.089      & 2.464           \\
N {\sc iv} $\lambda$1486                       & 0.000      & 0.000      & 0.002      & 0.047      & 0.241      & 0.431      & 0.515      & 0.547      & 0.570           \\
C {\sc iv} $\lambda$1550                       & 0.000      & 0.003      & 0.035      & 1.206      & 5.764      & 10.272     & 12.136     & 12.485     & 12.526          \\
He {\sc ii} $\lambda$1640                      & 0.263      & 1.425      & 3.617      & 2.246      & 2.091      & 2.358      & 2.625      & 3.033      & 3.162           \\
O {\sc iii}$]$ $\lambda$1664                   & 9.984e-5   & 0.005      & 0.030      & 0.150      & 0.558      & 1.040      & 1.430      & 1.638      & 1.892           \\
N {\sc iii} $\lambda$1750                      & 5.054e-5   & 0.001      & 0.006      & 0.021      & 0.054      & 0.076      & 0.080      & 0.065      & 0.058           \\
C {\sc iii}$]$ $\lambda$1909                   & 0.004      & 0.085      & 0.298      & 1.022      & 2.113      & 2.806      & 2.860      & 2.630      & 2.408           \\
C {\sc ii}$]$ $\lambda$2326                    & 0.012      & 0.036      & 0.063      & 0.137      & 0.210      & 0.266      & 0.293      & 0.324      & 0.381           \\
Mg {\sc ii} $\lambda$2800                      & 0.107      & 0.290      & 0.410      & 0.716      & 0.923      & 1.020      & 1.032      & 1.047      & 1.111           \\
$[$Ne {\sc v}$]$ $\lambda$3426                 & 0.000      & 0.001      & 0.004      & 0.147      & 0.686      & 1.484      & 2.089      & 2.631      & 2.825           \\
$[$Ne {\sc iii}$]$ $\lambda$3869               & 0.052      & 0.310      & 0.569      & 1.310      & 2.070      & 2.471      & 2.636      & 2.584      & 2.569           \\
$[$O {\sc ii}$]$ $\lambda\lambda$3727, 3729    & 1.119      & 1.241      & 1.065      & 1.160      & 1.316      & 1.416      & 1.415      & 1.416      & 1.526           \\
$[$S {\sc ii}$]$ $\lambda\lambda$4067, 4076    & 0.005      & 0.006      & 0.007      & 0.014      & 0.018      & 0.022      & 0.026      & 0.030      & 0.036           \\
$[$O {\sc iii}$]$ $\lambda$4363                & 0.000      & 0.009      & 0.034      & 0.120      & 0.298      & 0.448      & 0.549      & 0.586      & 0.638           \\
He {\sc ii} $\lambda$4686                      & 0.045      & 0.234      & 0.603      & 0.340      & 0.300      & 0.327      & 0.357      & 0.407      & 0.420           \\
H\ $\beta$\ $\lambda$4861                      & 1.000      & 1.000      & 1.000      & 1.000      & 1.000      & 1.000      & 1.000      & 1.000      & 1.000           \\
$[$O {\sc iii}$]$ $\lambda$5007                & 0.661      & 4.594      & 7.639      & 16.270     & 23.960     & 26.920     & 27.860     & 26.760     & 26.760          \\
$[$N {\sc i}$]$ $\lambda$5200                  & 0.004      & 0.021      & 0.057      & 0.142      & 0.203      & 0.255      & 0.294      & 0.343      & 0.420           \\
$[$Fe {\sc vii}$]$ $\lambda$6085               & 0.000      & 0.000      & 7.858e-5   & 0.003      & 0.021      & 0.078      & 0.142      & 0.199      & 0.205           \\
$[$O {\sc i}$]$ $\lambda$6300                  & 0.013      & 0.062      & 0.131      & 0.316      & 0.463      & 0.571      & 0.633      & 0.708      & 0.829           \\
H\ $\alpha$\ $\lambda$6563                     & 3.011      & 2.961      & 2.943      & 2.873      & 2.876      & 2.899      & 2.922      & 2.947      & 2.976           \\
$[$N {\sc ii}$]$ $\lambda$6583                 & 0.650      & 0.465      & 0.325      & 0.375      & 0.442      & 0.475      & 0.460      & 0.443      & 0.449           \\
$[$S {\sc ii}$]$ $\lambda\lambda$6716, 6731    & 0.244      & 0.165      & 0.180      & 0.334      & 0.426      & 0.526      & 0.613      & 0.715      & 0.855           \\
$[$O {\sc ii}$]$ $\lambda\lambda$7318, 7324    & 0.006      & 0.014      & 0.016      & 0.019      & 0.024      & 0.027      & 0.027      & 0.027      & 0.029           \\
$[$S {\sc iii}$]$ $\lambda\lambda$9069, 9532   & 1.056      & 1.299      & 0.965      & 1.014      & 1.025      & 1.003      & 0.950      & 0.903      & 0.922           \\
Br\ $\alpha$\ $\lambda$4.051$\micron$          & 0.102      & 0.094      & 0.092      & 0.081      & 0.076      & 0.073      & 0.072      & 0.070      & 0.069           \\
$[$Ar {\sc vi}$]$ $\lambda$4.530$\micron$      & 0.000      & 3.302e-5   & 0.001      & 0.003      & 0.008      & 0.019      & 0.028      & 0.033      & 0.034           \\
$[$Ar {\sc ii}$]$ $\lambda$6.983$\micron$      & 0.013      & 0.006      & 0.008      & 0.015      & 0.018      & 0.022      & 0.024      & 0.027      & 0.031           \\
$[$Ar {\sc iii}$]$ $\lambda$6.983$\micron$     & 0.264      & 0.282      & 0.159      & 0.130      & 0.109      & 0.090      & 0.075      & 0.060      & 0.051           \\
$[$Ne {\sc vi}$]$ $\lambda$7.652$\micron$      & 0.000      & 0.000      & 0.004      & 0.031      & 0.138      & 0.373      & 0.633      & 0.929      & 1.153           \\
$[$S {\sc iv}$]$ $\lambda$10.44$\micron$       & 0.144      & 1.993      & 3.005      & 3.303      & 3.149      & 2.851      & 2.588      & 2.298      & 2.167           \\
$[$Ne {\sc ii}$]$ $\lambda$12.8$\micron$       & 0.357      & 0.091      & 0.066      & 0.107      & 0.116      & 0.133      & 0.155      & 0.181      & 0.216           \\
$[$Ne {\sc v}$]$ $\lambda$14.5$\micron$        & 1.702e-5   & 0.007      & 0.036      & 0.098      & 0.196      & 0.302      & 0.363      & 0.417      & 0.419           \\
$[$Ne {\sc iii}$]$ $\lambda$15.5$\micron$      & 1.147      & 2.148      & 2.006      & 2.321      & 2.390      & 2.281      & 2.143      & 1.947      & 1.849           \\
$[$S {\sc iii}$]$ $\lambda$18.7$\micron$       & 0.848      & 0.667      & 0.367      & 0.334      & 0.315      & 0.300      & 0.284      & 0.273      & 0.283           \\
$[$O {\sc iv}$]$ $\lambda$25.9$\micron$        & 0.342      & 5.524      & 15.890     & 12.550     & 12.190     & 11.330     & 10.150     & 9.131      & 8.604           \\ \hline
log$_{10}$\ H$\beta$ (erg\ cm$^{-2}$\ s$^{-1}$)     & -4.622     & -4.130     & -3.769     & -3.599     & -3.464     & -3.359     & -3.261     & -3.164     & -3.082          \\
\enddata 
\end{deluxetable}

%% file: tab8.tex
\begin{deluxetable}{lrrrrrrrrr}
\tablewidth{0pc}
\tabletypesize{\scriptsize}
\tablecaption{Line Ratios for shock + precursor components of solar abundance models, n=1.0, B=3.23\label{lr_M_n1_be_sp}}
\tablehead{
\colhead{Line} & 
\multicolumn{9}{c}{Shock Velocity (km\ s$^{-1}$) } \\ \cline{2-10}
\colhead{}     &
\colhead{200 } &
\colhead{300 } &
\colhead{400 } &
\colhead{500 } &
\colhead{600 } &
\colhead{700 } &
\colhead{800 } &
\colhead{900 } &
\colhead{1000}}
%\rotate
\startdata 
C {\sc iii} $\lambda$977                                & 0.593      & 0.267      & 0.174      & 0.144      & 0.146      & 0.176      & 0.189      & 0.182      & 0.169         \\
N {\sc iii} $\lambda$991                                & 0.373      & 0.201      & 0.125      & 0.101      & 0.089      & 0.086      & 0.082      & 0.072      & 0.066         \\
O {\sc vi} $\lambda\lambda$1032, 1037                   & 30.700     & 16.634     & 12.392     & 10.507     & 8.917      & 9.088      & 10.818     & 13.538     & 15.709        \\
Ly\ $\alpha$\ $\lambda$1215                             & 25.055     & 24.923     & 27.313     & 29.194     & 32.916     & 38.213     & 43.695     & 49.091     & 53.477        \\
N {\sc v} $\lambda$1240                                 & 0.810      & 0.448      & 0.289      & 0.225      & 0.265      & 0.408      & 0.521      & 0.596      & 0.615         \\
Si {\sc iv} + O {\sc iv} $\lambda$1400                  & 1.448      & 0.750      & 0.517      & 0.454      & 0.621      & 0.830      & 0.941      & 0.993      & 1.079         \\
N {\sc iv} $\lambda$1486                                & 0.073      & 0.037      & 0.028      & 0.040      & 0.115      & 0.180      & 0.203      & 0.209      & 0.210         \\
C {\sc iv} $\lambda$1550                                & 2.759      & 1.306      & 0.791      & 1.113      & 2.843      & 4.366      & 4.866      & 4.841      & 4.674         \\
He {\sc ii} $\lambda$1640                               & 0.393      & 1.186      & 3.231      & 2.222      & 1.904      & 1.920      & 2.029      & 2.254      & 2.292         \\
O {\sc iii}$]$ $\lambda$1664                            & 0.427      & 0.247      & 0.182      & 0.208      & 0.354      & 0.511      & 0.632      & 0.688      & 0.752         \\
N {\sc iii} $\lambda$1750                               & 0.051      & 0.030      & 0.024      & 0.027      & 0.038      & 0.044      & 0.044      & 0.037      & 0.034         \\
C {\sc iii}$]$ $\lambda$1909                            & 0.436      & 0.367      & 0.586      & 0.925      & 1.345      & 1.583      & 1.612      & 1.581      & 1.511         \\
C {\sc ii}$]$ $\lambda$2326                             & 0.268      & 0.338      & 0.414      & 0.492      & 0.650      & 0.845      & 1.029      & 1.176      & 1.301         \\
Mg {\sc ii} $\lambda$2800                               & 0.381      & 0.898      & 1.149      & 1.426      & 1.851      & 2.193      & 2.561      & 2.832      & 3.199         \\
$[$Ne {\sc v}$]$ $\lambda$3426                          & 0.173      & 0.091      & 0.056      & 0.107      & 0.318      & 0.604      & 0.805      & 0.981      & 1.015         \\
$[$Ne {\sc iii}$]$ $\lambda$3869                        & 0.207      & 0.532      & 0.948      & 1.311      & 1.682      & 1.921      & 2.089      & 2.184      & 2.237         \\
$[$O {\sc ii}$]$ $\lambda\lambda$3727, 3729             & 4.030      & 6.897      & 8.321      & 8.882      & 8.902      & 9.304      & 9.815      & 10.352     & 10.567        \\
$[$S {\sc ii}$]$ $\lambda\lambda$4067, 4076             & 0.036      & 0.066      & 0.080      & 0.082      & 0.092      & 0.107      & 0.121      & 0.132      & 0.142         \\
$[$O {\sc iii}$]$ $\lambda$4363                         & 0.070      & 0.046      & 0.049      & 0.083      & 0.151      & 0.200      & 0.231      & 0.241      & 0.253         \\
He {\sc ii} $\lambda$4686                               & 0.054      & 0.188      & 0.519      & 0.339      & 0.282      & 0.279      & 0.292      & 0.322      & 0.325         \\
H\ $\beta$\ $\lambda$4861                               & 1.000      & 1.000      & 1.000      & 1.000      & 1.000      & 1.000      & 1.000      & 1.000      & 1.000         \\
$[$O {\sc iii}$]$ $\lambda$5007                         & 0.978      & 2.667      & 4.895      & 8.672      & 11.598     & 12.220     & 12.328     & 11.920     & 11.792        \\
$[$N {\sc i}$]$ $\lambda$5200                           & 0.050      & 0.148      & 0.187      & 0.326      & 0.655      & 0.977      & 1.237      & 1.403      & 1.580         \\
$[$Fe {\sc vii}$]$ $\lambda$6085                        & 0.008      & 0.009      & 0.005      & 0.005      & 0.012      & 0.033      & 0.056      & 0.075      & 0.074         \\
$[$O {\sc i}$]$ $\lambda$6300                           & 0.151      & 0.543      & 0.737      & 1.014      & 1.648      & 2.291      & 2.821      & 3.189      & 3.537         \\
H\ $\alpha$\ $\lambda$6563                              & 3.010      & 2.948      & 2.939      & 2.916      & 2.930      & 2.961      & 2.989      & 3.019      & 3.045         \\
$[$N {\sc ii}$]$ $\lambda$6583                          & 1.476      & 2.593      & 2.973      & 3.064      & 3.062      & 3.180      & 3.341      & 3.518      & 3.580         \\
$[$S {\sc ii}$]$ $\lambda\lambda$6716, 6731             & 1.284      & 1.754      & 1.854      & 1.895      & 2.246      & 2.705      & 3.085      & 3.321      & 3.595         \\
$[$O {\sc ii}$]$ $\lambda\lambda$7318, 7324             & 0.104      & 0.118      & 0.144      & 0.160      & 0.164      & 0.177      & 0.194      & 0.213      & 0.222         \\
$[$S {\sc iii}$]$ $\lambda\lambda$9069, 9532            & 0.926      & 1.946      & 2.219      & 2.670      & 3.300      & 3.877      & 4.322      & 4.640      & 4.899         \\
Br\ $\alpha$\ $\lambda$4.051$\micron$                   & 0.099      & 0.090      & 0.088      & 0.083      & 0.082      & 0.082      & 0.081      & 0.080      & 0.080         \\
$[$Ar {\sc vi}$]$ $\lambda$4.530$\micron$               & 0.000      & 0.000      & 0.001      & 0.001      & 0.003      & 0.008      & 0.011      & 0.012      & 0.012         \\
$[$Ar {\sc ii}$]$ $\lambda$6.983$\micron$               & 0.060      & 0.049      & 0.048      & 0.068      & 0.108      & 0.145      & 0.167      & 0.180      & 0.192         \\
$[$Ar {\sc iii}$]$ $\lambda$6.983$\micron$              & 0.174      & 0.292      & 0.259      & 0.247      & 0.223      & 0.203      & 0.189      & 0.179      & 0.173         \\
$[$Ne {\sc vi}$]$ $\lambda$7.652$\micron$               & 0.008      & 0.007      & 0.005      & 0.016      & 0.059      & 0.146      & 0.237      & 0.338      & 0.406         \\
$[$S {\sc iv}$]$ $\lambda$10.44$\micron$                & 0.064      & 0.852      & 1.362      & 1.436      & 1.324      & 1.143      & 1.018      & 0.904      & 0.844         \\
$[$Ne {\sc ii}$]$ $\lambda$12.8$\micron$                & 0.647      & 0.466      & 0.345      & 0.567      & 0.960      & 1.299      & 1.457      & 1.528      & 1.628         \\
$[$Ne {\sc v}$]$ $\lambda$14.5$\micron$                 & 0.008      & 0.007      & 0.018      & 0.044      & 0.082      & 0.117      & 0.136      & 0.152      & 0.148         \\
$[$Ne {\sc iii}$]$ $\lambda$15.5$\micron$               & 0.726      & 2.195      & 2.684      & 2.961      & 3.304      & 3.602      & 3.814      & 3.950      & 4.085         \\
$[$S {\sc iii}$]$ $\lambda$18.7$\micron$                & 0.603      & 0.889      & 0.883      & 1.088      & 1.367      & 1.594      & 1.737      & 1.828      & 1.907         \\
$[$O {\sc iv}$]$ $\lambda$25.9$\micron$                 & 0.217      & 2.393      & 7.275      & 5.538      & 5.186      & 4.584      & 4.038      & 3.662      & 3.441         \\ \hline 
log$_{10}$\ H$\beta$ (erg\ cm$^{-2}$\ s$^{-1}$)         & -4.223     & -3.757     & -3.418     & -3.228     & -3.075     & -2.944     & -2.831     & -2.723     & -2.627        \\
\enddata 
\end{deluxetable}